\documentclass[aps,twocolumn,superscriptaddress,floatfix]{revtex4-2}
\usepackage{graphicx}
\usepackage{latexsym}
\usepackage{amsmath}
\usepackage{amsfonts}
\usepackage{amssymb}
\usepackage{bm}
\usepackage{txfonts}
\newcommand{\fig}[2]{\includegraphics[width=#1]{#2}}
\newcommand{\llangle}{{\langle\!\langle}}
\newcommand{\rrangle}{{\rangle\!\rangle}}
\def\avs{{$A$V$_3$Sb$_5$}}
\def\bk{{\textbf{k}}}
\def\bM{{\textbf{M}}}
\def\br{{\textbf{r}}}
\def\bq{{\textbf{q}}}
\def\bQ{{\textbf{Q}}}
\def\bG{{\textbf{G}}}
\def\ba{{{\bm a}}}

\begin{document}
\title{Inter-Layer Correlation of Loop Current Charge Density Wave on the Bilayer Kagom\'e Lattice}

\author{Jin-Wei Dong}
\affiliation{Anhui Province Key Laboratory of Condensed Matter Physics at Extreme Conditions, High Magnetic Field Laboratory, Chinese Academy of Sciences, Hefei 230031, China}

\author{Yu-Han Lin}
\affiliation{CAS Key Laboratory of Theoretical Physics, Institute of Theoretical Physics, Chinese Academy of Sciences, Beijing 100190, China}
\affiliation{School of Physical Sciences, University of Chinese Academy of Sciences, Beijing 100049, China}

\author{Ruiqing Fu}
\affiliation{CAS Key Laboratory of Theoretical Physics, Institute of Theoretical Physics, Chinese Academy of Sciences, Beijing 100190, China}
\affiliation{School of Physical Sciences, University of Chinese Academy of Sciences, Beijing 100049, China}

\author{Gang Su}
\affiliation{School of Physical Sciences, University of Chinese Academy of Sciences, Beijing 100049, China}
\affiliation{CAS Center for Excellence in Topological Quantum Computation, University of Chinese Academy of Sciences, Beijing 100049, China}
\affiliation{Kavli Institute for Theoretical Sciences, University of Chinese Academy of Sciences, Beijing 100190, China}

\author{Ziqiang Wang}
\thanks{Corresponding author: wangzi@bc.edu}
\affiliation{Department of Physics, Boston College, Chestnut Hill, MA 02467, USA}

\author{Sen Zhou}
\thanks{Corresponding author: zhousen@itp.ac.cn}
\affiliation{CAS Key Laboratory of Theoretical Physics, Institute of Theoretical Physics, Chinese Academy of Sciences, Beijing 100190, China}
\affiliation{School of Physical Sciences, University of Chinese Academy of Sciences, Beijing 100049, China}
\affiliation{CAS Center for Excellence in Topological Quantum Computation, University of Chinese Academy of Sciences, Beijing 100049, China}

\begin{abstract}
Loop current order has been suggested as a promising candidate for the spontaneous time-reversal symmetry breaking $2a_0 \times 2a_0$ charge density wave (CDW) revealed in vanadium-based kagom\'e metals \avs\ ($A$ = K, Rb, Cs) near van Hove filling $n_\text{vH} = 5/12$.
Weak-coupling analyses and mean field calculations have demonstrated that nearest-neighbor Coulomb repulsion $V_1$ and next-nearest-neighbor Coulomb repulsion $V_2$ drives, respectively, real and imaginary bond-ordered CDW, with the latter corresponding to time-reversal symmetry breaking loop current CDW. 
It is important to understand the inter-layer correlation of these bond-ordered CDWs and its consequences in the bulk kagom\'e materials.
To provide physical insights, we investigate in this paper the $c$-axis stacking of them, loop current CDW in particular, on the minimal bilayer kagom\'e lattice.
The bare susceptibilities for stacking of real and imaginary bond orders are calculated for the free electrons on the bilayer kagom\'e lattice with inter-layer coupling $t_\perp=0.2t$, which splits the van Hove filling to $n_{+\text{vH}}=4.64/12$ and $n_{-\text{vH}}=5.44/12$.
While real and imaginary bond-ordered CDWs are still favored, respectively, by $V_1$ and $V_2$, their inter-layer coupling is sensitive to band filling $n$.
They tend to stack symmetrically near $n_{\pm\text{vH}}$ with identical bond orders in the two layers and give rise to a $2a_0 \times 2a_0 \times 1c_0$ CDW.
On the other hand, they prefer to stack antisymmetrically around $n_\text{vH}$ with opposite bond orders in the two layers and lead to a $2a_0 \times 2a_0 \times 2c_0$ CDW.
The concrete bilayer $t$-$t_\perp$-$V_1$-V$_2$ model is then studied.
We obtain the mean-field ground states and determine the inter-layer coupling as a function of band filling at various interactions.
The nontrivial topological properties of loop current CDWs are studied and possible connections to experiments are discussed.
\end{abstract}
\maketitle

\section{Introduction}
The vanadium-based kagom\'e metals \avs\ ($A$ = K, Rb, Cs) \cite{Ortiz-PRM19} have generated increasing interest in the community due to the discovery of a cascade of spontaneous symmetry-breaking and correlated electronic states.
With reducing temperatures, all \avs\ undergo charge density wave (CDW) transitions \cite{YinJiaxin-21NatMater, HeZhao-21Nat, Ortiz-PRX21, ChenXH-PRX21, Nana-PRB21, Ortiz-PRB21, LiHong-NatPhys22, Ortiz-PRL20} below $T_\text{cdw}$ (78-103 K) and superconducting (SC) transitions \cite{Ortiz-PRL20, Ortiz-PRM21, YinQiangwei-CPL21, ChenHui-Nat21} below $T_c$ (0.9-2.8 K).
Both CDW and SC possess highly nontrivial and intriguing properties.
The CDW states have $2a_0 \times 2a_0$ charge order in the kagom\'e plane stacked along the $c$-axis \cite{YinJiaxin-21NatMater, HeZhao-21Nat, ChenHui-Nat21, ChenXH-PRX21, YanBinghai-PRL21, LiHaoxiang-PRX21, Ortiz-PRX21,Nana-PRB21}, and breaks rotation symmetry \cite{ChenXinHui-Nat2022,LiHong-NatPhys22,JiangZhicheng-NanoLett23,Xing-arxiv23}.
Despite the absence of local moment \cite{Kenney-IOP21} and itinerant magnetism \cite{Ortiz-PRM19,Ortiz-PRM21}, the CDW states exhibit spontaneous time-reversal symmetry (TRS) breaking, as evidenced by scanning tunneling microscopy \cite{YinJiaxin-21NatMater, Xing-arxiv23}, muon relaxation \cite{Mielke-Nat22, YuLi-arxiv21}, optical Kerr rotation \cite{WuQiong-PRB22, XuYishuai-NatPhys22, HuYajian-arxiv22}, circular dichroism \cite{XuYishuai-NatPhys22}, nonlinear transport \cite{GuoChunyu-Nat22}, alongside a giant anomalous Hall effect \cite{YangShuoYing-SciAdv20, ChenXH-PRB21}.
Below $T_c$, a novel pair density wave (PDW) order with ${4 \over 3} a_0 \times {4 \over 3} a_0$ periodicity has been observed in the SC state \cite{ChenHui-Nat21}.
Remarkably, charge-4$e$ and charge-6$e$ flux quantization were observed in the magnetoresistance oscillations in mesoscopic ring structures of thin flake CsV$_3$Sb$_5$ samples, providing evidence for novel higher-charge SC \cite{GeJun-arxiv22}.

One of the central questions in \avs\ is the nature of the nonmagnetic TRS breaking CDW.
It has been conjectured \cite{YinJiaxin-21NatMater} that a promising candidate is an exotic loop-current (LC) order\cite{FengXilin-SciBull21, Denner-PRL21, SZ-NC21}, a long-sought after quantum state also relevant for the pseudogap phase in the high-T$_c$ cuprates\cite{Affleck-PRB1988, Varma-PRB1997,Varma-PRL1999, Chakravarty-PRB01, WenXG-RevMod06} and the quantum anomalous Hall insulators \cite{Haldane-PRL88}. 
Density functional theory (DFT) calculations revealed that the stable ground state of the \avs\ crystal has a breathing structural distortion of the kagom\'e lattice with $2a_0 \times 2a_0$ in-plane periodicity \cite{YanBinghai-PRL21}, corresponding to the Star-of-David (SD) or inverse SD (ISD) configuration, stacked with period $2c_0$ along the $c$-axis, in overall agreement with the observations of experiments \cite{YinJiaxin-21NatMater,HeZhao-21Nat,Ortiz-PRX21,ChenXH-PRX21,Wenzel-PRB22,LiHaoxiang-PRX21,Ratcliff-PRM21,XieYaofeng-PRB22,WuShangfei-PRB22,LiuGan-NC22}.
However, electron-phonon coupling cannot produce a CDW state that spontaneously breaks TRS \cite{YanBinghai-PRL21,Ferrari-PRB22}. 
Therefore, enormous efforts have devoted to investigate the possible origin of LC and its realization in microscopic models incorporating electron correlations on the kagom\'e lattice.

\begin{figure*}
\begin{center}
\fig{7.in}{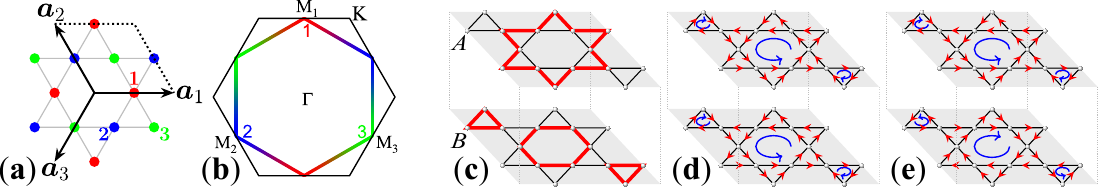}
\caption{(a) Kagom\'e lattice with three sublattices denoted by red (1), blue (2), and green (3) circles. 
The two basis lattice vectors are $\ba_1=(1,0)$, $\ba_2=(-{1\over 2}, {\sqrt{3}\over 2})$, and the third direction follows $\ba_3=-\ba_1-\ba_2$, with the lattice constant $a_0 \equiv 1$.
The corresponding reciprocal lattice vectors are given by $\bG_1=(0, {4\pi \over \sqrt{3}})$, $\bG_2=(-2\pi, -{2\pi \over \sqrt{3}})$, and $\bG_3=(2\pi, -{2\pi \over \sqrt{3}})$.
(b) FS at vH filling of free electrons on a single-layer kagom\'e lattice, with colors of lines displaying the sublattice contents.
Schematics of (c) SD-ISD, (d) FM-LC, and (e) AFM-LC CDW on the bilayer kagom\'e lattice.
The width of bonds and size/direction of arrows indicated the values of real and imaginary parts of nn bonds, respectively.
}\label{fig1}
\end{center}
\end{figure*}

The minimal model for \avs\ that captures the most essential physics of a vanadium $d$ band crossing the Fermi level near the van Hove (vH) points \cite{Ortiz-PRM19} is a single-orbital model on the kagom\'e lattice depicted in Fig. \ref{fig1}a, in which there are three sublattices in the unit cell at \br\ and their locations are given by $\br_1$=$\br -{1\over 2}\ba_3$, $\br_2$=$\br$, and $\br_3$=$\br+ {1\over 2}\ba_1$.
At vH filling, the vH points (labeled by $\bM_\alpha$, $\alpha=1,2,3$) with divergent density of states are perfectly nested by the $2a_0 \times 2a_0$ wave vectors $\bQ_\alpha= {1\over 2}\bG_\alpha$, where $\bG_\alpha$ is the reciprocal wave vector of the kagom\'e lattice, as illustrated by the hexagonal Fermi surface (FS) shown in Fig. \ref{fig1}b.
The corresponding instabilities in the presence of electron interactions have been analyzed using the weak-coupling renormalization group (RG) \cite{Park-PRB21, Christensen-PRB22}, and different real and complex CDW phases including LC have been proposed \cite{LinYuPing-PRB21, FengXilin-SciBull21, FengXilin-PRB21, Setty-arxiv21, Denner-PRL21, Yang-arxiv23, Mertz-Npj22, SZ-NC21, Rina-SciAdv22, Tazai-NC23}.

It is important to note that the Bloch states at the vH points are exclusively localized on the corresponding sublattices due to the sublattice quantum interference effect on the kagom\'e lattice \cite{Kiesel-PRL13, WuXX-PRL21, WuYiMing-PRB23}.
In order to utilize the nesting of the vH points, the $2a_0 \times 2a_0$ CDW is therefore believed to be associated with a bond order that connects different sublattices.
Consequently, intersite Coulomb interactions are expected to play an important role in triggering instabilities towards $2a_0 \times 2a_0$ orders, while the on-site Coulomb interaction $U$, which acts on the same sublattice, is obstructed and ineffective at producing such $2a_0 \times 2a_0$ CDWs.
Weak-coupling instability analyses show that nearest-neighbor (nn) Coulomb repulsion $V_1$ and next-nn (nnn) Coulomb repulsion $V_2$ drives, respectively, real and imaginary bond-ordered $2a_0 \times 2a_0$ CDW, with the latter corresponding to LC \cite{Dong-PRB23, RQFu-arXiv24}.
Mean-field calculations on the concrete single-orbital $t$-$V_1$-$V_2$ model found that LC CDW can indeed be stabilized by substantial nnn $V_2$ \cite{Dong-PRB23}.
We note that the single-orbital model is an oversimplified description for the multi-orbital kagom\'e metals and, consequently, the strengths of electron interactions can only be viewed as phenomenological input parameters.
Furthermore, it has been shown in transition metal dichalcogenides \cite{Ramezani-prb24} and black phosphorene \cite{Bagherpour-prb24} that unconventional screening of Coulomb interactions could result to a nnn $V_2$ comparable in strength to nn $V_1$.
Interestingly, it is recently argued that multiple vH singularities close in energy and with opposite mirror eigenvalues in a multi-orbital electronic structure can produce LC order solely by the nn $V_1$ \cite{Li-arXiv2023}.

The $2a_0 \times 2a_0$ bond-ordered CDW corresponds to the triple-$\bQ$ modulation of the breathing bonds on the kagom\'e lattice and is described by the Hamiltonian \cite{SZ-NC21}
\begin{equation}
\mathcal{H}_\text{cdw} = \sum_{\br,\alpha} \rho_\alpha\cos(\bQ_\alpha\cdot \br) \hat{\chiup}^\text{nn}_\alpha (\br) +h.c., \label{Hcdw}
\end{equation}
with the nn breathing bond
\begin{equation}
\hat{\chiup}^\text{nn}_{\alpha} (\br)  = c^\dagger_{\beta\br} c_{\gamma \br} - c^\dagger_{\beta\br} c_{\gamma \br-{\bm a}_\alpha} \label{breath}
\end{equation}
defined from the two independent nn bonds along each $\ba_\alpha$ direction within a unit cell, and $\rho_\alpha$ the corresponding ordering parameter.
Here $c^\dagger_{\alpha\br}$ creates an electron on sublattice $\alpha$ in unit cell $\br$, and the spin indices are left implicit.
The sublattice indices run over $(\alpha, \beta, \gamma)$ = $(1,2,3)$, $(2,3,1)$, and $(3,1,2)$.
Rotation and inversion symmetries are maintained when $\rho_\alpha =\rho$.
The real CDWs with real ordering $\rho$ exhibit the SD and ISD bond configurations for, respectively, $\rho>0$ and $\rho<0$, as displayed in the upper and lower layer in Fig. \ref{fig1}c.
In contrast, a imaginary $\rho$ leads to a LC CDW with circulating LCs, which produces orbital magnetic fluxes and breaks TRS, as depicted in each layer in Figs. \ref{fig1}d and \ref{fig1}e.

Given the $2a_0 \times 2a_0$ bond-ordered CDWs developed on the two-dimensional kagom\'e lattice, it is important to understand their inter-layer correlations, since the CDWs revealed in \avs\ are all three-dimensional with intricate correlation between the kagom\'e layers \cite{ChenXH-PRX21, Ortiz-PRX21, LiHaoxiang-PRX21, Ratcliff-PRM21, Wenzel-PRB22, WuShangfei-PRB22, XieYaofeng-PRB22}.
The inter-layer correlations of real CDWs have been investigated by DFT calculations \cite{YanBinghai-PRL21}, Landau free-energy analyses \cite{Park-PRB21, Christensen-PRB21}, and mean-field study based on a nine-orbital $p$-$d$ model \cite{HLi-prb23}.
In contrast, the inter-layer correlations of possible LC CDWs are only discussed by Landau free-energy analyses \cite{Park-PRB21} and less understood. 
In this work, we study the inter-layer correlations of all bond-ordered CDWs, LC CDWs in particular, by self-consistent mean-field theory based on the effective single-orbital model with electron correlation. 
We consider the minimal bilayer kagom\'e lattice consists of $A$ and $B$ layer displayed in Fig. \ref{fig1}c-\ref{fig1}d, which allows us to describe the $2a_0 \times 2a_0 \times 2c_0$ CDWs revealed in all three families of \avs\ at low temperatures \cite{Ratcliff-PRM21, Stahl-prb22, Xiao-prr23}.
To this end, we focus on rotation symmetry preserving states without a $\pi$-phase shift in Eq. (\ref{Hcdw}) between the two layers.
This implies that the $C_6$ rotation centers of the $2a_0 \times 2a_0$ CDWs on the two layers are lined up vertically in the bilayer stacking.

The inter-layer coupling is manifested by the relation between the CDW orders on these two kagom\'e layers, $\rho_A$ and $\rho_B$.
They are symmetrically stacked if $\rho_A =\rho_B$, whereas antisymmetrically stacked if $\rho_A = -\rho_B$.
The former leads to a $2a_0 \times 2a_0 \times 1c_0$ CDW as the two layers are identical to each other, and the latter is a $2a_0 \times 2a_0 \times 2c_0$ CDW.
Explicitly, when the CDW orders on the two layers are both real and opposite in values, it gives us a SD-ISD state sketched in Fig. \ref{fig1}c.
For imaginary $\rho_A =\rho_B$, LCs flow identically on the two layers, and the corresponding orbital magnetic fluxes are ferromagnetically (FM) coupled, as shown in Fig. \ref{fig1}d.
We call this state FM-LC from now on.
In contrast, LCs flow in opposite directions on the two layers with imaginary $\rho_A = -\rho_B$, resulting antiferromagnetically (AFM) coupled orbital magnetic fluxes depicted in Fig. \ref{fig1}e, and thus referred to as AFM-LC.

The rest of the paper is organized as follows.
In Sec. \ref{sec2}, we introduce the tight-binding (TB)  model for bilayer kagom\'e lattice with inter-layer coupling $t_\perp$ that splits the vH singularity into two locating at fillings $n_{\pm\text{vH}}$. 
The bare susceptibilities for bilayer stacking of nn and nnn bonds are computed for the free electrons. 
We find the behavior of stacking susceptibilities to be very sensitive to band filling $n$, with the leading instability lies in the symmetric channel near $n_{\pm\text{vH}}$ and in the antisymmetric channel at a filling around $n_\text{vH}$, for both nn and nnn bonds. 
On the other hand, the stacking susceptibilities for real components of nn bond orders are always greater than those for their imaginary counterparts, while the situation is exactly the opposite for nnn bonds.

In Sec. \ref{sec3}, we construct the concrete $t$-$t_\perp$-$V_1$-$V_2$ model on the bilayer kagom\'e lattice and describe the corresponding mean-field theory, in which the intersite Coulomb repulsions are decoupled into bond channels.
The mean-field results obtained via self-consistent calculations are presented in Sec. \ref{sec4}.
We determine the ground states and examine their evolutions as a function of band filling $n$.
Indeed, both real CDW and LC CDW can be stabilized, respectively, by the nn and nnn Coulomb repulsion, and their bilayer stacking pattern agrees remarkably well with the weak-coupling instability analysis provided in Sec. \ref{sec2}.
When both $V_1$ and $V_2$ are present and sufficiently large, besides the symmetrically stacked FM-LC and antisymmetrically stacked AFM-LC, ferrimagnetically (FI) stacked LC CDWs (FI-LC) can be stabilized, where the strength of LCs become different on the two layers.
This state is beyond the weak-coupling instability analysis.
The electronic structure, band topology, and Hall conductance of the mean-field states are investigated.
The paper is summarized in Sec. \ref{sec5}, together with some discussions in connection to experiments.

\section{Bilayer tight-binding model and stacking susceptibilities} \label{sec2}

\begin{figure}
	\begin{center}
		\fig{3.4in}{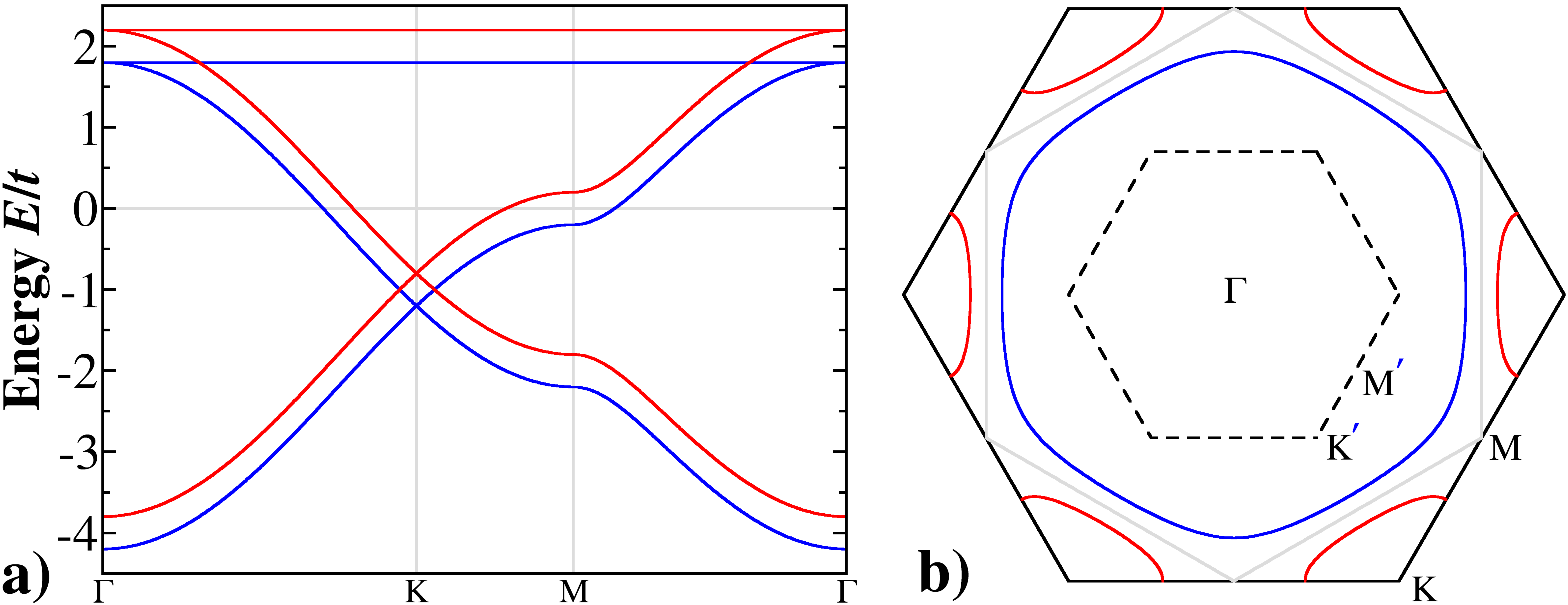}
		\caption{(a) TB dispersions and corresponding (b) FSs of the bilayer kagom\'e lattice with inter-layer coupling $t_\perp = 0.2$ at chemical potential $\mu = 0$. 
			Blue and red curves denote, respectively, the bonding and antibonding orbitals.}\label{TBtp20}
	\end{center}
\end{figure}

The TB Hamiltonian of the bilayer kagom\'e lattice reads
\begin{equation}
\mathcal{H}_0 = -t\sum_{\eta, \langle i,j\rangle} \left( c^\dagger_{\eta i} c_{\eta j} +h.c. \right)  -t_\perp \sum_{i} \left( c^\dagger_{Ai} c_{Bi} +h.c. \right), \label{Htb}
\end{equation}
where the layer index $\eta=\{A,B\}$, $t$ denotes the intra-layer nn hopping, and $t_\perp$ describes the inter-layer coupling.
Hereinafter, we set $t \equiv 1$ as the energy unit.
In terms of local bonding ($\tau=+$) and antibonding ($\tau=-$) orbitals
\begin{equation}
d_{i\tau} = {1\over \sqrt{2}} \left(c_{Ai} +\tau c_{Bi} \right),
\label{bondantibond}
\end{equation}
the TB Hamiltonian becomes
\begin{equation}
\mathcal{H}_0 = -t\sum_{\langle i,j\rangle, \tau}\left( d^\dagger_{i\tau} d_{j\tau} +h.c. \right) -t_\perp \sum_{i,\tau} \tau d^\dagger_{i\tau} d_{i\tau}. \label{Htb2}
\end{equation}
Clearly, the TB electronic structure of bilayer kagom\'e lattice consists of two copies of single-layer TB dispersions shifted by $\mp t_\perp$, respectively, for the bonding and antibonding orbitals.
They correspond to the $\bk_z =0$ and $\bk_z =\pi$ dispersion in the bulk limit.
We choose inter-layer coupling $t_\perp=0.2t$ to accommodate the energy separation between the vH points at $\bk_z =0$ and $\bk_z =\pi$ in the DFT band structure of \avs.
The TB dispersions at chemical potential $\mu = 0$ and the corresponding FS of the bilayer kagom\'e lattice are shown in Fig. \ref{TBtp20}.
When chemical potential is tuned to $\mu =- t_\perp$ ($+t_\perp$), the Fermi level passes the vH points of the bonding (antibonding) orbital, and gives rise to the vH singularity at filling $n_{+\text{vH}} = 4.64/12$ ($n_{-\text{vH}} = 5.44/12$).
In comparison, the vH filling of the single-layer kagom\'e lattice $n_\text{vH} = 5/12$.

The inter-layer coupling of $2a_0 \times 2a_0$ bond-ordered CDWs on the bilayer kagom\'e lattice can be described by the symmetric ($s=+$) and antisymmetric ($s=-$) stacking orders of the breathing bonds.
For nn bonds along $\ba_\alpha$ direction,
\begin{equation}
\hat{\chiup}^\text{nn}_{\alpha,s} (\br ) = \hat{\chiup}^\text{nn}_{A\alpha} (\br ) +s \hat{\chiup}^\text{nn}_{B \alpha} (\br), \label{bso}
\end{equation}
where $\hat{\chiup}^\text{nn}_{\eta \alpha}$ is the nn breathing bond defined in Eq. (\ref{breath}) on the $\eta$-layer.
In order to study separately the instabilities towards the real and imaginary CDW in the symmetric and antisymmetric stacking orders, we further divide $\hat{\chiup}^\text{nn}_{\alpha,s} (\br )$ into their real and imaginary components,
\begin{equation}
\hat{\chiup}^{\prime \text{nn}}_{\alpha,s} ={1\over 2} \left[ \hat{\chiup}^\text{nn}_{\alpha,s} + (\hat{\chiup}^\text{nn}_{\alpha,s})^\dagger \right] , \quad
\hat{\chiup}^{\prime \prime \text{nn}}_{\alpha,s} ={1\over 2 i} \left[ \hat{\chiup}^\text{nn}_{\alpha,s} - (\hat{\chiup}^\text{nn}_{\alpha,s})^\dagger \right],
\end{equation}
and compute the corresponding bilayer stacking susceptibilities $\Pi_{\hat{O}}(\bq)\equiv \Pi_{\hat{O}\hat{O}}(\bq)$,
\begin{equation}
\Pi_{\hat{O}\hat{O}'}(\bq) =\lim_{\omega=0} i\int^\infty_0 dt e^{i\omega t} \langle [ \hat{O} (\bq, t), \hat{O}'^{\dagger} (-\bq, 0)]\rangle,
\end{equation}
with $\hat{O}$, $\hat{O}'$  being one of the four stacking operators for nn bonds in each $\ba_\alpha$ direction, $\hat{\chiup}^{\prime \text{nn}}_{\alpha,\pm} $ and $\hat{\chiup}^{\prime \prime \text{nn}}_{\alpha,\pm}$. 
A temperature of $k_BT = 0.005t$ is applied in the calculation to avoid the logarithmic divergences when the vH singularities is present.

\begin{figure}
\begin{center}
\fig{3.4in}{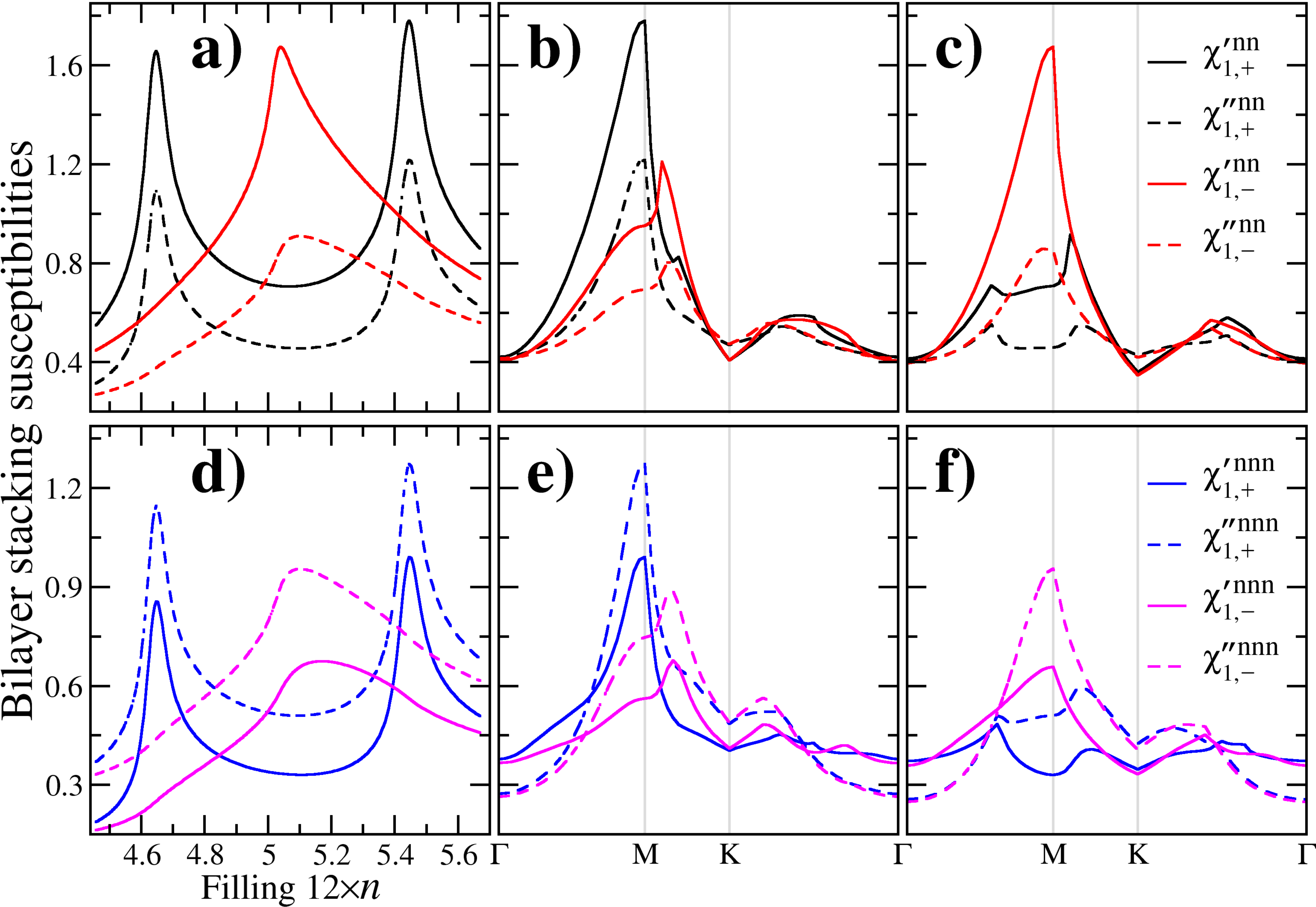}
\caption{Bilayer stacking susceptibilities. (a) Electron filling $n$ dependence of bilayer stacking susceptibilities at wavevector $\bM_1$ for nn bonds in the $\alpha=1$ direction. (b) Susceptibilities for nn bonds along the high-symmetry path at filling $n_{+\text{vH}}$. (c) Same as (b), but at filling $n_1$. (d-f) Similar to (a-c), but for nnn bonds. The fillings in (e) and (f) are, respectively, $n_{+\text{vH}}$ and $n_2$.} \label{sustp20}
\end{center}
\vskip-0.5cm
\end{figure}

The bare susceptibilities for the bilayer stacking orders of nn bonds in the $\alpha=1$ direction at the wavevector $\bM_1 \equiv \bQ_1$ are plotted and compared directly in Fig. \ref{sustp20}a as function of band filling $n$. 
The susceptibilities in the $\alpha\neq 1$ directions can be obtained from $\hat{\chiup}^{\prime \text{nn}}_{1,\pm} $ and $\hat{\chiup}^{\prime \prime \text{nn}}_{1,\pm}$ via $C_3$ rotations.
Clearly, the bilayer stacking susceptibilities are all strongly sensitive to band filling.
The susceptibilities for symmetric stacking, $\hat{\chiup}^{\prime \text{nn}}_{1,+} $ and $\hat{\chiup}^{\prime \prime \text{nn}}_{1,+}$ in solid and dashed black lines, peak at the two vH fillings $n_{\pm\text{vH}}$ of the bilayer kagom\'e lattice, while those for antisymmetric stacking, $\hat{\chiup}^{\prime \text{nn}}_{1,-} $ and $\hat{\chiup}^{\prime \prime \text{nn}}_{1,-}$ in solid and dashed red lines, are maximized at filling $n_1 \simeq 5.04/12$.
Furthermore, for nn bonds, the susceptibilities of real stacking orders are always greater than their imaginary counterparts, pointing to leading instabilities towards real CDW driven by nn Coulomb repulsion $V_1$.
Explicitly, one would expect $V_1$ to drive a symmetrically stacked real CDW, i.e., SD-SD or ISD-ISD, near the two vH fillings $n_{\pm \text{vH}}$, but a antisymmetrically stacked real CDW, i.e., SD-ISD shown in Fig. \ref{fig1}c, around filling $n_1$ on the bilayer kagom\'e lattice.
The momentum evolution of the bilayer stacking susceptibilities of nn bonds are displayed along the high-symmetry path in the BZ at filling $n_{+\text{vH}}$ in Fig. \ref{sustp20}b and $n_1$ in Fig. \ref{sustp20}c, both displaying a leading instability at $\bM$ tied to the formation of $2a_0 \times 2a_0$ CDW.

The behaviors of bilayer stacking susceptibilities can be better understood in the local bonding and antibonding basis where the TB Hamiltonian becomes block diagonal.
In this basis, we define intra-orbital ($\tau=\tau'$) and inter-orbital ($\tau\neq \tau'$) nn breathing bonds 
\begin{equation}
\hat{\chiup}^\text{nn}_{\tau\tau',\alpha} (\br) =d^\dagger_{\tau\beta\br} d_{\tau' \gamma\br} - d^\dagger_{\tau\beta\br} d_{\tau' \gamma\br-\ba_\alpha},
\end{equation}
in terms of which the bilayer stacking orders defined in Eq. (\ref{bso}) can be rewritten as
\begin{equation}
\hat{\chiup}^\text{nn}_{\alpha,+} (\br)=\sum_\tau \hat{\chiup}^\text{nn}_{\tau\tau} (\br),\quad 
\hat{\chiup}^\text{nn}_{\alpha,-} (\br)=\sum_\tau \hat{\chiup}^\text{nn}_{\tau\bar\tau} (\br),
\end{equation}
and, consequently, their susceptibilities becomes
\begin{equation}
\Pi_{\hat{\chiup}^\text{nn}_{\alpha,+}} = \sum_\tau \Pi_{\hat{\chiup}^\text{nn}_{\tau\tau,\alpha}}, \quad
\Pi_{\hat{\chiup}^\text{nn}_{\alpha,-}} = \sum_\tau \Pi_{\hat{\chiup}^\text{nn}_{\tau\bar\tau,\alpha}}.
\end{equation}
Here, we have used $\Pi_{\hat{\chiup}^\text{nn}_{\tau\tau,\alpha} \hat{\chiup}^\text{nn}_{\bar\tau\bar\tau,\alpha}} =\Pi_{\hat{\chiup}^\text{nn}_{\tau\bar\tau,\alpha} \hat{\chiup}^\text{nn}_{\bar\tau\tau,\alpha}}=0$ due to the block diagonalization of the TB Hamiltonian in the bonding and antibonding basis.
At vH filling $n_{\tau\text{vH}}$ with chemical potential $\mu=-\tau t_p$, the Fermi level passes through the vH points of the $d_\tau$ orbital, giving rise to a logarithmic divergence in $\Pi_{\hat{\chiup}^\text{nn}_{\tau \tau, \alpha}}$ at the zone boundary $\bM_\alpha$. 
Consequently, $\Pi_{\hat{\chiup}^\text{nn}_{\alpha,+}} (\bM_\alpha)$ diverges at these two vH fillings, as reflected by the two peaks shown in Fig. \ref{sustp20}a in the presence of finite thermal broadening.
In contrast, it can be shown that the susceptibilities for antisymmetrical stacked nn breathing bonds, $\Pi_{\hat{\chiup}^\text{nn}_{\alpha,-}}$ at $\bM_\alpha$,  would peaks (not diverges) at a filling in between the two vH fillings $n_{\pm\text{vH}}$.
Furthermore, owing to the unique lattice geometry and sublattice interference effect, it has been shown that nn bonds on the kagom\'e lattice are characterized by intrinsic real bond fluctuations \cite{Dong-PRB23, RQFu-arXiv24}.
Consequently, the susceptibilities of real part of nn bonds are greater than those of imaginary ones, \textit{i.e.}, $\Pi_{\hat{\chiup}^{\prime \text{nn}}_{\alpha,\pm}} > \Pi_{\hat{\chiup}^{\prime \prime \text{nn}}_{\alpha,\pm}}$ at all fillings, as shown in Fig. \ref{sustp20}a. 

In a similar vein, nnn breathing bonds can be defined from the two independent nnn bonds perpendicular to each basis vector $\ba_\alpha$ (i.e., along $\bQ_\alpha$ direction) of $\eta$ layer in a unit cell
\begin{equation}
\hat{\chiup}^\text{nnn}_{\eta\alpha} (\br) = c^\dagger_{\eta \beta \br} c_{\eta \gamma \br+\ba_\beta} -c^\dagger_{\eta \beta \br} c_{\eta \gamma \br+\ba_\gamma}.
\end{equation}
Consequently, the symmetric (+) and antisymmetric (-) stacking of breathing nnn bonds in the bilayer kagom\'e lattice is defined and divided into their real and imaginary components
\begin{equation}
\hat{\chiup}^\text{nnn}_{\alpha,\pm} (\br) = \hat{\chiup}^\text{nnn}_{A \alpha} (\br ) \pm \hat{\chiup}^\text{nnn}_{B \alpha} (\br) = \hat{\chiup}^{\prime \text{nnn}}_{\alpha,\pm} (\br) +i\hat{\chiup}^{\prime\prime \text{nnn}}_{\alpha,\pm} (\br).
\end{equation}
The calculated bilayer stacking susceptibilities of nnn bonds in the $\alpha=1$ direction at the wavevector $\bM_1$ are plotted in Fig. \ref{sustp20}c as functions of filling $n$.
They exhibit strong dependence on filling $n$, and as we argued above, the susceptibilities for symmetric stacking, $\hat{\chiup}^{\prime \text{nnn}}_{1,+} $ and $\hat{\chiup}^{\prime \prime \text{nnn}}_{1,+}$ in solid and dashed blue lines, peak at the two vH fillings $n_{\pm\text{vH}}$, while those for antisymmetric stacking, $\hat{\chiup}^{\prime \text{nnn}}_{1,-} $ and $\hat{\chiup}^{\prime \prime \text{nnn}}_{1,-}$ in solid and dashed magnet lines, are maximized at a filling in between, e.g., $n_2 \simeq 5.10/12$.
Intriguingly, for nnn bonds, the susceptibilities of real stacking orders are now weaker than their imaginary counterparts \cite{Dong-PRB23, RQFu-arXiv24}, pointing to leading instabilities towards LC CDWs with circulating LCs driven by nnn Coulomb repulsion $V_2$.
Explicitly, one would expect $V_2$ to drive a symmetrically stacked LC CDW, i.e., FM-LC displayed in Fig. \ref{fig1}d, near the two vH fillings $n_{\pm \text{vH}}$, but a antisymmetrically stacked LC CDW, i.e., AFM-LC depicted in Fig. \ref{fig1}e, around filling $n_2$ on the bilayer kagom\'e lattice.
The momentum evolution of the bilayer stacking susceptibilities of nnn bonds are displayed along the high-symmetry path in the BZ at filling $n_{+\text{vH}}$ in Fig. \ref{sustp20}d and at $n_2$ in Fig. \ref{sustp20}f, both exhibiting a pronounced peak at the $2a_0\times 2a_0$ wave vector $\bM$.

\section{The bilayer $\bm t$-$\bm t_\perp$-$\bm V_1$-$\bm V_2$ model and mean field theory}
\label{sec3}
Next, we investigate the inter-layer correlation of the $2a_0 \times 2a_0$ bond-ordered CDWs in the concrete $t$-$t_\perp$-$V_1$-$V_2$ model on the bilayer kagom\'e lattice
\begin{equation}
\mathcal{H}=\mathcal{H}_0 +V_1\sum_{\eta,\langle i,j\rangle} \hat{n}_{\eta i} \hat{n}_{\eta j} +V_2\sum_{\eta, \llangle i,j\rrangle} \hat{n}_{\eta i} \hat{n}_{\eta j} ,
\end{equation}
where $\mathcal{H}_0$ is the bilayer TB model given in Eq. (\ref{Htb}) and the density operators $\hat{n}_{\eta i}$ = $c^\dagger_{\eta i} c_{\eta i}$.
Decoupling the Coulomb repulsions $V_1$ and $V_2$ in terms of the nn and nnn bonds \cite{Dong-PRB23}, $\hat{\chiup}_{\eta,ij} = c^\dagger_{\eta i} c_{\eta j}$, the obtained mean-field Hamiltonian reads
\begin{align}
\mathcal{H}_\text{MF} = \mathcal{H}_0 &-V_1 \sum_{\eta, \langle i,j \rangle} \left( \chiup^*_{\eta, ij}  \hat{\chiup}_{\eta, ij} +h.c. - \left|\chiup_{\eta, ij}\right|^2 \right) \label{hmf} \\
&-V_2 \sum_{\eta, \llangle i,j \rrangle} \left( \chiup^*_{\eta, ij}  \hat{\chiup}_{\eta, ij} +h.c. - \left|\chiup_{\eta, ij}\right|^2 \right). \nonumber
\end{align}
The order parameters of the nn and nnn bonds $\chiup_{\eta, ij}=\langle \hat{\chiup}_{\eta, ij} \rangle$ are to be determined self-consistently by minimizing the state energy.
Note the direct Hatree terms from $V_1$ and $V_2$ depending on the electron density $n_{\eta i}=\langle \hat{n}_{\eta i} \rangle$ are neglected in Eq. (\ref{hmf}) to avoid double-counting, since their contributions are already included in the LDA \cite{JiangKun-PRB2016,Campo-IOP10,Belozerov-PRB12}.
Moreover, all symmetry invariant corrections to the real uniform order parameter are subtracted in the interactions, such that the vH points and the band structure are maintained without adjusting the bare band parameters \cite{Dong-PRB23}.
Implementing this subtraction scheme, the correlation effects in the self-consistent mean-field theory will only generate spontaneous symmetry-breaking states.

\begin{figure}
	\begin{center}
		\fig{3.4in}{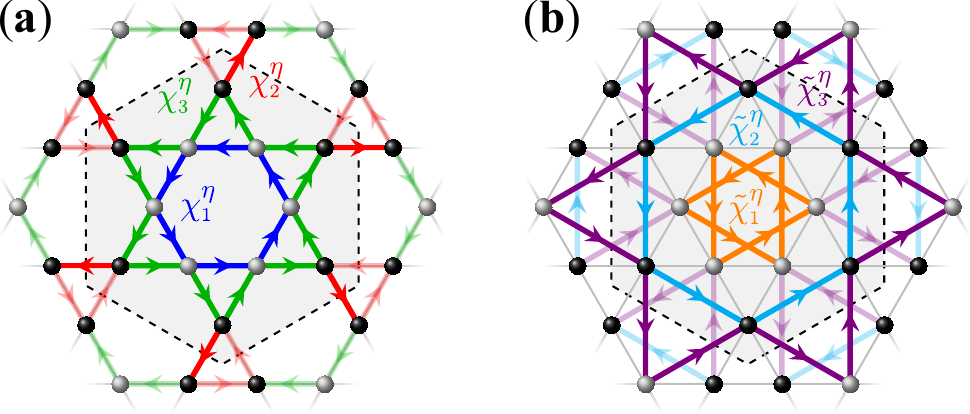}
		\caption{Schematic picture of (a) nn and (b) nnn bonds in the enlarged $2a_0\times 2a_0$ unit cell of each layer enclosed by dashed lines.
			$C_6$ symmetry and continuity condition dictate equality of sites, nn and nnn bonds, as indicated by the colors of the spheres and bonds.
			Arrows denote the flowing directions of currents.
			The colors of bonds replicated by $2a_0 \times 2a_0$ translation are weakened for a better visualization.}
		\label{pattern}
	\end{center}
	\vskip-0.65cm
\end{figure}

To utilize the leading instabilities at $\bM$ in the bare susceptibilities near vH filling, we consider periodic states with enlarged $2a_0 \times 2a_0$ unit cell, enclosed by black dashed lines in Fig. \ref{pattern}.
Furthermore, we focus on states invariant under $C_6$ rotation and satisfying the current continuity condition \cite{Dong-PRB23}.
Under these constraints, in each layer there are two independent sites with electron density $n^\eta_{1,2}$, three independent nn bonds labeled as $\chiup^{\eta}_{1,2,3}$, and three kinds of nnn bonds denoted by $\tilde{\chiup}^{\eta}_{1,2,3}$, as shown in Fig. \ref{pattern}.
The $C_6$ rotation centers of the two kaom\'e layers are lined up vertically, without any phase shift.
The flowing direction of LC on each bond is determined by the sign of the imaginary component of the corresponding bond.
In order to obtain all the possible states, we use different initial conditions for solving the self-consistency equations numerically.
When multiple converged states exist for a given set of parameters, $V=(V_1, V_2)$, we compare the state energies to determine the true ground state.

To this end, we note that, when LCs are developed, the converged bond orders are in general complex, with modulations in both real and imaginary components. 
For simplicity, we refer to these complex bond-ordered CDWs as FM-LC, AFM-LC, or FI-LC according to the pattern of LC order, without further specifying the real bond patterns. 
The topological property of these complex bond-ordered CDWs is solely determined by the pattern of LC order on the nn bonds \cite{Dong-PRB23}.
In the next section, we obtain the mean-field ground states and determine the inter-layer coupling as a function of band filling at various interactions.
The interaction parameters are chosen following two criteria. 
First, the interactions need to be large enough to drive the physics of interest.
Explicitly, unlike the susceptibilities for symmetric stacking that diverge at the two vH fillings, the susceptibilities for antisymmetric stacking only peak (not diverge) at a filling in between. 
Therefore, a sufficiently strong interaction is required to generate the antisymmetrically stacked CDWs, such as SD-ISD and AFM-LC.
Second, we prefer the interactions to be not too large. 
It’s because the CDW gap revealed experimentally in V-based kagom\'e metals is not too large. 
Furthermore, the numerical results obtained by mean-field calculations at these small but sufficient interactions would be more consistent with the instability analysis provided in Sec. \ref{sec2}.

\begin{figure}
\begin{center}
\fig{3.4in}{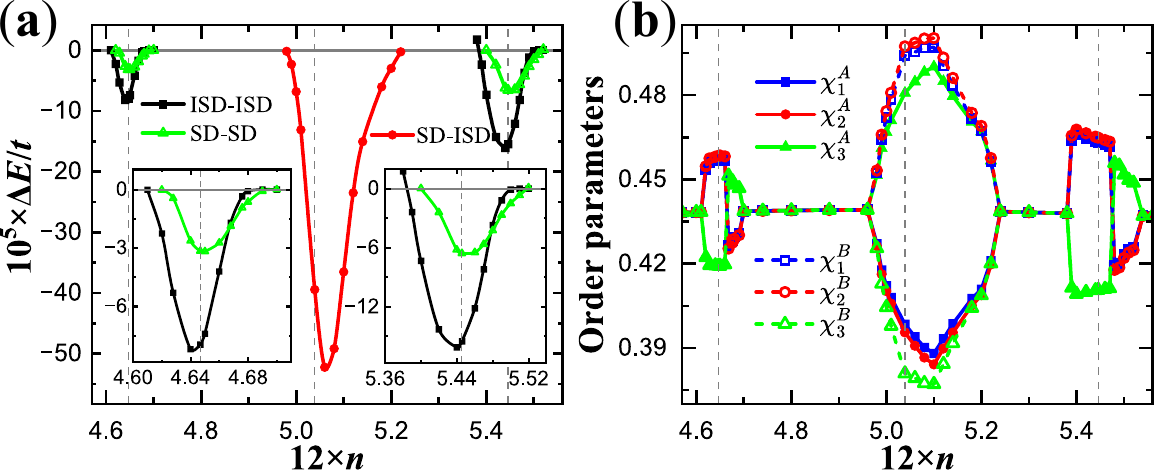}
\caption{(a) The state energy per site of real CDWs with respect to the PM phase at $V = (0.8t, 0)$, as a function of filling $n$. 
(b) Behavior of the real nn bonds in the ground state. 
Dashed vertical lines denote the two vH fillings $n_{\pm\text{vH}}$ and $n_1$ where the bare susceptibilities of nn bonds peak.
Insets in (a) zoom in fillings near $n_{\pm\text{vH}}$.} \label{v08v00}
\end{center}
\vskip-0.5cm
\end{figure}

\section{Bilayer stacking of bond-ordered CDW}
\label{sec4}
\subsection{Stacking of SD and ISD}
\label{subsec4a}

We first examine the inter-layer correlation of $2a_0\times 2a_0$ real CDWs, i.e., SD and ISD, in the presence of solely nn Coulomb repulsion $V_1$ by setting $V=(0.8t, 0)$.
Indeed, no LC CDWs could be stabilized and we managed to converge to three kinds of real CDW states on the bilayer kagom\'e lattice, \textit{i.e.}, symmetrically stacked SD-SD, ISD-ISD, and antisymmetrically stacked SD-ISD.
The state energies per site of these $2a_0\times 2a_0$ real CDWs are displayed in Fig. \ref{v08v00}a with respect to the paramagnetic (PM) phase.
Clearly, the symmetrically stacked SD-SD or ISD-ISD has the lowest energy in the filling regimes near $n_{\pm\text{vH}}$, \textit{e.g.}, $4.61/12 \lesssim n \lesssim 4.70/12$ and $5.38/12 \lesssim n \lesssim 5.52/12$, while the antisymmetrically stacked SD-ISD becomes the ground state at $4.98/12\lesssim n \lesssim 5.22/12$.
Furthermore, the energy gain of the $2a_0\times 2a_0$ real CDWs maximizes near $n_{\pm\text{vH}}$ and $n_1$, where the bare susceptibilities peak in Fig. \ref{sustp20}a, showing remarkable consistency with the weak-coupling instability analysis.
Interestingly, as one increases the filling near $n_{\pm\text{vH}}$, the symmetrically stacked real CDW undergoes a first-order transitions from ISD-ISD to SD-SD at roughly $n\simeq 4.66/12$ and $n\simeq 5.48/12$, as indicated by the level crossings shown in the insets of Fig. \ref{v08v00}a.
The inter-layer Coulomb repulsion studied in Ref. \cite{HLi-prb23} would further proliferate SD-ISD by saving potential energy.

The two kagom\'e layers are identical to each other with $\chiup^A_{1,2,3} =\chiup^B_{1,2,3}$ in the symmetrically stacked ISD-ISD and SD-SD, but become inequivalent in the antisymmetrically stacked SD-ISD, $\chiup^A_{1,2,3} \neq\chiup^B_{1,2,3}$.
Explicitly, at $n_{-\text{vH}}$, the nn bonds $\chiup^{A/B}$ = (0.463, 0.465, 0.411) in ISD-ISD with the stronger bonds $\chiup_1$ and $\chiup_2$ form the Triangle-and-Hexagon pattern, and $\chiup^{A/B}$ = (0.421, 0.419, 0.455) in SD-SD with the stronger bond $\chiup_3$ form the Star-of-David pattern. 
Meanwhile, the nn bonds in the SD-ISD at $n_1$ take the values of $\chiup^A = (0.398, 0.396, 0.481)$ and $\chiup^B = (0.494, 0.497, 0.381)$.
The nn bonds $\chiup^\eta_{1,2,3}$ of the real CDW with the lowest state energy, \textit{i.e.}, the ground state, are presented in Fig. \ref{v08v00}b as a function of filling $n$.
At fillings $n\simeq 4.66/12$ and $n\simeq 5.48/12$, there are abrupt changes in the nn bonds, indicating the first-order transition between the two symmetrically stacked $2a_0\times 2a_0$ real CDWs, ISD-ISD and SD-SD.
While the inversion symmetry on the bilayer lattice is preserved in the symmetrically stacked SD-SD and ISD-ISD, it is broken in the antisymmetrically stacked SD-ISD where the two layers become different.
We note that, however, the inversion symmetry is preserved in the SD-ISD configuration in 3D bulk materials.
Both ISD-ISD and SD-ISD CDWs have been revealed experimentally in vanadium-based kagom\'e metals \cite{YongHu-PRB22, Kang-NatMat22, Ritz-PRB23, Kautzsch-PRM23, Frassineti-PRR23, YuxinWang-PRB23, Scagnoli-JPCM24, wei2024three}.

\begin{figure}
\begin{center}
\fig{3.4in}{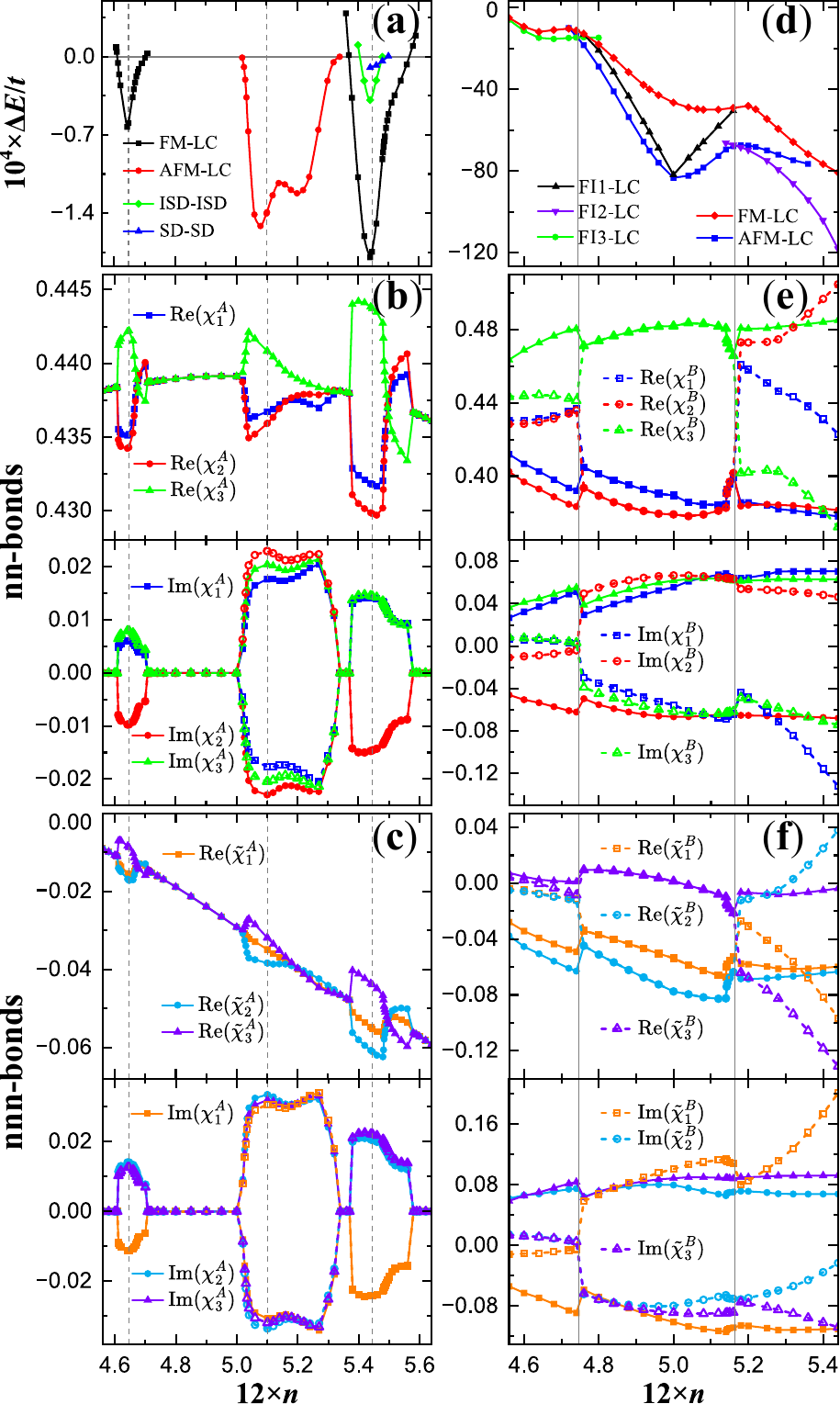}
\caption{(a) The state energy per site of converged bond-ordered CDWs with respect to PM phase, and the (b) nn bonds $\chiup^\eta_{1,2,3}$ and (c) nnn bonds $\tilde{\chiup}^\eta_{1,2,3}$ in the ground state at $V = (0, 1.2t)$. 
The three dashed vertical lines denote $n_{\pm\text{vH}}$ and $n_2$ at which the bare susceptibilities of nnn bonds peak. 
(d) The state energy per site of converged low-energy CDWs with respect to PM phase, and the (e) nn bonds $\chiup^\eta_{1,2,3}$ and (f) nnn bonds $\tilde{\chiup}^\eta_{1,2,3}$ in the ground state at $V = (0.8t, 1.6t)$. 
The two dashed vertical lines at fillings 4.74/12 and 5.16/12 denote the successive transitions from FI3-LC to AFM-LC and to FI2-LC. } \label{V2V12}
\end{center}
\end{figure}

\begin{table*}[htb]
\caption{The nn bonds $\chiup_{1,2,3}$ and nnn bonds $\tilde{\chiup}_{1,2,3}$ in the LC CDWs discussed in Subsection \ref{LCstates}. 
Only the bond orders in $A$-layer are given for FM-LC and AFM-LC, while the bond orders on both two layers are shown for the three FI-LCs, with the first and second row for $A$ and $B$ kagom\'e layer, respectively.}
\label{bonds}
\begin{ruledtabular}
\begin{tabular}{c|c|c|c|c|c|c|c|c}
$V$'s & State & $n$ & $\chiup_1$ & $\chiup_2$ & $\chiup_3$ & $\tilde{\chiup}_1$ & $\tilde{\chiup}_2$ & $\tilde{\chiup}_3$ \\
\hline
$V_1=0$ & FM-LC & $n_{-\text{vH}}$ & $0.432-0.014i$ & $0.430+0.015i$ & $0.444-0.014i$ & $-0.055+0.024i$ & $-0.061-0.020i$ & $-0.044-0.022i$ \\
\cline{2-9}
$V_2=1.2t$ & AFM-LC & $n_{2}$ & $0.437-0.018i$ & $0.436+0.023i$ & $0.441-0.020i$ & $-0.035+0.031i$ & $-0.038-0.033i$ & $-0.032-0.032i$ \\
\hline
 & FM-LC & $4.95/12$ & $0.388-0.057i$ & $0.380+0.068i$ & $0.483-0.061$ & $-0.052+0.099i$ & $-0.067-0.080i$ & $0.003-0.090i$ \\
\cline{2-9}
 & AFM-LC & $4.95/12$ & $0.390-0.052i$ & $0.380+0.066i$ & $0.482-0.059i$ & $-0.049+0.094i$ & $-0.071-0.080i$ & $0.004-0.087i$ \\
\cline{2-9}
 & FI1-LC & $4.95/12$ & $0.388-0.058i$ & $0.384+0.065i$ & $0.481-0.059i$ & $-0.051+0.097i$ & $-0.060-0.076i$ & $0.000-0.089i$ \\
$V_1=0.8t$ &   &  & $0.523-0.001i$ & $0.539+0.001i$ & $0.339-0.002i$ & $0.014+0.003i$ & $0.057-0.001i$ & $-0.091-0.002i$ \\
\cline{2-9}
$V_2=1.6t$ & FI2-LC & 5.32/12 & $0.381-0.070i$ & $0.384+0.066i$ & $0.482-0.063i$ & $-0.061+0.111i$ & $-0.067-0.068i$ & $-0.008-0.092i$ \\
 &  &  & $0.443+0.084i$ & $0.481-0.051i$ & $0.396+0.063i$ & $-0.056-0.128i$ & $0.003+0.054i$ & $-0.094+0.090i$ \\
\cline{2-9}
 & FI3-LC & 4.68/12 & $0.398-0.043i$ & $0.389+0.057i$ & $0.477-0.049i$ & $-0.043+0.077i$ & $-0.056-0.071i$ & $0.002-0.076i$ \\
 &  &  & $0.433-0.005i$ & $0.431+0.007i$ & $0.444-0.006i$ & $-0.009+0.009i$ & $-0.010-0.009i$ & $-0.003-0.009i$
\end{tabular}
\end{ruledtabular}
\end{table*}

\subsection{Stacking of loop current CDWs}
\label{LCstates}
Next, we change the interaction to $V=(0, 1.2t)$ to investigate the bilayer stacking of LC CDWs triggered by nnn coulomb repulsion $V_2$, in the absence of nn Coulomb repulsion $V_1$.
Here, the nnn $V_2$ is larger than the nn $V_1$ used in the last subsection because the susceptibility peak value for antisymmetric stacking on nnn bonds is lower than that on nn bonds, as shown in Fig. \ref{sustp20}d and \ref{sustp20}a. 
In fact, no AFM-LC could be obtained at any filling if $V_2$ is set to be 0.8$t$.
The energies of obtained $2a_0 \times 2a_0$ LC CDWs with respect to the PM phase are presented in Fig. \ref{V2V12}a as a function of filling $n$.
Indeed, FM-LC with identically circulating LCs on the two kagom\'e layers is stabilized near the two vH fillings $n_{\pm\text{vH}}$ where the leading instability peaks in the imaginary part of symmetric stacked nnn bonds, while AFM-LC with opposite circulating LCs on the two layers becomes the ground state around $n_2$ where the leading instability peaks in the imaginary part of antisymmetric stacked nnn bonds.
These mean-field results are in remarkable agreement with the bilayer stacking susceptibilities of nnn bonds shown in Fig. \ref{sustp20}d.
We note that, in addition to these LC CDWs, symmetrically stacked real CDWs, ISD-ISD and SD-SD, can also be obtained at fillings near $n_{-\text{vH}}$, but they are higher in energy and fail to present themselves as ground states in the phase diagram, as shown in Fig. \ref{V2V12}a.

Fig. \ref{V2V12}b and \ref{V2V12}c display, respectively, the nn bonds $\chiup^\eta_{1,2,3}$ and the nnn bonds $\tilde{\chiup}^{\eta}_{1,2,3}$ in the ground state with the lowest energy. 
In FM-LC obtained near $n_{\pm\text{vH}}$, $\chiup^A = \chiup^B$ and $\tilde{\chiup}^A = \tilde{\chiup}^B$, implying that the two kagom\'e layers are identical with LCs flowing in the same direction.
Interestingly, as one increases the filling near $n_{\pm\text{vH}}$, while the pattern of LCs remain unchanged, there is a continuous change in the configuration of real part of nn bonds, from SD to ISD, as shown by the line crossing at filling $n \simeq 4.68/12$ and $n \simeq 5.5/12$ in the upper panel of Fig. \ref{V2V12}b.
The real part of nnn bonds displayed in the upper panel of Fig. \ref{V2V12}c has a similar crossing behavior at these two fillings.
In AFM-LC converged near $n_2$, $\chiup^A =(\chiup^B)^*$ and $\tilde{\chiup}^A =(\tilde{\chiup}^B)^*$.
While the configuration of real part of bonds is the same on the two layers, the LCs flow in opposite direction with the same strength.
The bond orders in the FM-LC at $n_{-\text{vH}}$ and those in AFM-LC at $n_2$ are given explicitly in Table \ref{bonds}.
Clearly, the imaginary bond orders associated with LCs are much bigger than the modulation in real bond orders, consistent with the fact that the leading instability lies in the imaginary part of the bond operators, as shown in Fig. \ref{sustp20}d.

To investigate the inter-layer correlation of $2a_0\times 2a_0$ LC CDWs in the presence of both nn and nnn Coulomb repulsions, we move the interaction to $V=(0.8t, 1.6t)$ which is near the phase boundary and the behavior of single-layer kagom\'e lattice has been well studied at this interaction\cite{Dong-PRB23}.
Indeed, all the low energy states obtained here are LC CDWs, and their state energies per site with respect to the PM phase are displayed in Fig. \ref{V2V12}d, as a function of filling $n$.
Interestingly, besides the FM-LC and AFM-LC obtained in the absence of $V_1$ where the strength of LCs is identical on the two layers, we managed to converge to three kinds of FI-LC where the strength of LCs become different on the two kagom\'e layers. 
AFM-LC has the lowest state energy at a filling between $n_{1c} \simeq 4.74/12$ and $n_{2c} \simeq 5.16/12$, while FI2-LC and FI3-LC becomes the ground state at, respectively, $n>n_{2c}$ and $n<n_{1c}$.
These two transitions are both first-order, as indicated by the level crossing shown in Fig. \ref{V2V12}d.
FM-LC obtained at all fillings and FI1-LC converged at $4.76/12 \lesssim n \lesssim 5.16/12$ are always higher in energy and fail to present themselves as the ground state at $V=(0.8t, 1.6t)$.

The detail behavior of the nn and nnn bonds in the ground state are displayed in Figs. \ref{V2V12}e and \ref{V2V12}f as a function of filling $n$.
Clearly, the bond orders change abruptly at $n_{1c}$ and $n_{2c}$ where the first-order transitions take place.
The values of nn and nnn bonds are given explicitly in Table \ref{bonds} for FM-LC, AFM-LC, and FI1-LC at filling $4.95/12$, FI2-LC at filling $5.32/12$, and FI3-LC at $4.68/12$.
The bonds in the two kagom\'e layers are indentical to each other in FM-LC, and conjugate to each other in AFM-LC, with the real nn bonds forming SD-SD configuration.
LCs are of equivalent strength, though flowing in the same and opposite direction, respectively, in FM-LC and AFM-LC.
On the other hand, the strengths of LCs become different on the two layers in FI-LCs.
Explicitly, real nn bonds form SD-ISD configuration in FI1-LC obtained at $4.76/12\lesssim n \lesssim 5.16/12$, with LCs circulating mainly on the SD kaogom\'e layer while negligibly small on the ISD layer.
In FI2-LC obtained at $n\gtrsim 5.14/12$, real nn bonds form SD-ISD configuration as well, but LCs on the two layers are of similar strength, flowing in opposite directions.
The LCs in FI3-LC converged at $n\lesssim 4.80/12$ flow mainly on the SD layer, while the bond orders, both real and imaginary, are much smaller on the other layer.
We note that, while the mean-field results at interactions $V=(0.8t, 0)$ and $(0, 1.2t)$ can be well understood from the perspective of weak-coupling instability analysis provided in Section \ref{sec2}, the FI-LCs obtained at $V=(0.8t, 1.6t)$ are beyond the weak-coupling instability analysis.

\begin{figure}
\begin{center}
\fig{3.4in}{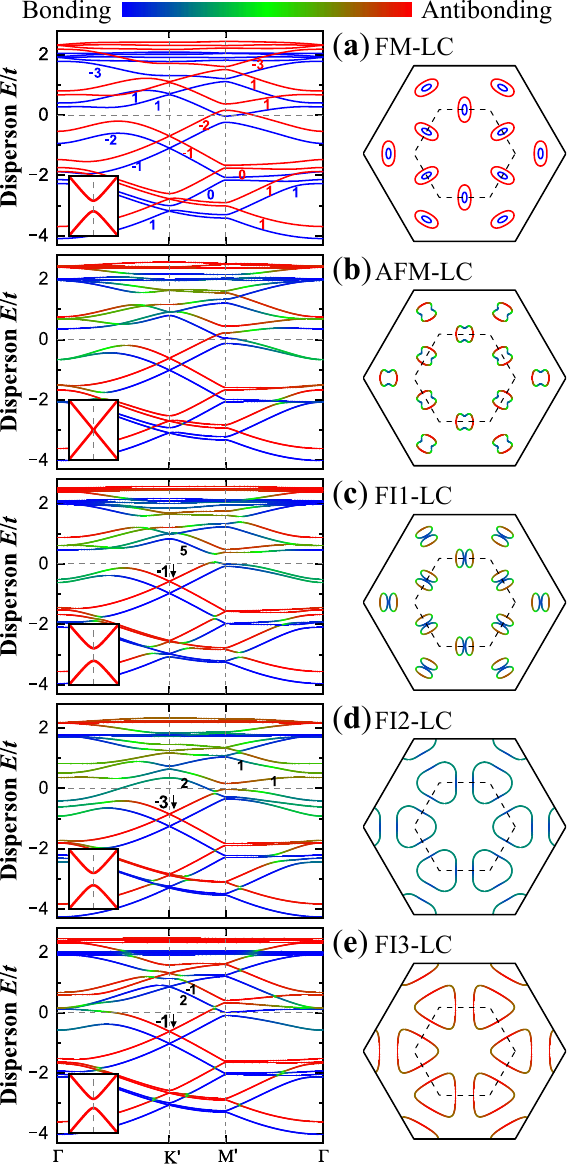}
\caption{The electronic structure and corresponding Fermi surfaces of the five bond-ordered CDWs listed in Table \ref{bonds} for $V=(0.8t, 1.6t)$: (a) FM-LC at 4.95/12, (b) AFM-LC at 4.95/12, (c) FI1-LC at 4.95/12, (d) FI2-LC at 5.32/12, and (e) FI3-LC at 4.68/12. 
Line color denotes the content of bonding and antibonding orbitals as indicated by the color bar at the top.
Insets zoom into the first Dirac cone underneath the Fermi level at K$'$ point, showing the preserving of Dirac cone in AFM-LC in (b) and its gaping in all other four states.
The gap size is roughly $3.2\times10^{-3}t$, 0, $3.7\times10^{-6}t$, $3.6\times10^{-3}t$, and $2.8\times10^{-3}t$, respectively, in figure (a-e).
Small numbers in right panels display the Chern numbers of the corresponding isolated bands, while the big numbers followed by a downarrow indicate the total Chern number of all the bands underneath it.} \label{dispfs}
\end{center}
\end{figure}

\subsection{Electronic structure and band topology}
We now investigate the electronic structure and topological properties of the $2a_0 \times 2a_0$ bond-ordered CDWs on the bilayer kagom\'e lattice obtained at $V=(0.8t, 1.6t)$ where LC is always involved in the low energy states.
The electronic structures and the corresponding Fermi surfaces of the five LC CDWs listed in Table \ref{bonds} for $V=(0.8t, 1.6t)$ are displayed in Fig. \ref{dispfs}, where the line colors denote the content of bonding (blue) and antibonding (red) orbitals defined in Eq. (\ref{bondantibond}).
Because the bond orders are identical on the two kagom\'e layers in FM-LC, its Hamiltonian matrix becomes block-diagonal in the bonding and antibonding orbitals basis.
As a result, the electronic structure of FM-LC consists of two separate copies from bonding and antibonding orbitals, as shown by blue and red lines in Fig. \ref{dispfs}a.
These two copies of bands are equivalent except for an energy shift of $2t_\perp$ between them.
On the other hand, the two kagom\'e layers become different in all other LC CDWs, AFM-LC and FI-LC, and thus the bonding and antibonding orbitals become mixed together, as illustrated by the color variation of lines in Fig. \ref{dispfs}b-e.

The emergence of LCs breaks the TRS in LC CDWs. 
Remarkably, we note that the combined symmetry of TRS and inversion is preserved in AFM-LC, though it is still broken in FM-LC and FI-LC.
Consequently, the Dirac points are preserved in AFM-LC but gaped out in all other four states, as illustrated clearly in the insets in Fig. \ref{dispfs}.
The opening of Dirac points would bring nontrivial topology to the electronic bands.
To examine the topological properties of these states, we calculate the Berry curvature of the $n$th quasiparticle band at momentum $\bk$,
\begin{equation}
\Omega_n(\bk)=i\langle \nabla_\bk u_{n\bk}| \times| \nabla_\bk u_{n\bk} \rangle \nonumber 
= i\left( \left\langle {\partial u_{n\bk} \over \partial k_x} \Big{|} {\partial u_{n\bk} \over \partial k_y} \right\rangle -\left\langle {\partial u_{n\bk} \over \partial k_y} \Big{|} {\partial u_{n\bk} \over \partial k_x} \right\rangle \right),
\end{equation}
where $u_{n\bk}$ is the periodic part of the Bloch state wave function.
The Chern number of $n$th band can be obtained by integrating the Berry curvature over the $2a_0 \times 2a_0$ reduced Brillouin zone (RBZ), $C_n ={1\over 2\pi} \int_\text{RBZ} \Omega_n(\bk) d\bk$.
Except for the AFM-LC state, the other four LC CDWs listed in Table \ref{bonds} are indeed all topologically nontrivial.
Each isolated quasiparticle band of them acquires an integer Chern number, as denoted by the corresponding numbers in Fig. \ref{dispfs}a and Figs. \ref{dispfs}c-e.
In the three FI-LC states, some of the bands are mixed up and very difficult to distinguish the Chern number of each individual band, we thus use a bigger number followed by a downarrow to indicate the total Chern number of all the bands underneath it.

In FM-LC, the completely-decoupled bonding and antibonding orbitals have the equivalent electronic structures, except for a shift of $2t_\perp$, and they share the same topological properties, as shown in Fig. \ref{dispfs}a.
Furthermore, their electronic structure and band topology are identical to that of a single-layer kagom\'e lattice with the same bond orders.
The $2a_0 \times 2a_0$ LC CDW opens up an energy gap of about $0.2t$, which is much smaller than the inter-layer splitting $2t_\perp=0.4t$ of the two copies of bands.
Therefore, the FM-LC obtained at $V=(0.8t, 1.6t)$ is metallic at all fillings near $n_\text{vH}$.
The corresponding Fermi surfaces of FM-LC at $n=4.95/12$ possess smaller electron pockets of bonding orbitals and larger hole pockets of antibonding orbitals, centered around the M$'$ points in the $2 \times 2$ reduced BZ, as displayed in right panel of Fig. \ref{dispfs}a.
Introducing extra electrons, the electron pockets would grow while the hole pockets shrink, and they would become degenerate at the filling $n_\text{vH}$.

The band dispersion and corresponding FS of AFM-LC at filling $n=4.95/12$ is presented in Fig. \ref{dispfs}b.
Clearly, the FS consists of a bone-shaped hole pocket centered around each M$'$ point.
The FS quasiparticles mainly come from the antibonding orbitals along M$'$-K$'$ direction, while there are strong hybridization between the bonding and antibonding orbitals along M$'$-$\Gamma$ direction. 
Despite the existence of LCs, the AFM-LC is topological trivial because of the preservation of the combined symmetry of TRS and inversion, as we discussed above.
Introducing extra electrons to this AFM-LC, the bone-shaped hole pockets break into two small hole pockets, and eventually shrink into two Dirac points at $n_\text{vH}$.
These Dirac points sit near M$'$ points and along the M$'$-K$'$ direction in the reduced BZ.
Therefore, AFM-LC is a Dirac semimetal at $n_\text{vH}$.

Figs. \ref{dispfs}c-\ref{dispfs}e show, respectively, the electronic structure and corresponding FS of FI1-LC at $n=4.95/12$, FI2-LC at $n=5.32/12$, and FI3-LC at $n=4.68/12$.
All FI-LCs are topologically nontrivial with the Chern numbers displayed in the figure.
The FS of FI1-LC at $n=4.95/12$ consists of two small and separated elliptic hole pockets near each M$'$ point and along M$'$-K$'$ direction.
Introducing extra electrons, FI1-LC would become a fully gaped Chern insulator at $n_\text{vH}$, with the total Chern number of all occupied bands $\mathcal{N}=1$.
The FS of FI2-LC at $n=5.32/12$ consists of large hole pockets centered around K$'$ points, while the FS of FI3-LC at $n=4.68/12$ consists of large electron pockets around K$'$ point in the reduced BZ. 

\subsection{Hall conductance under external magnetic field}
To this end, we comment on the Hall conductance of the bilayer kagom\'e lattice. 
In the absence of external field, the intrinsic contribution to the anomalous Hall conductivity (AHC) in a 2D system is given by 
\begin{equation}
\sigma_{xy} = -{e^2 \over 2\pi h} \sum_n \int_\text{RBZ} \Omega_n(\bk) f(\epsilon_{n\bk}) d\bk,
\end{equation}
where $\epsilon_{n\bk}$ is the energy dispersion of the $n$th band relative to the chemical potential, and $f$ is the Fermi distribution function.
Clearly, it contains contributions from both the fully and partially occupied Chern bands.
In the presence of external magnetic field $B$ along the $c$-axis, it couples to both the intrinsic spin and orbital angular momentum of the electrons via the Zeeman effect, which normalizes the energy dispersion of spin-$\sigma$ electrons to
\begin{equation}
\epsilon^\sigma_{n\bk}(B) = \epsilon_{n\bk}-\sigma B -m_n(\bk) B.
\end{equation}
Here, the orbital magnetic moment $m_n(\bk)$ induced by Berry curvature is given by
\begin{align}
m_n(\bk)&=-i{e\over 2\hbar} \langle \nabla_\bk u_{n\bk}| \times (H_\bk - \epsilon_{n\bk}) |\nabla_\bk u_{n\bk} \rangle \nonumber \\
&= -i {e\over 2\hbar} \left( \left\langle {\partial u_{n\bk} \over \partial k_x} \Big{|} (H_\bk - \epsilon_{n\bk}) \Big{|} {\partial u_{n\bk} \over \partial k_y} \right\rangle -c.c \right), \label{mk}
\end{align}
where $H_\bk$ is the Hamiltonian of the system.
An accurate description of CDWs in the presence of external magnetic field would requires self-consistent determination of the bond orders for any given magnetic field, which demands time-consuming calculations of the orbital magnetic moment $m_n(\bk)$ in each iteration.

\begin{table}[htb]
\caption{The state energy per site with respect to AFM-LC $\Delta E$, the AHC $\sigma_{xy}$ in the absence of external field, and the coefficients $\tilde{\alpha}$, $\tilde{\beta}$ of FI1-LC and FM-LC at band filling $n=4.95/12$ and interactions $V=(0.8t, 1.6t)$. 
Last row for the single-layer LC state at the same filling and interactions.
}\label{coeff}
\begin{ruledtabular}
\begin{tabular}{l|cc|cc}
State & $\Delta E$ $[10^{-3}t]$ & $\sigma_{xy}$ [$e^2/h$] & $\tilde{\alpha}$ $[ta^2_0e/2\hbar]$ & $\tilde{\beta}$ $[a^2_0e/2\hbar]$ \\
\hline
FI1-LC & $0.81$ & $0.41$ & $-5.11$ & $-16.53$ \\
FM-LC & $3.19$ & $1.74$ & $-5.51$ & $-31.04$ \\
Single-LC & / & 1.63 & $-5.88$ & $14.07$
\end{tabular}
\end{ruledtabular}
\end{table}

To a crude approximation in the linear response theory, the effects of Zeeman coupling with spin-up and spin-down electrons cancel each other, and all the effect comes from the coupling with orbital magnetic moments.
Explicitly, the the state energy per site and the AHC under the external field $B$ become
\begin{equation}
E(B) =E(0) -\tilde{\alpha} B, \qquad \sigma_{xy} (B) =\sigma_{xy} (0) +{e^2 \over h} \tilde{\beta} B, \label{linear}
\end{equation}
with the coefficient $\tilde{\alpha}$ and $\tilde{\beta}$ given by the band dispersion $\epsilon_{n\bk}$, Berry curvature $\Omega_n(\bk)$, and orbital magnetic moment $m_n(\bk)$ in the absence of external field,
\begin{align}
\tilde{\alpha}&=\sum_n \int_\text{RBZ} m_n(\bk) \left[ f(\epsilon_{n\bk}) +\epsilon_{n\bk} {\partial f \over \partial \epsilon } \right] d\bk, \label{alpha}\\
\tilde{\beta} &={1\over 2\pi} \sum_n \int_\text{RBZ} \Omega_n(\bk) m_n (\bk) {\partial f \over \partial \epsilon } d\bk. \label{beta}
\end{align}
At zero temperature, the derivative $\partial f /\partial \epsilon$ is nonzero only at FS with zero $\epsilon$.
The second term in the square bracket of Eq. (\ref{alpha}) thus vanishes.
Consequently, $\tilde{\alpha}=\sum_n \int_\text{RBZ} m_n(\bk) f(\epsilon_{n\bk})  d\bk$ which has contributions from all the states below Fermi level.
On the other hand, only the states on FS can contribute to the coefficient $\tilde{\beta}$ at zero temperature.

\begin{figure}
\begin{center}
\fig{3.4in}{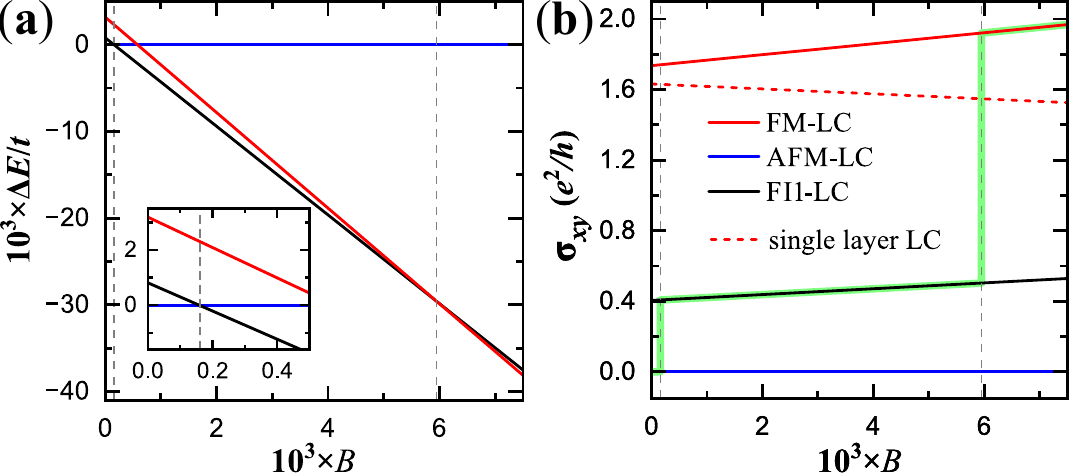}
\caption{Magnetic flux density $B$ dependence of (a) state energy per site with respect to AFM-LC $\Delta E$ and (b) the AHC $\sigma_{xy}$ of topologically nontrivial FI1-LC and FM-LC obtained at $V=(0.8t, 1.6t)$ and $n=4.95/12$.
The external field is applied along $-c$ direction and $B$ is measured in unit of $2\hbar/a^2_0 e \simeq 4521$ Tesla.
} \label{HallEng}
\end{center}
\end{figure}

We stay at band filling $n=4.95/12$ and interactions $V=(0.8t, 1.6t)$.
Let's first consider the single-layer kagom\'e lattice whose ground state is a LC CDW in the absence of external field.
Both TRS and its combination with inversion are broken in the single-layer LC CDW, thus it is topological nontrivial, giving rise to a nonzero AHC listed in Table \ref{coeff}.
The coefficients $\tilde{\alpha}$ and $\tilde{\beta}$ are also given in Table \ref{coeff}.
Clearly, $\tilde{\alpha}$ is negative, so its energy can be lowered by applying external field along $-c$ direction, $\textbf{B}=-B\hat{c}$.
The dashed red line in Fig. \ref{HallEng}b displays the AHC as a function of $B$, where $B$ is measured in unit of $2\hbar/a^2_0 e \simeq 4521$ Tesla, using the lattice constant $a_0 \simeq 5.4 \AA$.

Now we come to the bilayer case where the three converged LC CDWs in the absence of external field are AFM-LC, FI1-LC, and FM-LC.
The ground state AFM-LC preserves the combined symmetry of TRS and inversion, thus is topologically trivial with $\Omega_n(\bk) =m_n(\bk) =0$ for all $n$ and $\bk$. 
Consequently, the external field has no effect on it, its energy is unaltered and its AHC remains zero.
The metastable states, FI1-LC and FM-LC, are however topologically nontrivial.
Their state energies with respect to AFM-LC $\Delta E$ and AHC $\sigma_{xy}$ in the absence of external field are given in Table \ref{coeff}, together with the coefficients $\tilde{\alpha}$ and $\tilde{\beta}$.
The coefficient $\tilde{\alpha}$ is negative in both FI1-LC and FM-LC, so their energies in Eq. (\ref{linear}) would be lowered when the external field is applied along $-c$ direction, $\textbf{B}=-B\hat{c}$, as illustrated in Fig. \ref{HallEng}a.
Clearly, FI1-LC and FM-LC will successively take the place of ground state with the lowest energy at $B_1 \sim 0.73$ Tesla and $B_2 \sim 26.90$ Tesla.

The AHCs of AFM-LC, FI1-LC, and FM-LC on the bilayer kagom\'e lattice are plotted in Fig. \ref{HallEng}b as function of $B$ at $n$ = 4.95/12 and $V$ = $(0.8t$, $1.6t)$.
The ground state $\sigma_{xy}$ is displayed by the thick green line, which exhibits three segments of straight line with abrupt changes at $B_{1,2}$. 
As one switches on external field and increases its strength, the AHC of the bilayer kagom\'e lattice remains zero until the magnetic strength reaches $B_1$, where FI1-LC takes over as the ground state.
Its amplitude jumps to $\sim$0.41${e^2\over h}$ at $B_1$ and increases linearly with magnetic field until $B_2$, where it leaps to $\sim$1.92${e^2\over h}$ accompanying with the transition from FI1-LC to FM-LC.
Beyond $B_2$, the amplitude of AHC increases linearly with $B$.
If one use the line segment of $B>B_2$ to extract the intrinsic anomalous Hall conductivity, the obtained value would be that of FM-LC state in the absence of external field, which is estimated to be $\sigma_{xy}/c_0 \simeq -374$ $\Omega^{-1}\text{cm}^{-1}$, which accounts for the large intrinsic AHC (500 $\Omega^{-1}\text{cm}^{-1}$) observed in experiments \cite{YangShuoYing-SciAdv20, ChenXH-PRB21}.
Here, the $c$-axis lattice constant $c_0 \simeq 8.95\AA$ has been used in the estimation.

\section{Summary and Discussions}
\label{sec5}
In summary, we have investigated the inter-layer correlation of $2a_0 \times 2a_0$ bond-ordered CDWs on bilayer kagom\'e lattice.
The bilayer stacking susceptibilities of real and imaginary bond orders are calculated separately for the free electrons described by the bilayer TB model with inter-layer coupling $t_\perp=0.2t$, which splits the vH filling from $n_\text{vH}$ to $n_{\pm\text{vH}}$.
While the nn and nnn bonds have leading instabilities toward, respectively, real and imaginary bond-ordered CDWs, the preferred bilayer stacking pattern is found to be sensitive to the band filling $n$.
They tend to stack symmetrically near $n_{\pm\text{vH}}$ and antisymmetrically around $n_\text{vH}$, giving rise to, respectively, $2a_0 \times 2a_0 \times 1c_0$ and $2a_0 \times 2a_0 \times 2c_0$ CDWs.
The concrete single-orbital $t$-$t_\perp$-$V_1$-$V_2$ model is then studied on the bilayer kagom\'e lattice.
We obtained the mean-field ground states and determined the bilayer stacking pattern of the $2a_0 \times 2a_0$ CDWs as a function of filling.
At interaction $V=(0.8t, 0)$ and $V=(0, 1.2t)$ with solely nn or nnn Coulomb repulsion, the mean-field results agree remarkable with the weak-coupling instability analyses.
When both nn and nnn Coulomb repulsions are present, \text{e.g.}, $V=(0.8t, 1.6t)$, in addition to FM-LC and AFM-LC with equivalent strength of LCs on the two kagom\'e layers, we obtain three kinds of FI-LC where the strength of LCs on the two layers become different.
While TRS is broken in all LC CDWs due to the emergence of LCs, the combined symmetry of TRS and inversion is preserved in AFM-LC.
Consequently, AFM-LC is topologically trivial while FM-LC and FI-LC are topologically nontrivial with finite AHCs.
While the inter-layer coupling of CDWs is studied at these three sets of interactions, we note that the results are not very sensitive to the interactions.

The evolution of the ground state and corresponding AHC of the bilyaer kagom\'e lattice in the presence of external magnetic field is investigated within linear response theory at filling $n=4.95/12$ and interaction $V$ = $(0.8t, 1.6t)$.
The ground state is the topologically trivial AFM-LC in the absence of external field, and successively becomes FI1-LC and FM-LC at, respectively, $B_1 \sim 0.73$ Tesla and $B_2 \sim 26.90$ Tesla as increasing the magnetic flux density $B$.
The AHC $\sigma_{xy}$ possesses three line segments with abrupt changes at $B_{1,2}$.
Though the extracted intrinsic AHC from the behavior at $B>B_2$ agrees well with experiments, the discontinuous $\sigma_{xy}$ in Fig. \ref{HallEng}b show obvious discrepancy with experimental observations.
We expect it may partially caused by the fact that the feedback of the Zeeman coupling to the $2a_0 \times 2a_0$ CDWs is not considered in the linear response theory.
It would be desirable to carry out self-consistent calculations in the presence of external magnetic field, in which the bilayer kagom\'e lattice response to the external field.

We note in passing that this work has limited connections to the realistic \avs\ materials.
The CDWs considered in this work are restricted to those with $C_6$ rotation symmetry by lining up vertically the rotation center of the two kagom\'e layers, without any phase shift.
On the other hand, there are experimental and theoretical evidences \cite{YongHu-PRB22, Kang-NatMat22, Ritz-PRB23, Kautzsch-PRM23, Frassineti-PRR23, YuxinWang-PRB23, Scagnoli-JPCM24, wei2024three} for both with and without $\pi$-phase shift in the stacking of $2a_0 \times 2a_0$ CDWs in the real \avs\ material.
The $\pi$-phase shift would break the $C_6$ rotation symmetry and make it challenging to satisfy the continuity condition of LCs when developed.
Furthermore, a realistic model for \avs\ should be much more complicated than a single-orbital model, since its electronic structure contains three sets of vH singularity from V $d$-orbitals near the Fermi level and an electron pocket of Sb $p$-orbitals around the BZ center.
Therefore, we do not expect the effective single-orbital model studied in this work to provide any quantitative result capable of explaining the experimental observation of \avs.
On the other hand, this work provides theoretical insights on the inter-layer correlation of $2a_0 \times 2a_0$ CDWs, TRS breaking LC CDW in particular, on the bilayer kagom\'e lattice.
Such issues arise broadly in phenomena involving TRS breaking order parameters such as layered superfluid \cite{Leggett-prl91} and magnetoelectric states in bilayer graphene involving uniform LCs \cite{Varma-prb13}.

\section{Acknowledgments}
We thank Eduardo Fradkin for useful discussions.
This work is supported by the National Key R\&D Program of China (Grant No. 2022YFA1403800), the National Natural Science Foundation of China (Grant Nos. 12374153, 12047503, and 11974362), the Strategic Priority Research Program of CAS (Grant No. XDB28000000), the Basic Research Program of the CAS Based on Major Scientific Infrastructures (Grant No. JZHKYPT-2021-08) and the Postdoctoral Fellowship Program of CPSF (Grant No. GZC20241749).
Z.W. is supported by the U.S. Department of Energy, Basic Energy Sciences (Grant No. DE-FG02-99ER45747) and the Research Corporation for Science Advancement (Cottrell SEED Award No. 27856).
Numerical calculations in this work were performed on the HPC Cluster of ITP-CAS.

\bibliography{bilayerbib}

\begin{thebibliography}{76}%
\makeatletter
\providecommand \@ifxundefined [1]{%
 \@ifx{#1\undefined}
}%
\providecommand \@ifnum [1]{%
 \ifnum #1\expandafter \@firstoftwo
 \else \expandafter \@secondoftwo
 \fi
}%
\providecommand \@ifx [1]{%
 \ifx #1\expandafter \@firstoftwo
 \else \expandafter \@secondoftwo
 \fi
}%
\providecommand \natexlab [1]{#1}%
\providecommand \enquote  [1]{``#1''}%
\providecommand \bibnamefont  [1]{#1}%
\providecommand \bibfnamefont [1]{#1}%
\providecommand \citenamefont [1]{#1}%
\providecommand \href@noop [0]{\@secondoftwo}%
\providecommand \href [0]{\begingroup \@sanitize@url \@href}%
\providecommand \@href[1]{\@@startlink{#1}\@@href}%
\providecommand \@@href[1]{\endgroup#1\@@endlink}%
\providecommand \@sanitize@url [0]{\catcode `\\12\catcode `\$12\catcode
  `\&12\catcode `\#12\catcode `\^12\catcode `\_12\catcode `\%12\relax}%
\providecommand \@@startlink[1]{}%
\providecommand \@@endlink[0]{}%
\providecommand \url  [0]{\begingroup\@sanitize@url \@url }%
\providecommand \@url [1]{\endgroup\@href {#1}{\urlprefix }}%
\providecommand \urlprefix  [0]{URL }%
\providecommand \Eprint [0]{\href }%
\providecommand \doibase [0]{https://doi.org/}%
\providecommand \selectlanguage [0]{\@gobble}%
\providecommand \bibinfo  [0]{\@secondoftwo}%
\providecommand \bibfield  [0]{\@secondoftwo}%
\providecommand \translation [1]{[#1]}%
\providecommand \BibitemOpen [0]{}%
\providecommand \bibitemStop [0]{}%
\providecommand \bibitemNoStop [0]{.\EOS\space}%
\providecommand \EOS [0]{\spacefactor3000\relax}%
\providecommand \BibitemShut  [1]{\csname bibitem#1\endcsname}%
\let\auto@bib@innerbib\@empty
\bibitem [{\citenamefont {Ortiz}\ \emph {et~al.}(2019)\citenamefont {Ortiz},
  \citenamefont {Gomes}, \citenamefont {Morey}, \citenamefont {Winiarski},
  \citenamefont {Bordelon}, \citenamefont {Mangum}, \citenamefont {Oswald},
  \citenamefont {Rodriguez-Rivera}, \citenamefont {Neilson}, \citenamefont
  {Wilson}, \citenamefont {Ertekin}, \citenamefont {McQueen},\ and\
  \citenamefont {Toberer}}]{Ortiz-PRM19}%
  \BibitemOpen
  \bibfield  {author} {\bibinfo {author} {\bibfnamefont {B.~R.}\ \bibnamefont
  {Ortiz}}, \bibinfo {author} {\bibfnamefont {L.~C.}\ \bibnamefont {Gomes}},
  \bibinfo {author} {\bibfnamefont {J.~R.}\ \bibnamefont {Morey}}, \bibinfo
  {author} {\bibfnamefont {M.}~\bibnamefont {Winiarski}}, \bibinfo {author}
  {\bibfnamefont {M.}~\bibnamefont {Bordelon}}, \bibinfo {author}
  {\bibfnamefont {J.~S.}\ \bibnamefont {Mangum}}, \bibinfo {author}
  {\bibfnamefont {I.~W.~H.}\ \bibnamefont {Oswald}}, \bibinfo {author}
  {\bibfnamefont {J.~A.}\ \bibnamefont {Rodriguez-Rivera}}, \bibinfo {author}
  {\bibfnamefont {J.~R.}\ \bibnamefont {Neilson}}, \bibinfo {author}
  {\bibfnamefont {S.~D.}\ \bibnamefont {Wilson}}, \bibinfo {author}
  {\bibfnamefont {E.}~\bibnamefont {Ertekin}}, \bibinfo {author} {\bibfnamefont
  {T.~M.}\ \bibnamefont {McQueen}},\ and\ \bibinfo {author} {\bibfnamefont
  {E.~S.}\ \bibnamefont {Toberer}},\ }\bibfield  {title} {\bibinfo {title}
  {{New kagome prototype materials: discovery of
  ${\mathrm{KV}}_{3}{\mathrm{Sb}}_{5},{\mathrm{RbV}}_{3}{\mathrm{Sb}}_{5}$, and
  ${\mathrm{CsV}}_{3}{\mathrm{Sb}}_{5}$}},\ }\href
  {https://doi.org/10.1103/PhysRevMaterials.3.094407} {\bibfield  {journal}
  {\bibinfo  {journal} {Phys. Rev. Mater.}\ }\textbf {\bibinfo {volume} {3}},\
  \bibinfo {pages} {094407} (\bibinfo {year} {2019})}\BibitemShut {NoStop}%
\bibitem [{\citenamefont {Jiang}\ \emph {et~al.}(2021)\citenamefont {Jiang},
  \citenamefont {Yin}, \citenamefont {Denner}, \citenamefont {Shumiya},
  \citenamefont {Ortiz}, \citenamefont {Xu}, \citenamefont {Guguchia},
  \citenamefont {He}, \citenamefont {Hossain}, \citenamefont {Liu} \emph
  {et~al.}}]{YinJiaxin-21NatMater}%
  \BibitemOpen
  \bibfield  {author} {\bibinfo {author} {\bibfnamefont {Y.-X.}\ \bibnamefont
  {Jiang}}, \bibinfo {author} {\bibfnamefont {J.-X.}\ \bibnamefont {Yin}},
  \bibinfo {author} {\bibfnamefont {M.~M.}\ \bibnamefont {Denner}}, \bibinfo
  {author} {\bibfnamefont {N.}~\bibnamefont {Shumiya}}, \bibinfo {author}
  {\bibfnamefont {B.~R.}\ \bibnamefont {Ortiz}}, \bibinfo {author}
  {\bibfnamefont {G.}~\bibnamefont {Xu}}, \bibinfo {author} {\bibfnamefont
  {Z.}~\bibnamefont {Guguchia}}, \bibinfo {author} {\bibfnamefont
  {J.}~\bibnamefont {He}}, \bibinfo {author} {\bibfnamefont {M.~S.}\
  \bibnamefont {Hossain}}, \bibinfo {author} {\bibfnamefont {X.}~\bibnamefont
  {Liu}}, \emph {et~al.},\ }\bibfield  {title} {\bibinfo {title}
  {{Unconventional chiral charge order in kagome superconductor
  ${\mathrm{KV}}_{3}{\mathrm{Sb}}_{5}$}},\ }\href@noop {} {\bibfield  {journal}
  {\bibinfo  {journal} {Nature materials}\ }\textbf {\bibinfo {volume} {20}},\
  \bibinfo {pages} {1353} (\bibinfo {year} {2021})}\BibitemShut {NoStop}%
\bibitem [{\citenamefont {Zhao}\ \emph {et~al.}(2021)\citenamefont {Zhao},
  \citenamefont {Li}, \citenamefont {Ortiz}, \citenamefont {Teicher},
  \citenamefont {Park}, \citenamefont {Ye}, \citenamefont {Wang}, \citenamefont
  {Balents}, \citenamefont {Wilson},\ and\ \citenamefont
  {Zeljkovic}}]{HeZhao-21Nat}%
  \BibitemOpen
  \bibfield  {author} {\bibinfo {author} {\bibfnamefont {H.}~\bibnamefont
  {Zhao}}, \bibinfo {author} {\bibfnamefont {H.}~\bibnamefont {Li}}, \bibinfo
  {author} {\bibfnamefont {B.~R.}\ \bibnamefont {Ortiz}}, \bibinfo {author}
  {\bibfnamefont {S.~M.}\ \bibnamefont {Teicher}}, \bibinfo {author}
  {\bibfnamefont {T.}~\bibnamefont {Park}}, \bibinfo {author} {\bibfnamefont
  {M.}~\bibnamefont {Ye}}, \bibinfo {author} {\bibfnamefont {Z.}~\bibnamefont
  {Wang}}, \bibinfo {author} {\bibfnamefont {L.}~\bibnamefont {Balents}},
  \bibinfo {author} {\bibfnamefont {S.~D.}\ \bibnamefont {Wilson}},\ and\
  \bibinfo {author} {\bibfnamefont {I.}~\bibnamefont {Zeljkovic}},\ }\bibfield
  {title} {\bibinfo {title} {{Cascade of correlated electron states in the
  kagome superconductor ${\mathrm{CsV}}_{3}{\mathrm{Sb}}_{5}$}},\ }\href@noop
  {} {\bibfield  {journal} {\bibinfo  {journal} {Nature}\ }\textbf {\bibinfo
  {volume} {599}},\ \bibinfo {pages} {216} (\bibinfo {year}
  {2021})}\BibitemShut {NoStop}%
\bibitem [{\citenamefont {Ortiz}\ \emph
  {et~al.}(2021{\natexlab{a}})\citenamefont {Ortiz}, \citenamefont {Teicher},
  \citenamefont {Kautzsch}, \citenamefont {Sarte}, \citenamefont {Ratcliff},
  \citenamefont {Harter}, \citenamefont {Ruff}, \citenamefont {Seshadri},\ and\
  \citenamefont {Wilson}}]{Ortiz-PRX21}%
  \BibitemOpen
  \bibfield  {author} {\bibinfo {author} {\bibfnamefont {B.~R.}\ \bibnamefont
  {Ortiz}}, \bibinfo {author} {\bibfnamefont {S.~M.~L.}\ \bibnamefont
  {Teicher}}, \bibinfo {author} {\bibfnamefont {L.}~\bibnamefont {Kautzsch}},
  \bibinfo {author} {\bibfnamefont {P.~M.}\ \bibnamefont {Sarte}}, \bibinfo
  {author} {\bibfnamefont {N.}~\bibnamefont {Ratcliff}}, \bibinfo {author}
  {\bibfnamefont {J.}~\bibnamefont {Harter}}, \bibinfo {author} {\bibfnamefont
  {J.~P.~C.}\ \bibnamefont {Ruff}}, \bibinfo {author} {\bibfnamefont
  {R.}~\bibnamefont {Seshadri}},\ and\ \bibinfo {author} {\bibfnamefont
  {S.~D.}\ \bibnamefont {Wilson}},\ }\bibfield  {title} {\bibinfo {title}
  {{Fermi Surface Mapping and the Nature of Charge-Density-Wave Order in the
  Kagome Superconductor ${\mathrm{CsV}}_{3}{\mathrm{Sb}}_{5}$}},\ }\href
  {https://doi.org/10.1103/PhysRevX.11.041030} {\bibfield  {journal} {\bibinfo
  {journal} {Phys. Rev. X}\ }\textbf {\bibinfo {volume} {11}},\ \bibinfo
  {pages} {041030} (\bibinfo {year} {2021}{\natexlab{a}})}\BibitemShut
  {NoStop}%
\bibitem [{\citenamefont {Liang}\ \emph {et~al.}(2021)\citenamefont {Liang},
  \citenamefont {Hou}, \citenamefont {Zhang}, \citenamefont {Ma}, \citenamefont
  {Wu}, \citenamefont {Zhang}, \citenamefont {Yu}, \citenamefont {Ying},
  \citenamefont {Jiang}, \citenamefont {Shan}, \citenamefont {Wang},\ and\
  \citenamefont {Chen}}]{ChenXH-PRX21}%
  \BibitemOpen
  \bibfield  {author} {\bibinfo {author} {\bibfnamefont {Z.}~\bibnamefont
  {Liang}}, \bibinfo {author} {\bibfnamefont {X.}~\bibnamefont {Hou}}, \bibinfo
  {author} {\bibfnamefont {F.}~\bibnamefont {Zhang}}, \bibinfo {author}
  {\bibfnamefont {W.}~\bibnamefont {Ma}}, \bibinfo {author} {\bibfnamefont
  {P.}~\bibnamefont {Wu}}, \bibinfo {author} {\bibfnamefont {Z.}~\bibnamefont
  {Zhang}}, \bibinfo {author} {\bibfnamefont {F.}~\bibnamefont {Yu}}, \bibinfo
  {author} {\bibfnamefont {J.-J.}\ \bibnamefont {Ying}}, \bibinfo {author}
  {\bibfnamefont {K.}~\bibnamefont {Jiang}}, \bibinfo {author} {\bibfnamefont
  {L.}~\bibnamefont {Shan}}, \bibinfo {author} {\bibfnamefont {Z.}~\bibnamefont
  {Wang}},\ and\ \bibinfo {author} {\bibfnamefont {X.-H.}\ \bibnamefont
  {Chen}},\ }\bibfield  {title} {\bibinfo {title} {{Three-Dimensional Charge
  Density Wave and Surface-Dependent Vortex-Core States in a Kagome
  Superconductor ${\mathrm{CsV}}_{3}{\mathrm{Sb}}_{5}$}},\ }\href
  {https://doi.org/10.1103/PhysRevX.11.031026} {\bibfield  {journal} {\bibinfo
  {journal} {Phys. Rev. X}\ }\textbf {\bibinfo {volume} {11}},\ \bibinfo
  {pages} {031026} (\bibinfo {year} {2021})}\BibitemShut {NoStop}%
\bibitem [{\citenamefont {Shumiya}\ \emph {et~al.}(2021)\citenamefont
  {Shumiya}, \citenamefont {Hossain}, \citenamefont {Yin}, \citenamefont
  {Jiang}, \citenamefont {Ortiz}, \citenamefont {Liu}, \citenamefont {Shi},
  \citenamefont {Yin}, \citenamefont {Lei}, \citenamefont {Zhang},
  \citenamefont {Chang}, \citenamefont {Zhang}, \citenamefont {Cochran},
  \citenamefont {Multer}, \citenamefont {Litskevich}, \citenamefont {Cheng},
  \citenamefont {Yang}, \citenamefont {Guguchia}, \citenamefont {Wilson},\ and\
  \citenamefont {Hasan}}]{Nana-PRB21}%
  \BibitemOpen
  \bibfield  {author} {\bibinfo {author} {\bibfnamefont {N.}~\bibnamefont
  {Shumiya}}, \bibinfo {author} {\bibfnamefont {M.~S.}\ \bibnamefont
  {Hossain}}, \bibinfo {author} {\bibfnamefont {J.-X.}\ \bibnamefont {Yin}},
  \bibinfo {author} {\bibfnamefont {Y.-X.}\ \bibnamefont {Jiang}}, \bibinfo
  {author} {\bibfnamefont {B.~R.}\ \bibnamefont {Ortiz}}, \bibinfo {author}
  {\bibfnamefont {H.}~\bibnamefont {Liu}}, \bibinfo {author} {\bibfnamefont
  {Y.}~\bibnamefont {Shi}}, \bibinfo {author} {\bibfnamefont {Q.}~\bibnamefont
  {Yin}}, \bibinfo {author} {\bibfnamefont {H.}~\bibnamefont {Lei}}, \bibinfo
  {author} {\bibfnamefont {S.~S.}\ \bibnamefont {Zhang}}, \bibinfo {author}
  {\bibfnamefont {G.}~\bibnamefont {Chang}}, \bibinfo {author} {\bibfnamefont
  {Q.}~\bibnamefont {Zhang}}, \bibinfo {author} {\bibfnamefont {T.~A.}\
  \bibnamefont {Cochran}}, \bibinfo {author} {\bibfnamefont {D.}~\bibnamefont
  {Multer}}, \bibinfo {author} {\bibfnamefont {M.}~\bibnamefont {Litskevich}},
  \bibinfo {author} {\bibfnamefont {Z.-J.}\ \bibnamefont {Cheng}}, \bibinfo
  {author} {\bibfnamefont {X.~P.}\ \bibnamefont {Yang}}, \bibinfo {author}
  {\bibfnamefont {Z.}~\bibnamefont {Guguchia}}, \bibinfo {author}
  {\bibfnamefont {S.~D.}\ \bibnamefont {Wilson}},\ and\ \bibinfo {author}
  {\bibfnamefont {M.~Z.}\ \bibnamefont {Hasan}},\ }\bibfield  {title} {\bibinfo
  {title} {{Intrinsic nature of chiral charge order in the kagome
  superconductor $\mathrm{Rb}{\mathrm{V}}_{3}{\mathrm{Sb}}_{5}$}},\ }\href
  {https://doi.org/10.1103/PhysRevB.104.035131} {\bibfield  {journal} {\bibinfo
   {journal} {Phys. Rev. B}\ }\textbf {\bibinfo {volume} {104}},\ \bibinfo
  {pages} {035131} (\bibinfo {year} {2021})}\BibitemShut {NoStop}%
\bibitem [{\citenamefont {Uykur}\ \emph {et~al.}(2021)\citenamefont {Uykur},
  \citenamefont {Ortiz}, \citenamefont {Iakutkina}, \citenamefont {Wenzel},
  \citenamefont {Wilson}, \citenamefont {Dressel},\ and\ \citenamefont
  {Tsirlin}}]{Ortiz-PRB21}%
  \BibitemOpen
  \bibfield  {author} {\bibinfo {author} {\bibfnamefont {E.}~\bibnamefont
  {Uykur}}, \bibinfo {author} {\bibfnamefont {B.~R.}\ \bibnamefont {Ortiz}},
  \bibinfo {author} {\bibfnamefont {O.}~\bibnamefont {Iakutkina}}, \bibinfo
  {author} {\bibfnamefont {M.}~\bibnamefont {Wenzel}}, \bibinfo {author}
  {\bibfnamefont {S.~D.}\ \bibnamefont {Wilson}}, \bibinfo {author}
  {\bibfnamefont {M.}~\bibnamefont {Dressel}},\ and\ \bibinfo {author}
  {\bibfnamefont {A.~A.}\ \bibnamefont {Tsirlin}},\ }\bibfield  {title}
  {\bibinfo {title} {{Low-energy optical properties of the nonmagnetic kagome
  metal ${\mathrm{CsV}}_{3}{\mathrm{Sb}}_{5}$}},\ }\href
  {https://doi.org/10.1103/PhysRevB.104.045130} {\bibfield  {journal} {\bibinfo
   {journal} {Phys. Rev. B}\ }\textbf {\bibinfo {volume} {104}},\ \bibinfo
  {pages} {045130} (\bibinfo {year} {2021})}\BibitemShut {NoStop}%
\bibitem [{\citenamefont {Li}\ \emph {et~al.}(2022)\citenamefont {Li},
  \citenamefont {Zhao}, \citenamefont {Ortiz}, \citenamefont {Park},
  \citenamefont {Ye}, \citenamefont {Balents}, \citenamefont {Wang},
  \citenamefont {Wilson},\ and\ \citenamefont {Zeljkovic}}]{LiHong-NatPhys22}%
  \BibitemOpen
  \bibfield  {author} {\bibinfo {author} {\bibfnamefont {H.}~\bibnamefont
  {Li}}, \bibinfo {author} {\bibfnamefont {H.}~\bibnamefont {Zhao}}, \bibinfo
  {author} {\bibfnamefont {B.~R.}\ \bibnamefont {Ortiz}}, \bibinfo {author}
  {\bibfnamefont {T.}~\bibnamefont {Park}}, \bibinfo {author} {\bibfnamefont
  {M.}~\bibnamefont {Ye}}, \bibinfo {author} {\bibfnamefont {L.}~\bibnamefont
  {Balents}}, \bibinfo {author} {\bibfnamefont {Z.}~\bibnamefont {Wang}},
  \bibinfo {author} {\bibfnamefont {S.~D.}\ \bibnamefont {Wilson}},\ and\
  \bibinfo {author} {\bibfnamefont {I.}~\bibnamefont {Zeljkovic}},\ }\bibfield
  {title} {\bibinfo {title} {{Rotation symmetry breaking in the normal state of
  a kagome superconductor ${\mathrm{KV}}_{3}{\mathrm{Sb}}_{5}$}},\ }\href@noop
  {} {\bibfield  {journal} {\bibinfo  {journal} {Nature Physics}\ }\textbf
  {\bibinfo {volume} {18}},\ \bibinfo {pages} {265} (\bibinfo {year}
  {2022})}\BibitemShut {NoStop}%
\bibitem [{\citenamefont {Ortiz}\ \emph {et~al.}(2020)\citenamefont {Ortiz},
  \citenamefont {Teicher}, \citenamefont {Hu}, \citenamefont {Zuo},
  \citenamefont {Sarte}, \citenamefont {Schueller}, \citenamefont {Abeykoon},
  \citenamefont {Krogstad}, \citenamefont {Rosenkranz}, \citenamefont {Osborn},
  \citenamefont {Seshadri}, \citenamefont {Balents}, \citenamefont {He},\ and\
  \citenamefont {Wilson}}]{Ortiz-PRL20}%
  \BibitemOpen
  \bibfield  {author} {\bibinfo {author} {\bibfnamefont {B.~R.}\ \bibnamefont
  {Ortiz}}, \bibinfo {author} {\bibfnamefont {S.~M.~L.}\ \bibnamefont
  {Teicher}}, \bibinfo {author} {\bibfnamefont {Y.}~\bibnamefont {Hu}},
  \bibinfo {author} {\bibfnamefont {J.~L.}\ \bibnamefont {Zuo}}, \bibinfo
  {author} {\bibfnamefont {P.~M.}\ \bibnamefont {Sarte}}, \bibinfo {author}
  {\bibfnamefont {E.~C.}\ \bibnamefont {Schueller}}, \bibinfo {author}
  {\bibfnamefont {A.~M.~M.}\ \bibnamefont {Abeykoon}}, \bibinfo {author}
  {\bibfnamefont {M.~J.}\ \bibnamefont {Krogstad}}, \bibinfo {author}
  {\bibfnamefont {S.}~\bibnamefont {Rosenkranz}}, \bibinfo {author}
  {\bibfnamefont {R.}~\bibnamefont {Osborn}}, \bibinfo {author} {\bibfnamefont
  {R.}~\bibnamefont {Seshadri}}, \bibinfo {author} {\bibfnamefont
  {L.}~\bibnamefont {Balents}}, \bibinfo {author} {\bibfnamefont
  {J.}~\bibnamefont {He}},\ and\ \bibinfo {author} {\bibfnamefont {S.~D.}\
  \bibnamefont {Wilson}},\ }\bibfield  {title} {\bibinfo {title}
  {{$\mathrm{Cs}{\mathrm{V}}_{3}{\mathrm{Sb}}_{5}$: A ${\mathbb{Z}}_{2}$
  Topological Kagome Metal with a Superconducting Ground State}},\ }\href
  {https://doi.org/10.1103/PhysRevLett.125.247002} {\bibfield  {journal}
  {\bibinfo  {journal} {Phys. Rev. Lett.}\ }\textbf {\bibinfo {volume} {125}},\
  \bibinfo {pages} {247002} (\bibinfo {year} {2020})}\BibitemShut {NoStop}%
\bibitem [{\citenamefont {Ortiz}\ \emph
  {et~al.}(2021{\natexlab{b}})\citenamefont {Ortiz}, \citenamefont {Sarte},
  \citenamefont {Kenney}, \citenamefont {Graf}, \citenamefont {Teicher},
  \citenamefont {Seshadri},\ and\ \citenamefont {Wilson}}]{Ortiz-PRM21}%
  \BibitemOpen
  \bibfield  {author} {\bibinfo {author} {\bibfnamefont {B.~R.}\ \bibnamefont
  {Ortiz}}, \bibinfo {author} {\bibfnamefont {P.~M.}\ \bibnamefont {Sarte}},
  \bibinfo {author} {\bibfnamefont {E.~M.}\ \bibnamefont {Kenney}}, \bibinfo
  {author} {\bibfnamefont {M.~J.}\ \bibnamefont {Graf}}, \bibinfo {author}
  {\bibfnamefont {S.~M.~L.}\ \bibnamefont {Teicher}}, \bibinfo {author}
  {\bibfnamefont {R.}~\bibnamefont {Seshadri}},\ and\ \bibinfo {author}
  {\bibfnamefont {S.~D.}\ \bibnamefont {Wilson}},\ }\bibfield  {title}
  {\bibinfo {title} {{Superconductivity in the ${\mathbb{Z}}_{2}$ kagome metal
  ${\mathrm{KV}}_{3}{\mathrm{Sb}}_{5}$}},\ }\href
  {https://doi.org/10.1103/PhysRevMaterials.5.034801} {\bibfield  {journal}
  {\bibinfo  {journal} {Phys. Rev. Mater.}\ }\textbf {\bibinfo {volume} {5}},\
  \bibinfo {pages} {034801} (\bibinfo {year} {2021}{\natexlab{b}})}\BibitemShut
  {NoStop}%
\bibitem [{\citenamefont {Yin}\ \emph {et~al.}(2021)\citenamefont {Yin},
  \citenamefont {Tu}, \citenamefont {Gong}, \citenamefont {Fu}, \citenamefont
  {Yan},\ and\ \citenamefont {Lei}}]{YinQiangwei-CPL21}%
  \BibitemOpen
  \bibfield  {author} {\bibinfo {author} {\bibfnamefont {Q.}~\bibnamefont
  {Yin}}, \bibinfo {author} {\bibfnamefont {Z.}~\bibnamefont {Tu}}, \bibinfo
  {author} {\bibfnamefont {C.}~\bibnamefont {Gong}}, \bibinfo {author}
  {\bibfnamefont {Y.}~\bibnamefont {Fu}}, \bibinfo {author} {\bibfnamefont
  {S.}~\bibnamefont {Yan}},\ and\ \bibinfo {author} {\bibfnamefont
  {H.}~\bibnamefont {Lei}},\ }\bibfield  {title} {\bibinfo {title}
  {{Superconductivity and normal-state properties of kagome metal
  ${\mathrm{RbV}}_{3}{\mathrm{Sb}}_{5}$ single crystals}},\ }\href@noop {}
  {\bibfield  {journal} {\bibinfo  {journal} {Chinese Physics Letters}\
  }\textbf {\bibinfo {volume} {38}},\ \bibinfo {pages} {037403} (\bibinfo
  {year} {2021})}\BibitemShut {NoStop}%
\bibitem [{\citenamefont {Chen}\ \emph {et~al.}(2021)\citenamefont {Chen},
  \citenamefont {Yang}, \citenamefont {Hu}, \citenamefont {Zhao}, \citenamefont
  {Yuan}, \citenamefont {Xing}, \citenamefont {Qian}, \citenamefont {Huang},
  \citenamefont {Li}, \citenamefont {Ye} \emph {et~al.}}]{ChenHui-Nat21}%
  \BibitemOpen
  \bibfield  {author} {\bibinfo {author} {\bibfnamefont {H.}~\bibnamefont
  {Chen}}, \bibinfo {author} {\bibfnamefont {H.}~\bibnamefont {Yang}}, \bibinfo
  {author} {\bibfnamefont {B.}~\bibnamefont {Hu}}, \bibinfo {author}
  {\bibfnamefont {Z.}~\bibnamefont {Zhao}}, \bibinfo {author} {\bibfnamefont
  {J.}~\bibnamefont {Yuan}}, \bibinfo {author} {\bibfnamefont {Y.}~\bibnamefont
  {Xing}}, \bibinfo {author} {\bibfnamefont {G.}~\bibnamefont {Qian}}, \bibinfo
  {author} {\bibfnamefont {Z.}~\bibnamefont {Huang}}, \bibinfo {author}
  {\bibfnamefont {G.}~\bibnamefont {Li}}, \bibinfo {author} {\bibfnamefont
  {Y.}~\bibnamefont {Ye}}, \emph {et~al.},\ }\bibfield  {title} {\bibinfo
  {title} {{Roton pair density wave in a strong-coupling kagome
  superconductor}},\ }\href@noop {} {\bibfield  {journal} {\bibinfo  {journal}
  {Nature}\ }\textbf {\bibinfo {volume} {599}},\ \bibinfo {pages} {222}
  (\bibinfo {year} {2021})}\BibitemShut {NoStop}%
\bibitem [{\citenamefont {Tan}\ \emph {et~al.}(2021)\citenamefont {Tan},
  \citenamefont {Liu}, \citenamefont {Wang},\ and\ \citenamefont
  {Yan}}]{YanBinghai-PRL21}%
  \BibitemOpen
  \bibfield  {author} {\bibinfo {author} {\bibfnamefont {H.}~\bibnamefont
  {Tan}}, \bibinfo {author} {\bibfnamefont {Y.}~\bibnamefont {Liu}}, \bibinfo
  {author} {\bibfnamefont {Z.}~\bibnamefont {Wang}},\ and\ \bibinfo {author}
  {\bibfnamefont {B.}~\bibnamefont {Yan}},\ }\bibfield  {title} {\bibinfo
  {title} {{Charge Density Waves and Electronic Properties of Superconducting
  Kagome Metals}},\ }\href {https://doi.org/10.1103/PhysRevLett.127.046401}
  {\bibfield  {journal} {\bibinfo  {journal} {Phys. Rev. Lett.}\ }\textbf
  {\bibinfo {volume} {127}},\ \bibinfo {pages} {046401} (\bibinfo {year}
  {2021})}\BibitemShut {NoStop}%
\bibitem [{\citenamefont {Li}\ \emph {et~al.}(2021)\citenamefont {Li},
  \citenamefont {Zhang}, \citenamefont {Yilmaz}, \citenamefont {Pai},
  \citenamefont {Marvinney}, \citenamefont {Said}, \citenamefont {Yin},
  \citenamefont {Gong}, \citenamefont {Tu}, \citenamefont {Vescovo},
  \citenamefont {Nelson}, \citenamefont {Moore}, \citenamefont {Murakami},
  \citenamefont {Lei}, \citenamefont {Lee}, \citenamefont {Lawrie},\ and\
  \citenamefont {Miao}}]{LiHaoxiang-PRX21}%
  \BibitemOpen
  \bibfield  {author} {\bibinfo {author} {\bibfnamefont {H.}~\bibnamefont
  {Li}}, \bibinfo {author} {\bibfnamefont {T.~T.}\ \bibnamefont {Zhang}},
  \bibinfo {author} {\bibfnamefont {T.}~\bibnamefont {Yilmaz}}, \bibinfo
  {author} {\bibfnamefont {Y.~Y.}\ \bibnamefont {Pai}}, \bibinfo {author}
  {\bibfnamefont {C.~E.}\ \bibnamefont {Marvinney}}, \bibinfo {author}
  {\bibfnamefont {A.}~\bibnamefont {Said}}, \bibinfo {author} {\bibfnamefont
  {Q.~W.}\ \bibnamefont {Yin}}, \bibinfo {author} {\bibfnamefont {C.~S.}\
  \bibnamefont {Gong}}, \bibinfo {author} {\bibfnamefont {Z.~J.}\ \bibnamefont
  {Tu}}, \bibinfo {author} {\bibfnamefont {E.}~\bibnamefont {Vescovo}},
  \bibinfo {author} {\bibfnamefont {C.~S.}\ \bibnamefont {Nelson}}, \bibinfo
  {author} {\bibfnamefont {R.~G.}\ \bibnamefont {Moore}}, \bibinfo {author}
  {\bibfnamefont {S.}~\bibnamefont {Murakami}}, \bibinfo {author}
  {\bibfnamefont {H.~C.}\ \bibnamefont {Lei}}, \bibinfo {author} {\bibfnamefont
  {H.~N.}\ \bibnamefont {Lee}}, \bibinfo {author} {\bibfnamefont {B.~J.}\
  \bibnamefont {Lawrie}},\ and\ \bibinfo {author} {\bibfnamefont
  {H.}~\bibnamefont {Miao}},\ }\bibfield  {title} {\bibinfo {title}
  {{Observation of Unconventional Charge Density Wave without Acoustic Phonon
  Anomaly in Kagome Superconductors ${A\mathrm{V}}_{3}{\mathrm{Sb}}_{5}$
  ($A=\mathrm{Rb}$, Cs)}},\ }\href {https://doi.org/10.1103/PhysRevX.11.031050}
  {\bibfield  {journal} {\bibinfo  {journal} {Phys. Rev. X}\ }\textbf {\bibinfo
  {volume} {11}},\ \bibinfo {pages} {031050} (\bibinfo {year}
  {2021})}\BibitemShut {NoStop}%
\bibitem [{\citenamefont {Nie}\ \emph {et~al.}(2022)\citenamefont {Nie},
  \citenamefont {Sun}, \citenamefont {Ma}, \citenamefont {Song}, \citenamefont
  {Zheng}, \citenamefont {Liang}, \citenamefont {Wu}, \citenamefont {Yu},
  \citenamefont {Li}, \citenamefont {Shan} \emph
  {et~al.}}]{ChenXinHui-Nat2022}%
  \BibitemOpen
  \bibfield  {author} {\bibinfo {author} {\bibfnamefont {L.}~\bibnamefont
  {Nie}}, \bibinfo {author} {\bibfnamefont {K.}~\bibnamefont {Sun}}, \bibinfo
  {author} {\bibfnamefont {W.}~\bibnamefont {Ma}}, \bibinfo {author}
  {\bibfnamefont {D.}~\bibnamefont {Song}}, \bibinfo {author} {\bibfnamefont
  {L.}~\bibnamefont {Zheng}}, \bibinfo {author} {\bibfnamefont
  {Z.}~\bibnamefont {Liang}}, \bibinfo {author} {\bibfnamefont
  {P.}~\bibnamefont {Wu}}, \bibinfo {author} {\bibfnamefont {F.}~\bibnamefont
  {Yu}}, \bibinfo {author} {\bibfnamefont {J.}~\bibnamefont {Li}}, \bibinfo
  {author} {\bibfnamefont {M.}~\bibnamefont {Shan}}, \emph {et~al.},\
  }\bibfield  {title} {\bibinfo {title} {{Charge-density-wave-driven electronic
  nematicity in a kagome superconductor}},\ }\href@noop {} {\bibfield
  {journal} {\bibinfo  {journal} {Nature}\ }\textbf {\bibinfo {volume} {604}},\
  \bibinfo {pages} {59} (\bibinfo {year} {2022})}\BibitemShut {NoStop}%
\bibitem [{\citenamefont {Jiang}\ \emph {et~al.}(2023)\citenamefont {Jiang},
  \citenamefont {Ma}, \citenamefont {Xia}, \citenamefont {Liu}, \citenamefont
  {Xiao}, \citenamefont {Liu}, \citenamefont {Yang}, \citenamefont {Ding},
  \citenamefont {Huang}, \citenamefont {Liu} \emph
  {et~al.}}]{JiangZhicheng-NanoLett23}%
  \BibitemOpen
  \bibfield  {author} {\bibinfo {author} {\bibfnamefont {Z.}~\bibnamefont
  {Jiang}}, \bibinfo {author} {\bibfnamefont {H.}~\bibnamefont {Ma}}, \bibinfo
  {author} {\bibfnamefont {W.}~\bibnamefont {Xia}}, \bibinfo {author}
  {\bibfnamefont {Z.}~\bibnamefont {Liu}}, \bibinfo {author} {\bibfnamefont
  {Q.}~\bibnamefont {Xiao}}, \bibinfo {author} {\bibfnamefont {Z.}~\bibnamefont
  {Liu}}, \bibinfo {author} {\bibfnamefont {Y.}~\bibnamefont {Yang}}, \bibinfo
  {author} {\bibfnamefont {J.}~\bibnamefont {Ding}}, \bibinfo {author}
  {\bibfnamefont {Z.}~\bibnamefont {Huang}}, \bibinfo {author} {\bibfnamefont
  {J.}~\bibnamefont {Liu}}, \emph {et~al.},\ }\bibfield  {title} {\bibinfo
  {title} {{Observation of electronic nematicity driven by the
  three-dimensional charge density wave in kagome lattice
  ${\mathrm{KV}}_{3}{\mathrm{Sb}}_{5}$}},\ }\href@noop {} {\bibfield  {journal}
  {\bibinfo  {journal} {Nano Letters}\ }\textbf {\bibinfo {volume} {23}},\
  \bibinfo {pages} {5625} (\bibinfo {year} {2023})}\BibitemShut {NoStop}%
\bibitem [{\citenamefont {Xing}\ \emph {et~al.}(2023)\citenamefont {Xing},
  \citenamefont {Bae}, \citenamefont {Ritz}, \citenamefont {Yang},
  \citenamefont {Birol}, \citenamefont {Salinas}, \citenamefont {Ortiz},
  \citenamefont {Wilson}, \citenamefont {Wang}, \citenamefont {Fernandes} \emph
  {et~al.}}]{Xing-arxiv23}%
  \BibitemOpen
  \bibfield  {author} {\bibinfo {author} {\bibfnamefont {Y.}~\bibnamefont
  {Xing}}, \bibinfo {author} {\bibfnamefont {S.}~\bibnamefont {Bae}}, \bibinfo
  {author} {\bibfnamefont {E.}~\bibnamefont {Ritz}}, \bibinfo {author}
  {\bibfnamefont {F.}~\bibnamefont {Yang}}, \bibinfo {author} {\bibfnamefont
  {T.}~\bibnamefont {Birol}}, \bibinfo {author} {\bibfnamefont {A.~N.}\
  \bibnamefont {Salinas}}, \bibinfo {author} {\bibfnamefont {B.~R.}\
  \bibnamefont {Ortiz}}, \bibinfo {author} {\bibfnamefont {S.~D.}\ \bibnamefont
  {Wilson}}, \bibinfo {author} {\bibfnamefont {Z.}~\bibnamefont {Wang}},
  \bibinfo {author} {\bibfnamefont {R.~M.}\ \bibnamefont {Fernandes}}, \emph
  {et~al.},\ }\bibfield  {title} {\bibinfo {title} {{Optical manipulation of
  the charge density wave state in ${\mathrm{RbV}}_{3}{\mathrm{Sb}}_{5}$}},\
  }\href@noop {} {\bibfield  {journal} {\bibinfo  {journal} {arXiv preprint
  arXiv:2308.04128}\ } (\bibinfo {year} {2023})}\BibitemShut {NoStop}%
\bibitem [{\citenamefont {Kenney}\ \emph {et~al.}(2021)\citenamefont {Kenney},
  \citenamefont {Ortiz}, \citenamefont {Wang}, \citenamefont {Wilson},\ and\
  \citenamefont {Graf}}]{Kenney-IOP21}%
  \BibitemOpen
  \bibfield  {author} {\bibinfo {author} {\bibfnamefont {E.~M.}\ \bibnamefont
  {Kenney}}, \bibinfo {author} {\bibfnamefont {B.~R.}\ \bibnamefont {Ortiz}},
  \bibinfo {author} {\bibfnamefont {C.}~\bibnamefont {Wang}}, \bibinfo {author}
  {\bibfnamefont {S.~D.}\ \bibnamefont {Wilson}},\ and\ \bibinfo {author}
  {\bibfnamefont {M.~J.}\ \bibnamefont {Graf}},\ }\bibfield  {title} {\bibinfo
  {title} {{Absence of local moments in the kagome metal
  ${\mathrm{KV}}_{3}{\mathrm{Sb}}_{5}$ as determined by muon spin
  spectroscopy}},\ }\href {https://doi.org/10.1088/1361-648X/abe8f9} {\bibfield
   {journal} {\bibinfo  {journal} {Journal of Physics: Condensed Matter}\
  }\textbf {\bibinfo {volume} {33}},\ \bibinfo {pages} {235801} (\bibinfo
  {year} {2021})}\BibitemShut {NoStop}%
\bibitem [{\citenamefont {Mielke~III}\ \emph {et~al.}(2022)\citenamefont
  {Mielke~III}, \citenamefont {Das}, \citenamefont {Yin}, \citenamefont {Liu},
  \citenamefont {Gupta}, \citenamefont {Jiang}, \citenamefont {Medarde},
  \citenamefont {Wu}, \citenamefont {Lei}, \citenamefont {Chang} \emph
  {et~al.}}]{Mielke-Nat22}%
  \BibitemOpen
  \bibfield  {author} {\bibinfo {author} {\bibfnamefont {C.}~\bibnamefont
  {Mielke~III}}, \bibinfo {author} {\bibfnamefont {D.}~\bibnamefont {Das}},
  \bibinfo {author} {\bibfnamefont {J.-X.}\ \bibnamefont {Yin}}, \bibinfo
  {author} {\bibfnamefont {H.}~\bibnamefont {Liu}}, \bibinfo {author}
  {\bibfnamefont {R.}~\bibnamefont {Gupta}}, \bibinfo {author} {\bibfnamefont
  {Y.-X.}\ \bibnamefont {Jiang}}, \bibinfo {author} {\bibfnamefont
  {M.}~\bibnamefont {Medarde}}, \bibinfo {author} {\bibfnamefont
  {X.}~\bibnamefont {Wu}}, \bibinfo {author} {\bibfnamefont {H.~C.}\
  \bibnamefont {Lei}}, \bibinfo {author} {\bibfnamefont {J.}~\bibnamefont
  {Chang}}, \emph {et~al.},\ }\bibfield  {title} {\bibinfo {title}
  {{Time-reversal symmetry-breaking charge order in a kagome superconductor}},\
  }\href@noop {} {\bibfield  {journal} {\bibinfo  {journal} {Nature}\ }\textbf
  {\bibinfo {volume} {602}},\ \bibinfo {pages} {245} (\bibinfo {year}
  {2022})}\BibitemShut {NoStop}%
\bibitem [{\citenamefont {Yu}\ \emph {et~al.}(2021{\natexlab{a}})\citenamefont
  {Yu}, \citenamefont {Wang}, \citenamefont {Zhang}, \citenamefont {Sander},
  \citenamefont {Ni}, \citenamefont {Lu}, \citenamefont {Ma}, \citenamefont
  {Wang}, \citenamefont {Zhao}, \citenamefont {Chen} \emph
  {et~al.}}]{YuLi-arxiv21}%
  \BibitemOpen
  \bibfield  {author} {\bibinfo {author} {\bibfnamefont {L.}~\bibnamefont
  {Yu}}, \bibinfo {author} {\bibfnamefont {C.}~\bibnamefont {Wang}}, \bibinfo
  {author} {\bibfnamefont {Y.}~\bibnamefont {Zhang}}, \bibinfo {author}
  {\bibfnamefont {M.}~\bibnamefont {Sander}}, \bibinfo {author} {\bibfnamefont
  {S.}~\bibnamefont {Ni}}, \bibinfo {author} {\bibfnamefont {Z.}~\bibnamefont
  {Lu}}, \bibinfo {author} {\bibfnamefont {S.}~\bibnamefont {Ma}}, \bibinfo
  {author} {\bibfnamefont {Z.}~\bibnamefont {Wang}}, \bibinfo {author}
  {\bibfnamefont {Z.}~\bibnamefont {Zhao}}, \bibinfo {author} {\bibfnamefont
  {H.}~\bibnamefont {Chen}}, \emph {et~al.},\ }\bibfield  {title} {\bibinfo
  {title} {{Evidence of a hidden flux phase in the topological kagome metal
  ${\mathrm{CsV}}_{3}{\mathrm{Sb}}_{5}$}},\ }\href@noop {} {\bibfield
  {journal} {\bibinfo  {journal} {arXiv preprint arXiv:2107.10714}\ } (\bibinfo
  {year} {2021}{\natexlab{a}})}\BibitemShut {NoStop}%
\bibitem [{\citenamefont {Wu}\ \emph {et~al.}(2022{\natexlab{a}})\citenamefont
  {Wu}, \citenamefont {Wang}, \citenamefont {Liu}, \citenamefont {Li},
  \citenamefont {Xu}, \citenamefont {Yin}, \citenamefont {Gong}, \citenamefont
  {Tu}, \citenamefont {Lei}, \citenamefont {Dong},\ and\ \citenamefont
  {Wang}}]{WuQiong-PRB22}%
  \BibitemOpen
  \bibfield  {author} {\bibinfo {author} {\bibfnamefont {Q.}~\bibnamefont
  {Wu}}, \bibinfo {author} {\bibfnamefont {Z.~X.}\ \bibnamefont {Wang}},
  \bibinfo {author} {\bibfnamefont {Q.~M.}\ \bibnamefont {Liu}}, \bibinfo
  {author} {\bibfnamefont {R.~S.}\ \bibnamefont {Li}}, \bibinfo {author}
  {\bibfnamefont {S.~X.}\ \bibnamefont {Xu}}, \bibinfo {author} {\bibfnamefont
  {Q.~W.}\ \bibnamefont {Yin}}, \bibinfo {author} {\bibfnamefont {C.~S.}\
  \bibnamefont {Gong}}, \bibinfo {author} {\bibfnamefont {Z.~J.}\ \bibnamefont
  {Tu}}, \bibinfo {author} {\bibfnamefont {H.~C.}\ \bibnamefont {Lei}},
  \bibinfo {author} {\bibfnamefont {T.}~\bibnamefont {Dong}},\ and\ \bibinfo
  {author} {\bibfnamefont {N.~L.}\ \bibnamefont {Wang}},\ }\bibfield  {title}
  {\bibinfo {title} {{Simultaneous formation of two-fold rotation symmetry with
  charge order in the kagome superconductor
  ${\mathrm{CsV}}_{3}{\mathrm{Sb}}_{5}$ by optical polarization rotation
  measurement}},\ }\href {https://doi.org/10.1103/PhysRevB.106.205109}
  {\bibfield  {journal} {\bibinfo  {journal} {Phys. Rev. B}\ }\textbf {\bibinfo
  {volume} {106}},\ \bibinfo {pages} {205109} (\bibinfo {year}
  {2022}{\natexlab{a}})}\BibitemShut {NoStop}%
\bibitem [{\citenamefont {Xu}\ \emph {et~al.}(2022)\citenamefont {Xu},
  \citenamefont {Ni}, \citenamefont {Liu}, \citenamefont {Ortiz}, \citenamefont
  {Deng}, \citenamefont {Wilson}, \citenamefont {Yan}, \citenamefont
  {Balents},\ and\ \citenamefont {Wu}}]{XuYishuai-NatPhys22}%
  \BibitemOpen
  \bibfield  {author} {\bibinfo {author} {\bibfnamefont {Y.}~\bibnamefont
  {Xu}}, \bibinfo {author} {\bibfnamefont {Z.}~\bibnamefont {Ni}}, \bibinfo
  {author} {\bibfnamefont {Y.}~\bibnamefont {Liu}}, \bibinfo {author}
  {\bibfnamefont {B.~R.}\ \bibnamefont {Ortiz}}, \bibinfo {author}
  {\bibfnamefont {Q.}~\bibnamefont {Deng}}, \bibinfo {author} {\bibfnamefont
  {S.~D.}\ \bibnamefont {Wilson}}, \bibinfo {author} {\bibfnamefont
  {B.}~\bibnamefont {Yan}}, \bibinfo {author} {\bibfnamefont {L.}~\bibnamefont
  {Balents}},\ and\ \bibinfo {author} {\bibfnamefont {L.}~\bibnamefont {Wu}},\
  }\bibfield  {title} {\bibinfo {title} {{Three-state nematicity and
  magneto-optical Kerr effect in the charge density waves in kagome
  superconductors}},\ }\href@noop {} {\bibfield  {journal} {\bibinfo  {journal}
  {Nature physics}\ }\textbf {\bibinfo {volume} {18}},\ \bibinfo {pages} {1470}
  (\bibinfo {year} {2022})}\BibitemShut {NoStop}%
\bibitem [{\citenamefont {Hu}\ \emph {et~al.}(2022{\natexlab{a}})\citenamefont
  {Hu}, \citenamefont {Yamane}, \citenamefont {Mattoni}, \citenamefont {Yada},
  \citenamefont {Obata}, \citenamefont {Li}, \citenamefont {Yao}, \citenamefont
  {Wang}, \citenamefont {Wang}, \citenamefont {Farhang} \emph
  {et~al.}}]{HuYajian-arxiv22}%
  \BibitemOpen
  \bibfield  {author} {\bibinfo {author} {\bibfnamefont {Y.}~\bibnamefont
  {Hu}}, \bibinfo {author} {\bibfnamefont {S.}~\bibnamefont {Yamane}}, \bibinfo
  {author} {\bibfnamefont {G.}~\bibnamefont {Mattoni}}, \bibinfo {author}
  {\bibfnamefont {K.}~\bibnamefont {Yada}}, \bibinfo {author} {\bibfnamefont
  {K.}~\bibnamefont {Obata}}, \bibinfo {author} {\bibfnamefont
  {Y.}~\bibnamefont {Li}}, \bibinfo {author} {\bibfnamefont {Y.}~\bibnamefont
  {Yao}}, \bibinfo {author} {\bibfnamefont {Z.}~\bibnamefont {Wang}}, \bibinfo
  {author} {\bibfnamefont {J.}~\bibnamefont {Wang}}, \bibinfo {author}
  {\bibfnamefont {C.}~\bibnamefont {Farhang}}, \emph {et~al.},\ }\bibfield
  {title} {\bibinfo {title} {{Time-reversal symmetry breaking in charge density
  wave of ${\mathrm{CsV}}_{3}{\mathrm{Sb}}_{5}$ detected by polar Kerr
  effect}},\ }\href@noop {} {\bibfield  {journal} {\bibinfo  {journal} {arXiv
  preprint arXiv:2208.08036}\ } (\bibinfo {year}
  {2022}{\natexlab{a}})}\BibitemShut {NoStop}%
\bibitem [{\citenamefont {Guo}\ \emph {et~al.}(2022)\citenamefont {Guo},
  \citenamefont {Putzke}, \citenamefont {Konyzheva}, \citenamefont {Huang},
  \citenamefont {Gutierrez-Amigo}, \citenamefont {Errea}, \citenamefont {Chen},
  \citenamefont {Vergniory}, \citenamefont {Felser}, \citenamefont {Fischer}
  \emph {et~al.}}]{GuoChunyu-Nat22}%
  \BibitemOpen
  \bibfield  {author} {\bibinfo {author} {\bibfnamefont {C.}~\bibnamefont
  {Guo}}, \bibinfo {author} {\bibfnamefont {C.}~\bibnamefont {Putzke}},
  \bibinfo {author} {\bibfnamefont {S.}~\bibnamefont {Konyzheva}}, \bibinfo
  {author} {\bibfnamefont {X.}~\bibnamefont {Huang}}, \bibinfo {author}
  {\bibfnamefont {M.}~\bibnamefont {Gutierrez-Amigo}}, \bibinfo {author}
  {\bibfnamefont {I.}~\bibnamefont {Errea}}, \bibinfo {author} {\bibfnamefont
  {D.}~\bibnamefont {Chen}}, \bibinfo {author} {\bibfnamefont {M.~G.}\
  \bibnamefont {Vergniory}}, \bibinfo {author} {\bibfnamefont {C.}~\bibnamefont
  {Felser}}, \bibinfo {author} {\bibfnamefont {M.~H.}\ \bibnamefont {Fischer}},
  \emph {et~al.},\ }\bibfield  {title} {\bibinfo {title} {{Switchable chiral
  transport in charge-ordered kagome metal
  ${\mathrm{CsV}}_{3}{\mathrm{Sb}}_{5}$}},\ }\href@noop {} {\bibfield
  {journal} {\bibinfo  {journal} {Nature}\ }\textbf {\bibinfo {volume} {611}},\
  \bibinfo {pages} {461} (\bibinfo {year} {2022})}\BibitemShut {NoStop}%
\bibitem [{\citenamefont {Yang}\ \emph {et~al.}(2020)\citenamefont {Yang},
  \citenamefont {Wang}, \citenamefont {Ortiz}, \citenamefont {Liu},
  \citenamefont {Gayles}, \citenamefont {Derunova}, \citenamefont
  {Gonzalez-Hernandez}, \citenamefont {Šmejkal}, \citenamefont {Chen},
  \citenamefont {Parkin}, \citenamefont {Wilson}, \citenamefont {Toberer},
  \citenamefont {McQueen},\ and\ \citenamefont {Ali}}]{YangShuoYing-SciAdv20}%
  \BibitemOpen
  \bibfield  {author} {\bibinfo {author} {\bibfnamefont {S.-Y.}\ \bibnamefont
  {Yang}}, \bibinfo {author} {\bibfnamefont {Y.}~\bibnamefont {Wang}}, \bibinfo
  {author} {\bibfnamefont {B.~R.}\ \bibnamefont {Ortiz}}, \bibinfo {author}
  {\bibfnamefont {D.}~\bibnamefont {Liu}}, \bibinfo {author} {\bibfnamefont
  {J.}~\bibnamefont {Gayles}}, \bibinfo {author} {\bibfnamefont
  {E.}~\bibnamefont {Derunova}}, \bibinfo {author} {\bibfnamefont
  {R.}~\bibnamefont {Gonzalez-Hernandez}}, \bibinfo {author} {\bibfnamefont
  {L.}~\bibnamefont {Šmejkal}}, \bibinfo {author} {\bibfnamefont
  {Y.}~\bibnamefont {Chen}}, \bibinfo {author} {\bibfnamefont {S.~S.~P.}\
  \bibnamefont {Parkin}}, \bibinfo {author} {\bibfnamefont {S.~D.}\
  \bibnamefont {Wilson}}, \bibinfo {author} {\bibfnamefont {E.~S.}\
  \bibnamefont {Toberer}}, \bibinfo {author} {\bibfnamefont {T.}~\bibnamefont
  {McQueen}},\ and\ \bibinfo {author} {\bibfnamefont {M.~N.}\ \bibnamefont
  {Ali}},\ }\bibfield  {title} {\bibinfo {title} {{Giant, unconventional
  anomalous Hall effect in the metallic frustrated magnet candidate,
  ${\mathrm{KV}}_{3}{\mathrm{Sb}}_{5}$}},\ }\href
  {https://doi.org/10.1126/sciadv.abb6003} {\bibfield  {journal} {\bibinfo
  {journal} {Science Advances}\ }\textbf {\bibinfo {volume} {6}},\ \bibinfo
  {pages} {eabb6003} (\bibinfo {year} {2020})},\ \Eprint
  {https://arxiv.org/abs/https://www.science.org/doi/pdf/10.1126/sciadv.abb6003}
  {https://www.science.org/doi/pdf/10.1126/sciadv.abb6003} \BibitemShut
  {NoStop}%
\bibitem [{\citenamefont {Yu}\ \emph {et~al.}(2021{\natexlab{b}})\citenamefont
  {Yu}, \citenamefont {Wu}, \citenamefont {Wang}, \citenamefont {Lei},
  \citenamefont {Zhuo}, \citenamefont {Ying},\ and\ \citenamefont
  {Chen}}]{ChenXH-PRB21}%
  \BibitemOpen
  \bibfield  {author} {\bibinfo {author} {\bibfnamefont {F.~H.}\ \bibnamefont
  {Yu}}, \bibinfo {author} {\bibfnamefont {T.}~\bibnamefont {Wu}}, \bibinfo
  {author} {\bibfnamefont {Z.~Y.}\ \bibnamefont {Wang}}, \bibinfo {author}
  {\bibfnamefont {B.}~\bibnamefont {Lei}}, \bibinfo {author} {\bibfnamefont
  {W.~Z.}\ \bibnamefont {Zhuo}}, \bibinfo {author} {\bibfnamefont {J.~J.}\
  \bibnamefont {Ying}},\ and\ \bibinfo {author} {\bibfnamefont {X.~H.}\
  \bibnamefont {Chen}},\ }\bibfield  {title} {\bibinfo {title} {{Concurrence of
  anomalous Hall effect and charge density wave in a superconducting
  topological kagome metal}},\ }\href
  {https://doi.org/10.1103/PhysRevB.104.L041103} {\bibfield  {journal}
  {\bibinfo  {journal} {Phys. Rev. B}\ }\textbf {\bibinfo {volume} {104}},\
  \bibinfo {pages} {L041103} (\bibinfo {year}
  {2021}{\natexlab{b}})}\BibitemShut {NoStop}%
\bibitem [{\citenamefont {Ge}\ \emph {et~al.}(2022)\citenamefont {Ge},
  \citenamefont {Wang}, \citenamefont {Xing}, \citenamefont {Yin},
  \citenamefont {Lei}, \citenamefont {Wang},\ and\ \citenamefont
  {Wang}}]{GeJun-arxiv22}%
  \BibitemOpen
  \bibfield  {author} {\bibinfo {author} {\bibfnamefont {J.}~\bibnamefont
  {Ge}}, \bibinfo {author} {\bibfnamefont {P.}~\bibnamefont {Wang}}, \bibinfo
  {author} {\bibfnamefont {Y.}~\bibnamefont {Xing}}, \bibinfo {author}
  {\bibfnamefont {Q.}~\bibnamefont {Yin}}, \bibinfo {author} {\bibfnamefont
  {H.}~\bibnamefont {Lei}}, \bibinfo {author} {\bibfnamefont {Z.}~\bibnamefont
  {Wang}},\ and\ \bibinfo {author} {\bibfnamefont {J.}~\bibnamefont {Wang}},\
  }\bibfield  {title} {\bibinfo {title} {{Discovery of charge-4e and charge-6e
  superconductivity in kagome superconductor
  ${\mathrm{CsV}}_{3}{\mathrm{Sb}}_{5}$}},\ }\href@noop {} {\bibfield
  {journal} {\bibinfo  {journal} {arXiv preprint arXiv:2201.10352}\ } (\bibinfo
  {year} {2022})}\BibitemShut {NoStop}%
\bibitem [{\citenamefont {Feng}\ \emph
  {et~al.}(2021{\natexlab{a}})\citenamefont {Feng}, \citenamefont {Jiang},
  \citenamefont {Wang},\ and\ \citenamefont {Hu}}]{FengXilin-SciBull21}%
  \BibitemOpen
  \bibfield  {author} {\bibinfo {author} {\bibfnamefont {X.}~\bibnamefont
  {Feng}}, \bibinfo {author} {\bibfnamefont {K.}~\bibnamefont {Jiang}},
  \bibinfo {author} {\bibfnamefont {Z.}~\bibnamefont {Wang}},\ and\ \bibinfo
  {author} {\bibfnamefont {J.}~\bibnamefont {Hu}},\ }\bibfield  {title}
  {\bibinfo {title} {{Chiral flux phase in the Kagome superconductor AV3Sb5}},\
  }\href {https://doi.org/https://doi.org/10.1016/j.scib.2021.04.043}
  {\bibfield  {journal} {\bibinfo  {journal} {Science Bulletin}\ }\textbf
  {\bibinfo {volume} {66}},\ \bibinfo {pages} {1384} (\bibinfo {year}
  {2021}{\natexlab{a}})}\BibitemShut {NoStop}%
\bibitem [{\citenamefont {Denner}\ \emph {et~al.}(2021)\citenamefont {Denner},
  \citenamefont {Thomale},\ and\ \citenamefont {Neupert}}]{Denner-PRL21}%
  \BibitemOpen
  \bibfield  {author} {\bibinfo {author} {\bibfnamefont {M.~M.}\ \bibnamefont
  {Denner}}, \bibinfo {author} {\bibfnamefont {R.}~\bibnamefont {Thomale}},\
  and\ \bibinfo {author} {\bibfnamefont {T.}~\bibnamefont {Neupert}},\
  }\bibfield  {title} {\bibinfo {title} {{Analysis of Charge Order in the
  Kagome Metal $A{\mathrm{V}}_{3}{\mathrm{Sb}}_{5}$
  ($A=\mathrm{K},\mathrm{Rb},\mathrm{Cs}$)}},\ }\href
  {https://doi.org/10.1103/PhysRevLett.127.217601} {\bibfield  {journal}
  {\bibinfo  {journal} {Phys. Rev. Lett.}\ }\textbf {\bibinfo {volume} {127}},\
  \bibinfo {pages} {217601} (\bibinfo {year} {2021})}\BibitemShut {NoStop}%
\bibitem [{\citenamefont {Zhou}\ and\ \citenamefont {Wang}(2022)}]{SZ-NC21}%
  \BibitemOpen
  \bibfield  {author} {\bibinfo {author} {\bibfnamefont {S.}~\bibnamefont
  {Zhou}}\ and\ \bibinfo {author} {\bibfnamefont {Z.}~\bibnamefont {Wang}},\
  }\bibfield  {title} {\bibinfo {title} {{Chern Fermi pocket, topological pair
  density wave, and charge-4e and charge-6e superconductivity in kagom\'e
  superconductors}},\ }\href {https://doi.org/10.1038/s41467-022-34832-2}
  {\bibfield  {journal} {\bibinfo  {journal} {Nat. Commun.}\ }\textbf {\bibinfo
  {volume} {13}},\ \bibinfo {pages} {7288} (\bibinfo {year}
  {2022})}\BibitemShut {NoStop}%
\bibitem [{\citenamefont {Affleck}\ and\ \citenamefont
  {Marston}(1988)}]{Affleck-PRB1988}%
  \BibitemOpen
  \bibfield  {author} {\bibinfo {author} {\bibfnamefont {I.}~\bibnamefont
  {Affleck}}\ and\ \bibinfo {author} {\bibfnamefont {J.~B.}\ \bibnamefont
  {Marston}},\ }\bibfield  {title} {\bibinfo {title} {{Large-n limit of the
  Heisenberg-Hubbard model: Implications for high-${T}_{c}$ superconductors}},\
  }\href {https://doi.org/10.1103/PhysRevB.37.3774} {\bibfield  {journal}
  {\bibinfo  {journal} {Phys. Rev. B}\ }\textbf {\bibinfo {volume} {37}},\
  \bibinfo {pages} {3774} (\bibinfo {year} {1988})}\BibitemShut {NoStop}%
\bibitem [{\citenamefont {Varma}(1997)}]{Varma-PRB1997}%
  \BibitemOpen
  \bibfield  {author} {\bibinfo {author} {\bibfnamefont {C.~M.}\ \bibnamefont
  {Varma}},\ }\bibfield  {title} {\bibinfo {title} {{Non-Fermi-liquid states
  and pairing instability of a general model of copper oxide metals}},\ }\href
  {https://doi.org/10.1103/PhysRevB.55.14554} {\bibfield  {journal} {\bibinfo
  {journal} {Phys. Rev. B}\ }\textbf {\bibinfo {volume} {55}},\ \bibinfo
  {pages} {14554} (\bibinfo {year} {1997})}\BibitemShut {NoStop}%
\bibitem [{\citenamefont {Varma}(1999)}]{Varma-PRL1999}%
  \BibitemOpen
  \bibfield  {author} {\bibinfo {author} {\bibfnamefont {C.~M.}\ \bibnamefont
  {Varma}},\ }\bibfield  {title} {\bibinfo {title} {{Pseudogap Phase and the
  Quantum-Critical Point in Copper-Oxide Metals}},\ }\href
  {https://doi.org/10.1103/PhysRevLett.83.3538} {\bibfield  {journal} {\bibinfo
   {journal} {Phys. Rev. Lett.}\ }\textbf {\bibinfo {volume} {83}},\ \bibinfo
  {pages} {3538} (\bibinfo {year} {1999})}\BibitemShut {NoStop}%
\bibitem [{\citenamefont {Chakravarty}\ \emph {et~al.}(2001)\citenamefont
  {Chakravarty}, \citenamefont {Laughlin}, \citenamefont {Morr},\ and\
  \citenamefont {Nayak}}]{Chakravarty-PRB01}%
  \BibitemOpen
  \bibfield  {author} {\bibinfo {author} {\bibfnamefont {S.}~\bibnamefont
  {Chakravarty}}, \bibinfo {author} {\bibfnamefont {R.~B.}\ \bibnamefont
  {Laughlin}}, \bibinfo {author} {\bibfnamefont {D.~K.}\ \bibnamefont {Morr}},\
  and\ \bibinfo {author} {\bibfnamefont {C.}~\bibnamefont {Nayak}},\ }\bibfield
   {title} {\bibinfo {title} {{Hidden order in the cuprates}},\ }\href
  {https://doi.org/10.1103/PhysRevB.63.094503} {\bibfield  {journal} {\bibinfo
  {journal} {Phys. Rev. B}\ }\textbf {\bibinfo {volume} {63}},\ \bibinfo
  {pages} {094503} (\bibinfo {year} {2001})}\BibitemShut {NoStop}%
\bibitem [{\citenamefont {Lee}\ \emph {et~al.}(2006)\citenamefont {Lee},
  \citenamefont {Nagaosa},\ and\ \citenamefont {Wen}}]{WenXG-RevMod06}%
  \BibitemOpen
  \bibfield  {author} {\bibinfo {author} {\bibfnamefont {P.~A.}\ \bibnamefont
  {Lee}}, \bibinfo {author} {\bibfnamefont {N.}~\bibnamefont {Nagaosa}},\ and\
  \bibinfo {author} {\bibfnamefont {X.-G.}\ \bibnamefont {Wen}},\ }\bibfield
  {title} {\bibinfo {title} {{Doping a Mott insulator: Physics of
  high-temperature superconductivity}},\ }\href
  {https://doi.org/10.1103/RevModPhys.78.17} {\bibfield  {journal} {\bibinfo
  {journal} {Rev. Mod. Phys.}\ }\textbf {\bibinfo {volume} {78}},\ \bibinfo
  {pages} {17} (\bibinfo {year} {2006})}\BibitemShut {NoStop}%
\bibitem [{\citenamefont {Haldane}(1988)}]{Haldane-PRL88}%
  \BibitemOpen
  \bibfield  {author} {\bibinfo {author} {\bibfnamefont {F.~D.~M.}\
  \bibnamefont {Haldane}},\ }\bibfield  {title} {\bibinfo {title} {Model for a
  quantum hall effect without landau levels: Condensed-matter realization of
  the "parity anomaly"},\ }\href {https://doi.org/10.1103/PhysRevLett.61.2015}
  {\bibfield  {journal} {\bibinfo  {journal} {Phys. Rev. Lett.}\ }\textbf
  {\bibinfo {volume} {61}},\ \bibinfo {pages} {2015} (\bibinfo {year}
  {1988})}\BibitemShut {NoStop}%
\bibitem [{\citenamefont {Wenzel}\ \emph {et~al.}(2022)\citenamefont {Wenzel},
  \citenamefont {Ortiz}, \citenamefont {Wilson}, \citenamefont {Dressel},
  \citenamefont {Tsirlin},\ and\ \citenamefont {Uykur}}]{Wenzel-PRB22}%
  \BibitemOpen
  \bibfield  {author} {\bibinfo {author} {\bibfnamefont {M.}~\bibnamefont
  {Wenzel}}, \bibinfo {author} {\bibfnamefont {B.~R.}\ \bibnamefont {Ortiz}},
  \bibinfo {author} {\bibfnamefont {S.~D.}\ \bibnamefont {Wilson}}, \bibinfo
  {author} {\bibfnamefont {M.}~\bibnamefont {Dressel}}, \bibinfo {author}
  {\bibfnamefont {A.~A.}\ \bibnamefont {Tsirlin}},\ and\ \bibinfo {author}
  {\bibfnamefont {E.}~\bibnamefont {Uykur}},\ }\bibfield  {title} {\bibinfo
  {title} {{Optical study of ${\mathrm{RbV}}_{3}{\mathrm{Sb}}_{5}$: Multiple
  density-wave gaps and phonon anomalies}},\ }\href
  {https://doi.org/10.1103/PhysRevB.105.245123} {\bibfield  {journal} {\bibinfo
   {journal} {Phys. Rev. B}\ }\textbf {\bibinfo {volume} {105}},\ \bibinfo
  {pages} {245123} (\bibinfo {year} {2022})}\BibitemShut {NoStop}%
\bibitem [{\citenamefont {Ratcliff}\ \emph {et~al.}(2021)\citenamefont
  {Ratcliff}, \citenamefont {Hallett}, \citenamefont {Ortiz}, \citenamefont
  {Wilson},\ and\ \citenamefont {Harter}}]{Ratcliff-PRM21}%
  \BibitemOpen
  \bibfield  {author} {\bibinfo {author} {\bibfnamefont {N.}~\bibnamefont
  {Ratcliff}}, \bibinfo {author} {\bibfnamefont {L.}~\bibnamefont {Hallett}},
  \bibinfo {author} {\bibfnamefont {B.~R.}\ \bibnamefont {Ortiz}}, \bibinfo
  {author} {\bibfnamefont {S.~D.}\ \bibnamefont {Wilson}},\ and\ \bibinfo
  {author} {\bibfnamefont {J.~W.}\ \bibnamefont {Harter}},\ }\bibfield  {title}
  {\bibinfo {title} {{Coherent phonon spectroscopy and interlayer modulation of
  charge density wave order in the kagome metal
  ${\mathrm{CsV}}_{3}{\mathrm{Sb}}_{5}$}},\ }\href
  {https://doi.org/10.1103/PhysRevMaterials.5.L111801} {\bibfield  {journal}
  {\bibinfo  {journal} {Phys. Rev. Mater.}\ }\textbf {\bibinfo {volume} {5}},\
  \bibinfo {pages} {L111801} (\bibinfo {year} {2021})}\BibitemShut {NoStop}%
\bibitem [{\citenamefont {Xie}\ \emph {et~al.}(2022)\citenamefont {Xie},
  \citenamefont {Li}, \citenamefont {Bourges}, \citenamefont {Ivanov},
  \citenamefont {Ye}, \citenamefont {Yin}, \citenamefont {Hasan}, \citenamefont
  {Luo}, \citenamefont {Yao}, \citenamefont {Wang}, \citenamefont {Xu},\ and\
  \citenamefont {Dai}}]{XieYaofeng-PRB22}%
  \BibitemOpen
  \bibfield  {author} {\bibinfo {author} {\bibfnamefont {Y.}~\bibnamefont
  {Xie}}, \bibinfo {author} {\bibfnamefont {Y.}~\bibnamefont {Li}}, \bibinfo
  {author} {\bibfnamefont {P.}~\bibnamefont {Bourges}}, \bibinfo {author}
  {\bibfnamefont {A.}~\bibnamefont {Ivanov}}, \bibinfo {author} {\bibfnamefont
  {Z.}~\bibnamefont {Ye}}, \bibinfo {author} {\bibfnamefont {J.-X.}\
  \bibnamefont {Yin}}, \bibinfo {author} {\bibfnamefont {M.~Z.}\ \bibnamefont
  {Hasan}}, \bibinfo {author} {\bibfnamefont {A.}~\bibnamefont {Luo}}, \bibinfo
  {author} {\bibfnamefont {Y.}~\bibnamefont {Yao}}, \bibinfo {author}
  {\bibfnamefont {Z.}~\bibnamefont {Wang}}, \bibinfo {author} {\bibfnamefont
  {G.}~\bibnamefont {Xu}},\ and\ \bibinfo {author} {\bibfnamefont
  {P.}~\bibnamefont {Dai}},\ }\bibfield  {title} {\bibinfo {title}
  {{Electron-phonon coupling in the charge density wave state of
  ${\mathrm{CsV}}_{3}{\mathrm{Sb}}_{5}$}},\ }\href
  {https://doi.org/10.1103/PhysRevB.105.L140501} {\bibfield  {journal}
  {\bibinfo  {journal} {Phys. Rev. B}\ }\textbf {\bibinfo {volume} {105}},\
  \bibinfo {pages} {L140501} (\bibinfo {year} {2022})}\BibitemShut {NoStop}%
\bibitem [{\citenamefont {Wu}\ \emph {et~al.}(2022{\natexlab{b}})\citenamefont
  {Wu}, \citenamefont {Ortiz}, \citenamefont {Tan}, \citenamefont {Wilson},
  \citenamefont {Yan}, \citenamefont {Birol},\ and\ \citenamefont
  {Blumberg}}]{WuShangfei-PRB22}%
  \BibitemOpen
  \bibfield  {author} {\bibinfo {author} {\bibfnamefont {S.}~\bibnamefont
  {Wu}}, \bibinfo {author} {\bibfnamefont {B.~R.}\ \bibnamefont {Ortiz}},
  \bibinfo {author} {\bibfnamefont {H.}~\bibnamefont {Tan}}, \bibinfo {author}
  {\bibfnamefont {S.~D.}\ \bibnamefont {Wilson}}, \bibinfo {author}
  {\bibfnamefont {B.}~\bibnamefont {Yan}}, \bibinfo {author} {\bibfnamefont
  {T.}~\bibnamefont {Birol}},\ and\ \bibinfo {author} {\bibfnamefont
  {G.}~\bibnamefont {Blumberg}},\ }\bibfield  {title} {\bibinfo {title}
  {{Charge density wave order in the kagome metal
  $A{\mathrm{V}}_{3}{\mathrm{Sb}}_{5}$
  $(A=\mathrm{Cs},\mathrm{Rb},\mathrm{K})$}},\ }\href
  {https://doi.org/10.1103/PhysRevB.105.155106} {\bibfield  {journal} {\bibinfo
   {journal} {Phys. Rev. B}\ }\textbf {\bibinfo {volume} {105}},\ \bibinfo
  {pages} {155106} (\bibinfo {year} {2022}{\natexlab{b}})}\BibitemShut
  {NoStop}%
\bibitem [{\citenamefont {Liu}\ \emph {et~al.}(2022)\citenamefont {Liu},
  \citenamefont {Ma}, \citenamefont {He}, \citenamefont {Li}, \citenamefont
  {Tan}, \citenamefont {Liu}, \citenamefont {Xu}, \citenamefont {Tang},
  \citenamefont {Watanabe}, \citenamefont {Taniguchi} \emph
  {et~al.}}]{LiuGan-NC22}%
  \BibitemOpen
  \bibfield  {author} {\bibinfo {author} {\bibfnamefont {G.}~\bibnamefont
  {Liu}}, \bibinfo {author} {\bibfnamefont {X.}~\bibnamefont {Ma}}, \bibinfo
  {author} {\bibfnamefont {K.}~\bibnamefont {He}}, \bibinfo {author}
  {\bibfnamefont {Q.}~\bibnamefont {Li}}, \bibinfo {author} {\bibfnamefont
  {H.}~\bibnamefont {Tan}}, \bibinfo {author} {\bibfnamefont {Y.}~\bibnamefont
  {Liu}}, \bibinfo {author} {\bibfnamefont {J.}~\bibnamefont {Xu}}, \bibinfo
  {author} {\bibfnamefont {W.}~\bibnamefont {Tang}}, \bibinfo {author}
  {\bibfnamefont {K.}~\bibnamefont {Watanabe}}, \bibinfo {author}
  {\bibfnamefont {T.}~\bibnamefont {Taniguchi}}, \emph {et~al.},\ }\bibfield
  {title} {\bibinfo {title} {{Observation of anomalous amplitude modes in the
  kagome metal CsV3Sb5}},\ }\href@noop {} {\bibfield  {journal} {\bibinfo
  {journal} {Nature communications}\ }\textbf {\bibinfo {volume} {13}},\
  \bibinfo {pages} {3461} (\bibinfo {year} {2022})}\BibitemShut {NoStop}%
\bibitem [{\citenamefont {Ferrari}\ \emph {et~al.}(2022)\citenamefont
  {Ferrari}, \citenamefont {Becca},\ and\ \citenamefont
  {Valent\'{\i}}}]{Ferrari-PRB22}%
  \BibitemOpen
  \bibfield  {author} {\bibinfo {author} {\bibfnamefont {F.}~\bibnamefont
  {Ferrari}}, \bibinfo {author} {\bibfnamefont {F.}~\bibnamefont {Becca}},\
  and\ \bibinfo {author} {\bibfnamefont {R.}~\bibnamefont {Valent\'{\i}}},\
  }\bibfield  {title} {\bibinfo {title} {{Charge density waves in
  kagome-lattice extended Hubbard models at the van Hove filling}},\ }\href
  {https://doi.org/10.1103/PhysRevB.106.L081107} {\bibfield  {journal}
  {\bibinfo  {journal} {Phys. Rev. B}\ }\textbf {\bibinfo {volume} {106}},\
  \bibinfo {pages} {L081107} (\bibinfo {year} {2022})}\BibitemShut {NoStop}%
\bibitem [{\citenamefont {Park}\ \emph {et~al.}(2021)\citenamefont {Park},
  \citenamefont {Ye},\ and\ \citenamefont {Balents}}]{Park-PRB21}%
  \BibitemOpen
  \bibfield  {author} {\bibinfo {author} {\bibfnamefont {T.}~\bibnamefont
  {Park}}, \bibinfo {author} {\bibfnamefont {M.}~\bibnamefont {Ye}},\ and\
  \bibinfo {author} {\bibfnamefont {L.}~\bibnamefont {Balents}},\ }\bibfield
  {title} {\bibinfo {title} {{Electronic instabilities of kagome metals: Saddle
  points and Landau theory}},\ }\href
  {https://doi.org/10.1103/PhysRevB.104.035142} {\bibfield  {journal} {\bibinfo
   {journal} {Phys. Rev. B}\ }\textbf {\bibinfo {volume} {104}},\ \bibinfo
  {pages} {035142} (\bibinfo {year} {2021})}\BibitemShut {NoStop}%
\bibitem [{\citenamefont {Christensen}\ \emph {et~al.}(2022)\citenamefont
  {Christensen}, \citenamefont {Birol}, \citenamefont {Andersen},\ and\
  \citenamefont {Fernandes}}]{Christensen-PRB22}%
  \BibitemOpen
  \bibfield  {author} {\bibinfo {author} {\bibfnamefont {M.~H.}\ \bibnamefont
  {Christensen}}, \bibinfo {author} {\bibfnamefont {T.}~\bibnamefont {Birol}},
  \bibinfo {author} {\bibfnamefont {B.~M.}\ \bibnamefont {Andersen}},\ and\
  \bibinfo {author} {\bibfnamefont {R.~M.}\ \bibnamefont {Fernandes}},\
  }\bibfield  {title} {\bibinfo {title} {{Loop currents in
  $A{\mathrm{V}}_{3}{\mathrm{Sb}}_{5}$ kagome metals: Multipolar and toroidal
  magnetic orders}},\ }\href {https://doi.org/10.1103/PhysRevB.106.144504}
  {\bibfield  {journal} {\bibinfo  {journal} {Phys. Rev. B}\ }\textbf {\bibinfo
  {volume} {106}},\ \bibinfo {pages} {144504} (\bibinfo {year}
  {2022})}\BibitemShut {NoStop}%
\bibitem [{\citenamefont {Lin}\ and\ \citenamefont
  {Nandkishore}(2021)}]{LinYuPing-PRB21}%
  \BibitemOpen
  \bibfield  {author} {\bibinfo {author} {\bibfnamefont {Y.-P.}\ \bibnamefont
  {Lin}}\ and\ \bibinfo {author} {\bibfnamefont {R.~M.}\ \bibnamefont
  {Nandkishore}},\ }\bibfield  {title} {\bibinfo {title} {{Complex charge
  density waves at Van Hove singularity on hexagonal lattices: Haldane-model
  phase diagram and potential realization in the kagome metals
  $A{V}_{3}{\mathrm{Sb}}_{5}$ ($A$=K, Rb, Cs)}},\ }\href
  {https://doi.org/10.1103/PhysRevB.104.045122} {\bibfield  {journal} {\bibinfo
   {journal} {Phys. Rev. B}\ }\textbf {\bibinfo {volume} {104}},\ \bibinfo
  {pages} {045122} (\bibinfo {year} {2021})}\BibitemShut {NoStop}%
\bibitem [{\citenamefont {Feng}\ \emph
  {et~al.}(2021{\natexlab{b}})\citenamefont {Feng}, \citenamefont {Zhang},
  \citenamefont {Jiang},\ and\ \citenamefont {Hu}}]{FengXilin-PRB21}%
  \BibitemOpen
  \bibfield  {author} {\bibinfo {author} {\bibfnamefont {X.}~\bibnamefont
  {Feng}}, \bibinfo {author} {\bibfnamefont {Y.}~\bibnamefont {Zhang}},
  \bibinfo {author} {\bibfnamefont {K.}~\bibnamefont {Jiang}},\ and\ \bibinfo
  {author} {\bibfnamefont {J.}~\bibnamefont {Hu}},\ }\bibfield  {title}
  {\bibinfo {title} {{Low-energy effective theory and symmetry classification
  of flux phases on the kagome lattice}},\ }\href
  {https://doi.org/10.1103/PhysRevB.104.165136} {\bibfield  {journal} {\bibinfo
   {journal} {Phys. Rev. B}\ }\textbf {\bibinfo {volume} {104}},\ \bibinfo
  {pages} {165136} (\bibinfo {year} {2021}{\natexlab{b}})}\BibitemShut
  {NoStop}%
\bibitem [{\citenamefont {Setty}\ \emph {et~al.}(2021)\citenamefont {Setty},
  \citenamefont {Hu}, \citenamefont {Chen},\ and\ \citenamefont
  {Si}}]{Setty-arxiv21}%
  \BibitemOpen
  \bibfield  {author} {\bibinfo {author} {\bibfnamefont {C.}~\bibnamefont
  {Setty}}, \bibinfo {author} {\bibfnamefont {H.}~\bibnamefont {Hu}}, \bibinfo
  {author} {\bibfnamefont {L.}~\bibnamefont {Chen}},\ and\ \bibinfo {author}
  {\bibfnamefont {Q.}~\bibnamefont {Si}},\ }\bibfield  {title} {\bibinfo
  {title} {{Electron correlations and T-breaking density wave order in a
  ${\mathbb{Z}}_{2}$ kagome metal}},\ }\href@noop {} {\bibfield  {journal}
  {\bibinfo  {journal} {arXiv preprint arXiv:2105.15204}\ } (\bibinfo {year}
  {2021})}\BibitemShut {NoStop}%
\bibitem [{\citenamefont {Yang}\ \emph {et~al.}(2023)\citenamefont {Yang},
  \citenamefont {Kim}, \citenamefont {Jeong}, \citenamefont {Kim},
  \citenamefont {Han},\ and\ \citenamefont {Lee}}]{Yang-arxiv23}%
  \BibitemOpen
  \bibfield  {author} {\bibinfo {author} {\bibfnamefont {H.-J.}\ \bibnamefont
  {Yang}}, \bibinfo {author} {\bibfnamefont {H.~S.}\ \bibnamefont {Kim}},
  \bibinfo {author} {\bibfnamefont {M.~Y.}\ \bibnamefont {Jeong}}, \bibinfo
  {author} {\bibfnamefont {Y.~B.}\ \bibnamefont {Kim}}, \bibinfo {author}
  {\bibfnamefont {M.~J.}\ \bibnamefont {Han}},\ and\ \bibinfo {author}
  {\bibfnamefont {S.}~\bibnamefont {Lee}},\ }\bibfield  {title} {\bibinfo
  {title} {{Intertwining orbital current order and superconductivity in kagome
  metal}},\ }\href@noop {} {\bibfield  {journal} {\bibinfo  {journal} {SciPost
  Physics Core}\ }\textbf {\bibinfo {volume} {6}},\ \bibinfo {pages} {008}
  (\bibinfo {year} {2023})}\BibitemShut {NoStop}%
\bibitem [{\citenamefont {Mertz}\ \emph {et~al.}(2022)\citenamefont {Mertz},
  \citenamefont {Wunderlich}, \citenamefont {Bhattacharyya}, \citenamefont
  {Ferrari},\ and\ \citenamefont {Valent{\'\i}}}]{Mertz-Npj22}%
  \BibitemOpen
  \bibfield  {author} {\bibinfo {author} {\bibfnamefont {T.}~\bibnamefont
  {Mertz}}, \bibinfo {author} {\bibfnamefont {P.}~\bibnamefont {Wunderlich}},
  \bibinfo {author} {\bibfnamefont {S.}~\bibnamefont {Bhattacharyya}}, \bibinfo
  {author} {\bibfnamefont {F.}~\bibnamefont {Ferrari}},\ and\ \bibinfo {author}
  {\bibfnamefont {R.}~\bibnamefont {Valent{\'\i}}},\ }\bibfield  {title}
  {\bibinfo {title} {{Statistical learning of engineered topological phases in
  the kagome superlattice of $A{\mathrm{V}}_{3}{\mathrm{Sb}}_{5}$}},\
  }\href@noop {} {\bibfield  {journal} {\bibinfo  {journal} {npj computational
  materials}\ }\textbf {\bibinfo {volume} {8}},\ \bibinfo {pages} {66}
  (\bibinfo {year} {2022})}\BibitemShut {NoStop}%
\bibitem [{\citenamefont {Tazai}\ \emph {et~al.}(2022)\citenamefont {Tazai},
  \citenamefont {Yamakawa}, \citenamefont {Onari},\ and\ \citenamefont
  {Kontani}}]{Rina-SciAdv22}%
  \BibitemOpen
  \bibfield  {author} {\bibinfo {author} {\bibfnamefont {R.}~\bibnamefont
  {Tazai}}, \bibinfo {author} {\bibfnamefont {Y.}~\bibnamefont {Yamakawa}},
  \bibinfo {author} {\bibfnamefont {S.}~\bibnamefont {Onari}},\ and\ \bibinfo
  {author} {\bibfnamefont {H.}~\bibnamefont {Kontani}},\ }\bibfield  {title}
  {\bibinfo {title} {{Mechanism of exotic density-wave and beyond-Migdal
  unconventional superconductivity in kagome metal
  $A{\mathrm{V}}_{3}{\mathrm{Sb}}_{5}$
  ($A=\mathrm{K},\mathrm{Rb},\mathrm{Cs}$)}},\ }\href
  {https://doi.org/10.1126/sciadv.abl4108} {\bibfield  {journal} {\bibinfo
  {journal} {Science Advances}\ }\textbf {\bibinfo {volume} {8}},\ \bibinfo
  {pages} {eabl4108} (\bibinfo {year} {2022})}\BibitemShut {NoStop}%
\bibitem [{\citenamefont {Tazai}\ \emph {et~al.}(2023)\citenamefont {Tazai},
  \citenamefont {Yamakawa},\ and\ \citenamefont {Kontani}}]{Tazai-NC23}%
  \BibitemOpen
  \bibfield  {author} {\bibinfo {author} {\bibfnamefont {R.}~\bibnamefont
  {Tazai}}, \bibinfo {author} {\bibfnamefont {Y.}~\bibnamefont {Yamakawa}},\
  and\ \bibinfo {author} {\bibfnamefont {H.}~\bibnamefont {Kontani}},\
  }\bibfield  {title} {\bibinfo {title} {Charge-loop current order and z 3
  nematicity mediated by bond order fluctuations in kagome metals},\
  }\href@noop {} {\bibfield  {journal} {\bibinfo  {journal} {Nature
  Communications}\ }\textbf {\bibinfo {volume} {14}},\ \bibinfo {pages} {7845}
  (\bibinfo {year} {2023})}\BibitemShut {NoStop}%
\bibitem [{\citenamefont {Kiesel}\ \emph {et~al.}(2013)\citenamefont {Kiesel},
  \citenamefont {Platt},\ and\ \citenamefont {Thomale}}]{Kiesel-PRL13}%
  \BibitemOpen
  \bibfield  {author} {\bibinfo {author} {\bibfnamefont {M.~L.}\ \bibnamefont
  {Kiesel}}, \bibinfo {author} {\bibfnamefont {C.}~\bibnamefont {Platt}},\ and\
  \bibinfo {author} {\bibfnamefont {R.}~\bibnamefont {Thomale}},\ }\bibfield
  {title} {\bibinfo {title} {Unconventional fermi surface instabilities in the
  kagome hubbard model},\ }\href
  {https://doi.org/10.1103/PhysRevLett.110.126405} {\bibfield  {journal}
  {\bibinfo  {journal} {Phys. Rev. Lett.}\ }\textbf {\bibinfo {volume} {110}},\
  \bibinfo {pages} {126405} (\bibinfo {year} {2013})}\BibitemShut {NoStop}%
\bibitem [{\citenamefont {Wu}\ \emph {et~al.}(2021)\citenamefont {Wu},
  \citenamefont {Schwemmer}, \citenamefont {M\"uller}, \citenamefont
  {Consiglio}, \citenamefont {Sangiovanni}, \citenamefont {Di~Sante},
  \citenamefont {Iqbal}, \citenamefont {Hanke}, \citenamefont {Schnyder},
  \citenamefont {Denner}, \citenamefont {Fischer}, \citenamefont {Neupert},\
  and\ \citenamefont {Thomale}}]{WuXX-PRL21}%
  \BibitemOpen
  \bibfield  {author} {\bibinfo {author} {\bibfnamefont {X.}~\bibnamefont
  {Wu}}, \bibinfo {author} {\bibfnamefont {T.}~\bibnamefont {Schwemmer}},
  \bibinfo {author} {\bibfnamefont {T.}~\bibnamefont {M\"uller}}, \bibinfo
  {author} {\bibfnamefont {A.}~\bibnamefont {Consiglio}}, \bibinfo {author}
  {\bibfnamefont {G.}~\bibnamefont {Sangiovanni}}, \bibinfo {author}
  {\bibfnamefont {D.}~\bibnamefont {Di~Sante}}, \bibinfo {author}
  {\bibfnamefont {Y.}~\bibnamefont {Iqbal}}, \bibinfo {author} {\bibfnamefont
  {W.}~\bibnamefont {Hanke}}, \bibinfo {author} {\bibfnamefont {A.~P.}\
  \bibnamefont {Schnyder}}, \bibinfo {author} {\bibfnamefont {M.~M.}\
  \bibnamefont {Denner}}, \bibinfo {author} {\bibfnamefont {M.~H.}\
  \bibnamefont {Fischer}}, \bibinfo {author} {\bibfnamefont {T.}~\bibnamefont
  {Neupert}},\ and\ \bibinfo {author} {\bibfnamefont {R.}~\bibnamefont
  {Thomale}},\ }\bibfield  {title} {\bibinfo {title} {Nature of unconventional
  pairing in the kagome superconductors $a{\mathrm{v}}_{3}{\mathrm{sb}}_{5}$
  ($a=\mathrm{K},\mathrm{Rb},\mathrm{Cs}$)},\ }\href
  {https://doi.org/10.1103/PhysRevLett.127.177001} {\bibfield  {journal}
  {\bibinfo  {journal} {Phys. Rev. Lett.}\ }\textbf {\bibinfo {volume} {127}},\
  \bibinfo {pages} {177001} (\bibinfo {year} {2021})}\BibitemShut {NoStop}%
\bibitem [{\citenamefont {Wu}\ \emph {et~al.}(2023)\citenamefont {Wu},
  \citenamefont {Thomale},\ and\ \citenamefont {Raghu}}]{WuYiMing-PRB23}%
  \BibitemOpen
  \bibfield  {author} {\bibinfo {author} {\bibfnamefont {Y.-M.}\ \bibnamefont
  {Wu}}, \bibinfo {author} {\bibfnamefont {R.}~\bibnamefont {Thomale}},\ and\
  \bibinfo {author} {\bibfnamefont {S.}~\bibnamefont {Raghu}},\ }\bibfield
  {title} {\bibinfo {title} {Sublattice interference promotes pair density wave
  order in kagome metals},\ }\href
  {https://doi.org/10.1103/PhysRevB.108.L081117} {\bibfield  {journal}
  {\bibinfo  {journal} {Phys. Rev. B}\ }\textbf {\bibinfo {volume} {108}},\
  \bibinfo {pages} {L081117} (\bibinfo {year} {2023})}\BibitemShut {NoStop}%
\bibitem [{\citenamefont {Dong}\ \emph {et~al.}(2023)\citenamefont {Dong},
  \citenamefont {Wang},\ and\ \citenamefont {Zhou}}]{Dong-PRB23}%
  \BibitemOpen
  \bibfield  {author} {\bibinfo {author} {\bibfnamefont {J.-W.}\ \bibnamefont
  {Dong}}, \bibinfo {author} {\bibfnamefont {Z.}~\bibnamefont {Wang}},\ and\
  \bibinfo {author} {\bibfnamefont {S.}~\bibnamefont {Zhou}},\ }\bibfield
  {title} {\bibinfo {title} {{Loop-current charge density wave driven by
  long-range Coulomb repulsion on the kagom\'e lattice}},\ }\href
  {https://doi.org/10.1103/PhysRevB.107.045127} {\bibfield  {journal} {\bibinfo
   {journal} {Phys. Rev. B}\ }\textbf {\bibinfo {volume} {107}},\ \bibinfo
  {pages} {045127} (\bibinfo {year} {2023})}\BibitemShut {NoStop}%
\bibitem [{\citenamefont {Fu}\ \emph {et~al.}(2024)\citenamefont {Fu},
  \citenamefont {Zhan}, \citenamefont {Dürrnagel}, \citenamefont {Hohmann},
  \citenamefont {Thomale}, \citenamefont {Hu}, \citenamefont {Wang},
  \citenamefont {Zhou},\ and\ \citenamefont {Wu}}]{RQFu-arXiv24}%
  \BibitemOpen
  \bibfield  {author} {\bibinfo {author} {\bibfnamefont {R.-Q.}\ \bibnamefont
  {Fu}}, \bibinfo {author} {\bibfnamefont {J.}~\bibnamefont {Zhan}}, \bibinfo
  {author} {\bibfnamefont {M.}~\bibnamefont {Dürrnagel}}, \bibinfo {author}
  {\bibfnamefont {H.}~\bibnamefont {Hohmann}}, \bibinfo {author} {\bibfnamefont
  {R.}~\bibnamefont {Thomale}}, \bibinfo {author} {\bibfnamefont
  {J.}~\bibnamefont {Hu}}, \bibinfo {author} {\bibfnamefont {Z.}~\bibnamefont
  {Wang}}, \bibinfo {author} {\bibfnamefont {S.}~\bibnamefont {Zhou}},\ and\
  \bibinfo {author} {\bibfnamefont {X.}~\bibnamefont {Wu}},\ }\href
  {https://arxiv.org/abs/2405.09451} {\bibinfo {title} {Exotic charge density
  waves and superconductivity on the kagome lattice}} (\bibinfo {year}
  {2024}),\ \Eprint {https://arxiv.org/abs/2405.09451} {arXiv:2405.09451
  [cond-mat.str-el]} \BibitemShut {NoStop}%
\bibitem [{\citenamefont {Ramezani}\ \emph {et~al.}(2024)\citenamefont
  {Ramezani}, \citenamefont {\ifmmode \mbox{\c{S}}\else \c{S}\fi{}a\ifmmode
  \mbox{\c{s}}\else \c{s}\fi{}\ifmmode \imath \else \i
  \fi{}o\ifmmode~\breve{g}\else \u{g}\fi{}lu}, \citenamefont {Hadipour},
  \citenamefont {Soleimani}, \citenamefont {Friedrich}, \citenamefont
  {Bl\"ugel},\ and\ \citenamefont {Mertig}}]{Ramezani-prb24}%
  \BibitemOpen
  \bibfield  {author} {\bibinfo {author} {\bibfnamefont {H.~R.}\ \bibnamefont
  {Ramezani}}, \bibinfo {author} {\bibfnamefont {E.}~\bibnamefont {\ifmmode
  \mbox{\c{S}}\else \c{S}\fi{}a\ifmmode \mbox{\c{s}}\else \c{s}\fi{}\ifmmode
  \imath \else \i \fi{}o\ifmmode~\breve{g}\else \u{g}\fi{}lu}}, \bibinfo
  {author} {\bibfnamefont {H.}~\bibnamefont {Hadipour}}, \bibinfo {author}
  {\bibfnamefont {H.~R.}\ \bibnamefont {Soleimani}}, \bibinfo {author}
  {\bibfnamefont {C.}~\bibnamefont {Friedrich}}, \bibinfo {author}
  {\bibfnamefont {S.}~\bibnamefont {Bl\"ugel}},\ and\ \bibinfo {author}
  {\bibfnamefont {I.}~\bibnamefont {Mertig}},\ }\bibfield  {title} {\bibinfo
  {title} {Nonconventional screening of coulomb interaction in two-dimensional
  semiconductors and metals: A comprehensive constrained random phase
  approximation study of $mx_2$ ($m$ = mo, w, nb, ta; $x$ = s, se, te)},\
  }\href {https://doi.org/10.1103/PhysRevB.109.125108} {\bibfield  {journal}
  {\bibinfo  {journal} {Phys. Rev. B}\ }\textbf {\bibinfo {volume} {109}},\
  \bibinfo {pages} {125108} (\bibinfo {year} {2024})}\BibitemShut {NoStop}%
\bibitem [{\citenamefont {Bagherpour}\ \emph {et~al.}(2024)\citenamefont
  {Bagherpour}, \citenamefont {Mahdavifar}, \citenamefont {Lapasar},\ and\
  \citenamefont {Hadipour}}]{Bagherpour-prb24}%
  \BibitemOpen
  \bibfield  {author} {\bibinfo {author} {\bibfnamefont {F.}~\bibnamefont
  {Bagherpour}}, \bibinfo {author} {\bibfnamefont {S.}~\bibnamefont
  {Mahdavifar}}, \bibinfo {author} {\bibfnamefont {E.~H.}\ \bibnamefont
  {Lapasar}},\ and\ \bibinfo {author} {\bibfnamefont {H.}~\bibnamefont
  {Hadipour}},\ }\bibfield  {title} {\bibinfo {title} {Electron screening and
  strength of long-range coulomb interactions in black phosphorene: From bulk
  to nanoribbon},\ }\href {https://doi.org/10.1103/PhysRevB.109.165115}
  {\bibfield  {journal} {\bibinfo  {journal} {Phys. Rev. B}\ }\textbf {\bibinfo
  {volume} {109}},\ \bibinfo {pages} {165115} (\bibinfo {year}
  {2024})}\BibitemShut {NoStop}%
\bibitem [{\citenamefont {Li}\ \emph {et~al.}(2024)\citenamefont {Li},
  \citenamefont {Kim},\ and\ \citenamefont {Kee}}]{Li-arXiv2023}%
  \BibitemOpen
  \bibfield  {author} {\bibinfo {author} {\bibfnamefont {H.}~\bibnamefont
  {Li}}, \bibinfo {author} {\bibfnamefont {Y.~B.}\ \bibnamefont {Kim}},\ and\
  \bibinfo {author} {\bibfnamefont {H.-Y.}\ \bibnamefont {Kee}},\ }\bibfield
  {title} {\bibinfo {title} {Intertwined van hove singularities as a mechanism
  for loop current order in kagome metals},\ }\href
  {https://doi.org/10.1103/PhysRevLett.132.146501} {\bibfield  {journal}
  {\bibinfo  {journal} {Phys. Rev. Lett.}\ }\textbf {\bibinfo {volume} {132}},\
  \bibinfo {pages} {146501} (\bibinfo {year} {2024})}\BibitemShut {NoStop}%
\bibitem [{\citenamefont {Christensen}\ \emph {et~al.}(2021)\citenamefont
  {Christensen}, \citenamefont {Birol}, \citenamefont {Andersen},\ and\
  \citenamefont {Fernandes}}]{Christensen-PRB21}%
  \BibitemOpen
  \bibfield  {author} {\bibinfo {author} {\bibfnamefont {M.~H.}\ \bibnamefont
  {Christensen}}, \bibinfo {author} {\bibfnamefont {T.}~\bibnamefont {Birol}},
  \bibinfo {author} {\bibfnamefont {B.~M.}\ \bibnamefont {Andersen}},\ and\
  \bibinfo {author} {\bibfnamefont {R.~M.}\ \bibnamefont {Fernandes}},\
  }\bibfield  {title} {\bibinfo {title} {{Theory of the charge density wave in
  $A{\mathrm{V}}_{3}{\mathrm{Sb}}_{5}$ kagome metals}},\ }\href
  {https://doi.org/10.1103/PhysRevB.104.214513} {\bibfield  {journal} {\bibinfo
   {journal} {Phys. Rev. B}\ }\textbf {\bibinfo {volume} {104}},\ \bibinfo
  {pages} {214513} (\bibinfo {year} {2021})}\BibitemShut {NoStop}%
\bibitem [{\citenamefont {Li}\ \emph {et~al.}(2023)\citenamefont {Li},
  \citenamefont {Liu}, \citenamefont {Kim},\ and\ \citenamefont
  {Kee}}]{HLi-prb23}%
  \BibitemOpen
  \bibfield  {author} {\bibinfo {author} {\bibfnamefont {H.}~\bibnamefont
  {Li}}, \bibinfo {author} {\bibfnamefont {X.}~\bibnamefont {Liu}}, \bibinfo
  {author} {\bibfnamefont {Y.~B.}\ \bibnamefont {Kim}},\ and\ \bibinfo {author}
  {\bibfnamefont {H.-Y.}\ \bibnamefont {Kee}},\ }\bibfield  {title} {\bibinfo
  {title} {Origin of $\ensuremath{\pi}$-shifted three-dimensional charge
  density waves in the kagom\'e metal ${A\mathrm{V}}_{3}{\mathrm{sb}}_{5}$
  $(a=\mathrm{Cs}, \mathrm{Rb}, \mathrm{K})$},\ }\href
  {https://doi.org/10.1103/PhysRevB.108.075102} {\bibfield  {journal} {\bibinfo
   {journal} {Phys. Rev. B}\ }\textbf {\bibinfo {volume} {108}},\ \bibinfo
  {pages} {075102} (\bibinfo {year} {2023})}\BibitemShut {NoStop}%
\bibitem [{\citenamefont {Stahl}\ \emph {et~al.}(2022)\citenamefont {Stahl},
  \citenamefont {Chen}, \citenamefont {Ritschel}, \citenamefont {Shekhar},
  \citenamefont {Sadrollahi}, \citenamefont {Rahn}, \citenamefont {Ivashko},
  \citenamefont {Zimmermann}, \citenamefont {Felser},\ and\ \citenamefont
  {Geck}}]{Stahl-prb22}%
  \BibitemOpen
  \bibfield  {author} {\bibinfo {author} {\bibfnamefont {Q.}~\bibnamefont
  {Stahl}}, \bibinfo {author} {\bibfnamefont {D.}~\bibnamefont {Chen}},
  \bibinfo {author} {\bibfnamefont {T.}~\bibnamefont {Ritschel}}, \bibinfo
  {author} {\bibfnamefont {C.}~\bibnamefont {Shekhar}}, \bibinfo {author}
  {\bibfnamefont {E.}~\bibnamefont {Sadrollahi}}, \bibinfo {author}
  {\bibfnamefont {M.~C.}\ \bibnamefont {Rahn}}, \bibinfo {author}
  {\bibfnamefont {O.}~\bibnamefont {Ivashko}}, \bibinfo {author} {\bibfnamefont
  {M.~v.}\ \bibnamefont {Zimmermann}}, \bibinfo {author} {\bibfnamefont
  {C.}~\bibnamefont {Felser}},\ and\ \bibinfo {author} {\bibfnamefont
  {J.}~\bibnamefont {Geck}},\ }\bibfield  {title} {\bibinfo {title}
  {Temperature-driven reorganization of electronic order in
  ${\mathrm{csv}}_{3}{\mathrm{sb}}_{5}$},\ }\href
  {https://doi.org/10.1103/PhysRevB.105.195136} {\bibfield  {journal} {\bibinfo
   {journal} {Phys. Rev. B}\ }\textbf {\bibinfo {volume} {105}},\ \bibinfo
  {pages} {195136} (\bibinfo {year} {2022})}\BibitemShut {NoStop}%
\bibitem [{\citenamefont {Xiao}\ \emph {et~al.}(2023)\citenamefont {Xiao},
  \citenamefont {Lin}, \citenamefont {Li}, \citenamefont {Zheng}, \citenamefont
  {Francoual}, \citenamefont {Plueckthun}, \citenamefont {Xia}, \citenamefont
  {Qiu}, \citenamefont {Zhang}, \citenamefont {Guo}, \citenamefont {Feng},\
  and\ \citenamefont {Peng}}]{Xiao-prr23}%
  \BibitemOpen
  \bibfield  {author} {\bibinfo {author} {\bibfnamefont {Q.}~\bibnamefont
  {Xiao}}, \bibinfo {author} {\bibfnamefont {Y.}~\bibnamefont {Lin}}, \bibinfo
  {author} {\bibfnamefont {Q.}~\bibnamefont {Li}}, \bibinfo {author}
  {\bibfnamefont {X.}~\bibnamefont {Zheng}}, \bibinfo {author} {\bibfnamefont
  {S.}~\bibnamefont {Francoual}}, \bibinfo {author} {\bibfnamefont
  {C.}~\bibnamefont {Plueckthun}}, \bibinfo {author} {\bibfnamefont
  {W.}~\bibnamefont {Xia}}, \bibinfo {author} {\bibfnamefont {Q.}~\bibnamefont
  {Qiu}}, \bibinfo {author} {\bibfnamefont {S.}~\bibnamefont {Zhang}}, \bibinfo
  {author} {\bibfnamefont {Y.}~\bibnamefont {Guo}}, \bibinfo {author}
  {\bibfnamefont {J.}~\bibnamefont {Feng}},\ and\ \bibinfo {author}
  {\bibfnamefont {Y.}~\bibnamefont {Peng}},\ }\bibfield  {title} {\bibinfo
  {title} {Coexistence of multiple stacking charge density waves in kagome
  superconductor ${\mathrm{csv}}_{3}{\mathrm{sb}}_{5}$},\ }\href
  {https://doi.org/10.1103/PhysRevResearch.5.L012032} {\bibfield  {journal}
  {\bibinfo  {journal} {Phys. Rev. Res.}\ }\textbf {\bibinfo {volume} {5}},\
  \bibinfo {pages} {L012032} (\bibinfo {year} {2023})}\BibitemShut {NoStop}%
\bibitem [{\citenamefont {Jiang}\ \emph {et~al.}(2016)\citenamefont {Jiang},
  \citenamefont {Hu}, \citenamefont {Ding},\ and\ \citenamefont
  {Wang}}]{JiangKun-PRB2016}%
  \BibitemOpen
  \bibfield  {author} {\bibinfo {author} {\bibfnamefont {K.}~\bibnamefont
  {Jiang}}, \bibinfo {author} {\bibfnamefont {J.}~\bibnamefont {Hu}}, \bibinfo
  {author} {\bibfnamefont {H.}~\bibnamefont {Ding}},\ and\ \bibinfo {author}
  {\bibfnamefont {Z.}~\bibnamefont {Wang}},\ }\bibfield  {title} {\bibinfo
  {title} {Interatomic coulomb interaction and electron nematic bond order in
  \text{FeSe}},\ }\href {https://doi.org/10.1103/PhysRevB.93.115138} {\bibfield
   {journal} {\bibinfo  {journal} {Phys. Rev. B}\ }\textbf {\bibinfo {volume}
  {93}},\ \bibinfo {pages} {115138} (\bibinfo {year} {2016})}\BibitemShut
  {NoStop}%
\bibitem [{\citenamefont {Campo}\ and\ \citenamefont
  {Cococcioni}(2010)}]{Campo-IOP10}%
  \BibitemOpen
  \bibfield  {author} {\bibinfo {author} {\bibfnamefont {V.~L.}\ \bibnamefont
  {Campo}}\ and\ \bibinfo {author} {\bibfnamefont {M.}~\bibnamefont
  {Cococcioni}},\ }\bibfield  {title} {\bibinfo {title} {Extended
  \text{DFT+U+V} method with on-site and inter-site electronic interactions},\
  }\href@noop {} {\bibfield  {journal} {\bibinfo  {journal} {Journal of
  Physics: Condensed Matter}\ }\textbf {\bibinfo {volume} {22}},\ \bibinfo
  {pages} {055602} (\bibinfo {year} {2010})}\BibitemShut {NoStop}%
\bibitem [{\citenamefont {Belozerov}\ \emph {et~al.}(2012)\citenamefont
  {Belozerov}, \citenamefont {Korotin}, \citenamefont {Anisimov},\ and\
  \citenamefont {Poteryaev}}]{Belozerov-PRB12}%
  \BibitemOpen
  \bibfield  {author} {\bibinfo {author} {\bibfnamefont {A.~S.}\ \bibnamefont
  {Belozerov}}, \bibinfo {author} {\bibfnamefont {M.~A.}\ \bibnamefont
  {Korotin}}, \bibinfo {author} {\bibfnamefont {V.~I.}\ \bibnamefont
  {Anisimov}},\ and\ \bibinfo {author} {\bibfnamefont {A.~I.}\ \bibnamefont
  {Poteryaev}},\ }\bibfield  {title} {\bibinfo {title} {Monoclinic
  $\textit{M}_1$ phase of $\text{VO}_2$: Mott-hubbard versus band insulator},\
  }\href {https://doi.org/10.1103/PhysRevB.85.045109} {\bibfield  {journal}
  {\bibinfo  {journal} {Phys. Rev. B}\ }\textbf {\bibinfo {volume} {85}},\
  \bibinfo {pages} {045109} (\bibinfo {year} {2012})}\BibitemShut {NoStop}%
\bibitem [{\citenamefont {Hu}\ \emph {et~al.}(2022{\natexlab{b}})\citenamefont
  {Hu}, \citenamefont {Wu}, \citenamefont {Ortiz}, \citenamefont {Han},
  \citenamefont {Plumb}, \citenamefont {Wilson}, \citenamefont {Schnyder},\
  and\ \citenamefont {Shi}}]{YongHu-PRB22}%
  \BibitemOpen
  \bibfield  {author} {\bibinfo {author} {\bibfnamefont {Y.}~\bibnamefont
  {Hu}}, \bibinfo {author} {\bibfnamefont {X.}~\bibnamefont {Wu}}, \bibinfo
  {author} {\bibfnamefont {B.~R.}\ \bibnamefont {Ortiz}}, \bibinfo {author}
  {\bibfnamefont {X.}~\bibnamefont {Han}}, \bibinfo {author} {\bibfnamefont
  {N.~C.}\ \bibnamefont {Plumb}}, \bibinfo {author} {\bibfnamefont {S.~D.}\
  \bibnamefont {Wilson}}, \bibinfo {author} {\bibfnamefont {A.~P.}\
  \bibnamefont {Schnyder}},\ and\ \bibinfo {author} {\bibfnamefont
  {M.}~\bibnamefont {Shi}},\ }\bibfield  {title} {\bibinfo {title} {Coexistence
  of trihexagonal and star-of-david pattern in the charge density wave of the
  kagome superconductor $a{\mathrm{v}}_{3}{\mathrm{sb}}_{5}$},\ }\href
  {https://doi.org/10.1103/PhysRevB.106.L241106} {\bibfield  {journal}
  {\bibinfo  {journal} {Phys. Rev. B}\ }\textbf {\bibinfo {volume} {106}},\
  \bibinfo {pages} {L241106} (\bibinfo {year}
  {2022}{\natexlab{b}})}\BibitemShut {NoStop}%
\bibitem [{\citenamefont {Kang}\ \emph {et~al.}(2022)\citenamefont {Kang},
  \citenamefont {Fang}, \citenamefont {Yoo}, \citenamefont {Ortiz},
  \citenamefont {Oey}, \citenamefont {Choi}, \citenamefont {Ryu}, \citenamefont
  {Kim}, \citenamefont {Jozwiak}, \citenamefont {Bostwick}, \citenamefont
  {Rotenberg}, \citenamefont {Kaxiras}, \citenamefont {Checkelsky},
  \citenamefont {Wilson}, \citenamefont {Park},\ and\ \citenamefont
  {Comin}}]{Kang-NatMat22}%
  \BibitemOpen
  \bibfield  {author} {\bibinfo {author} {\bibfnamefont {M.}~\bibnamefont
  {Kang}}, \bibinfo {author} {\bibfnamefont {S.}~\bibnamefont {Fang}}, \bibinfo
  {author} {\bibfnamefont {J.}~\bibnamefont {Yoo}}, \bibinfo {author}
  {\bibfnamefont {B.~R.}\ \bibnamefont {Ortiz}}, \bibinfo {author}
  {\bibfnamefont {Y.~M.}\ \bibnamefont {Oey}}, \bibinfo {author} {\bibfnamefont
  {J.}~\bibnamefont {Choi}}, \bibinfo {author} {\bibfnamefont {S.~H.}\
  \bibnamefont {Ryu}}, \bibinfo {author} {\bibfnamefont {J.}~\bibnamefont
  {Kim}}, \bibinfo {author} {\bibfnamefont {C.}~\bibnamefont {Jozwiak}},
  \bibinfo {author} {\bibfnamefont {A.}~\bibnamefont {Bostwick}}, \bibinfo
  {author} {\bibfnamefont {E.}~\bibnamefont {Rotenberg}}, \bibinfo {author}
  {\bibfnamefont {E.}~\bibnamefont {Kaxiras}}, \bibinfo {author} {\bibfnamefont
  {J.~G.}\ \bibnamefont {Checkelsky}}, \bibinfo {author} {\bibfnamefont
  {S.~D.}\ \bibnamefont {Wilson}}, \bibinfo {author} {\bibfnamefont {J.-H.}\
  \bibnamefont {Park}},\ and\ \bibinfo {author} {\bibfnamefont
  {R.}~\bibnamefont {Comin}},\ }\bibfield  {title} {\bibinfo {title} {Charge
  order landscape and competition with superconductivity in kagome metals},\
  }\href {https://doi.org/10.1038/s41563-022-01375-2} {\bibfield  {journal}
  {\bibinfo  {journal} {Nat. Mater.}\ }\textbf {\bibinfo {volume} {22}},\
  \bibinfo {pages} {186} (\bibinfo {year} {2022})}\BibitemShut {NoStop}%
\bibitem [{\citenamefont {Ritz}\ \emph {et~al.}(2023)\citenamefont {Ritz},
  \citenamefont {Fernandes},\ and\ \citenamefont {Birol}}]{Ritz-PRB23}%
  \BibitemOpen
  \bibfield  {author} {\bibinfo {author} {\bibfnamefont {E.~T.}\ \bibnamefont
  {Ritz}}, \bibinfo {author} {\bibfnamefont {R.~M.}\ \bibnamefont
  {Fernandes}},\ and\ \bibinfo {author} {\bibfnamefont {T.}~\bibnamefont
  {Birol}},\ }\bibfield  {title} {\bibinfo {title} {Impact of sb degrees of
  freedom on the charge density wave phase diagram of the kagome metal
  ${\mathrm{csv}}_{3}{\mathrm{sb}}_{5}$},\ }\href
  {https://doi.org/10.1103/PhysRevB.107.205131} {\bibfield  {journal} {\bibinfo
   {journal} {Phys. Rev. B}\ }\textbf {\bibinfo {volume} {107}},\ \bibinfo
  {pages} {205131} (\bibinfo {year} {2023})}\BibitemShut {NoStop}%
\bibitem [{\citenamefont {Kautzsch}\ \emph {et~al.}(2023)\citenamefont
  {Kautzsch}, \citenamefont {Ortiz}, \citenamefont {Mallayya}, \citenamefont
  {Plumb}, \citenamefont {Pokharel}, \citenamefont {Ruff}, \citenamefont
  {Islam}, \citenamefont {Kim}, \citenamefont {Seshadri},\ and\ \citenamefont
  {Wilson}}]{Kautzsch-PRM23}%
  \BibitemOpen
  \bibfield  {author} {\bibinfo {author} {\bibfnamefont {L.}~\bibnamefont
  {Kautzsch}}, \bibinfo {author} {\bibfnamefont {B.~R.}\ \bibnamefont {Ortiz}},
  \bibinfo {author} {\bibfnamefont {K.}~\bibnamefont {Mallayya}}, \bibinfo
  {author} {\bibfnamefont {J.}~\bibnamefont {Plumb}}, \bibinfo {author}
  {\bibfnamefont {G.}~\bibnamefont {Pokharel}}, \bibinfo {author}
  {\bibfnamefont {J.~P.~C.}\ \bibnamefont {Ruff}}, \bibinfo {author}
  {\bibfnamefont {Z.}~\bibnamefont {Islam}}, \bibinfo {author} {\bibfnamefont
  {E.-A.}\ \bibnamefont {Kim}}, \bibinfo {author} {\bibfnamefont
  {R.}~\bibnamefont {Seshadri}},\ and\ \bibinfo {author} {\bibfnamefont
  {S.~D.}\ \bibnamefont {Wilson}},\ }\bibfield  {title} {\bibinfo {title}
  {Structural evolution of the kagome superconductors
  $a{\mathrm{v}}_{3}{\mathrm{sb}}_{5}$ (a = k, rb, and cs) through charge
  density wave order},\ }\href
  {https://doi.org/10.1103/PhysRevMaterials.7.024806} {\bibfield  {journal}
  {\bibinfo  {journal} {Phys. Rev. Mater.}\ }\textbf {\bibinfo {volume} {7}},\
  \bibinfo {pages} {024806} (\bibinfo {year} {2023})}\BibitemShut {NoStop}%
\bibitem [{\citenamefont {Frassineti}\ \emph {et~al.}(2023)\citenamefont
  {Frassineti}, \citenamefont {Bonf\`a}, \citenamefont {Allodi}, \citenamefont
  {Garcia}, \citenamefont {Cong}, \citenamefont {Ortiz}, \citenamefont
  {Wilson}, \citenamefont {De~Renzi}, \citenamefont
  {Mitrovi\ifmmode~\acute{c}\else \'{c}\fi{}},\ and\ \citenamefont
  {Sanna}}]{Frassineti-PRR23}%
  \BibitemOpen
  \bibfield  {author} {\bibinfo {author} {\bibfnamefont {J.}~\bibnamefont
  {Frassineti}}, \bibinfo {author} {\bibfnamefont {P.}~\bibnamefont {Bonf\`a}},
  \bibinfo {author} {\bibfnamefont {G.}~\bibnamefont {Allodi}}, \bibinfo
  {author} {\bibfnamefont {E.}~\bibnamefont {Garcia}}, \bibinfo {author}
  {\bibfnamefont {R.}~\bibnamefont {Cong}}, \bibinfo {author} {\bibfnamefont
  {B.~R.}\ \bibnamefont {Ortiz}}, \bibinfo {author} {\bibfnamefont {S.~D.}\
  \bibnamefont {Wilson}}, \bibinfo {author} {\bibfnamefont {R.}~\bibnamefont
  {De~Renzi}}, \bibinfo {author} {\bibfnamefont {V.~F.}\ \bibnamefont
  {Mitrovi\ifmmode~\acute{c}\else \'{c}\fi{}}},\ and\ \bibinfo {author}
  {\bibfnamefont {S.}~\bibnamefont {Sanna}},\ }\bibfield  {title} {\bibinfo
  {title} {Microscopic nature of the charge-density wave in the kagome
  superconductor ${\mathrm{rbv}}_{3}{\mathrm{sb}}_{5}$},\ }\href
  {https://doi.org/10.1103/PhysRevResearch.5.L012017} {\bibfield  {journal}
  {\bibinfo  {journal} {Phys. Rev. Res.}\ }\textbf {\bibinfo {volume} {5}},\
  \bibinfo {pages} {L012017} (\bibinfo {year} {2023})}\BibitemShut {NoStop}%
\bibitem [{\citenamefont {Wang}\ \emph {et~al.}(2023)\citenamefont {Wang},
  \citenamefont {Wu}, \citenamefont {Li}, \citenamefont {Jiang},\ and\
  \citenamefont {Hu}}]{YuxinWang-PRB23}%
  \BibitemOpen
  \bibfield  {author} {\bibinfo {author} {\bibfnamefont {Y.}~\bibnamefont
  {Wang}}, \bibinfo {author} {\bibfnamefont {T.}~\bibnamefont {Wu}}, \bibinfo
  {author} {\bibfnamefont {Z.}~\bibnamefont {Li}}, \bibinfo {author}
  {\bibfnamefont {K.}~\bibnamefont {Jiang}},\ and\ \bibinfo {author}
  {\bibfnamefont {J.}~\bibnamefont {Hu}},\ }\bibfield  {title} {\bibinfo
  {title} {Structure of the kagome superconductor
  ${\mathrm{csv}}_{3}{\mathrm{sb}}_{5}$ in the charge density wave state},\
  }\href {https://doi.org/10.1103/PhysRevB.107.184106} {\bibfield  {journal}
  {\bibinfo  {journal} {Phys. Rev. B}\ }\textbf {\bibinfo {volume} {107}},\
  \bibinfo {pages} {184106} (\bibinfo {year} {2023})}\BibitemShut {NoStop}%
\bibitem [{\citenamefont {Scagnoli}\ \emph {et~al.}(2024)\citenamefont
  {Scagnoli}, \citenamefont {Riddiford}, \citenamefont {Huang}, \citenamefont
  {Shi}, \citenamefont {Tu}, \citenamefont {Lei}, \citenamefont {Bombardi},
  \citenamefont {Nisbet},\ and\ \citenamefont {Guguchia}}]{Scagnoli-JPCM24}%
  \BibitemOpen
  \bibfield  {author} {\bibinfo {author} {\bibfnamefont {V.}~\bibnamefont
  {Scagnoli}}, \bibinfo {author} {\bibfnamefont {L.~J.}\ \bibnamefont
  {Riddiford}}, \bibinfo {author} {\bibfnamefont {S.~W.}\ \bibnamefont
  {Huang}}, \bibinfo {author} {\bibfnamefont {Y.-G.}\ \bibnamefont {Shi}},
  \bibinfo {author} {\bibfnamefont {Z.}~\bibnamefont {Tu}}, \bibinfo {author}
  {\bibfnamefont {H.}~\bibnamefont {Lei}}, \bibinfo {author} {\bibfnamefont
  {A.}~\bibnamefont {Bombardi}}, \bibinfo {author} {\bibfnamefont
  {G.}~\bibnamefont {Nisbet}},\ and\ \bibinfo {author} {\bibfnamefont
  {Z.}~\bibnamefont {Guguchia}},\ }\bibfield  {title} {\bibinfo {title}
  {Resonant x-ray diffraction measurements in charge ordered kagome
  superconductors kv$_3$sb$_5$ and rbv$_3$sb$_5$},\ }\href
  {https://doi.org/10.1088/1361-648X/ad20a2} {\bibfield  {journal} {\bibinfo
  {journal} {Journal of Physics: Condensed Matter}\ }\textbf {\bibinfo {volume}
  {36}},\ \bibinfo {pages} {185604} (\bibinfo {year} {2024})}\BibitemShut
  {NoStop}%
\bibitem [{\citenamefont {Wei}\ \emph {et~al.}(2024)\citenamefont {Wei},
  \citenamefont {Tian}, \citenamefont {Cui}, \citenamefont {Zhai},
  \citenamefont {Li}, \citenamefont {Liu}, \citenamefont {Song}, \citenamefont
  {Feng}, \citenamefont {Huang}, \citenamefont {Wang} \emph
  {et~al.}}]{wei2024three}%
  \BibitemOpen
  \bibfield  {author} {\bibinfo {author} {\bibfnamefont {X.}~\bibnamefont
  {Wei}}, \bibinfo {author} {\bibfnamefont {C.}~\bibnamefont {Tian}}, \bibinfo
  {author} {\bibfnamefont {H.}~\bibnamefont {Cui}}, \bibinfo {author}
  {\bibfnamefont {Y.}~\bibnamefont {Zhai}}, \bibinfo {author} {\bibfnamefont
  {Y.}~\bibnamefont {Li}}, \bibinfo {author} {\bibfnamefont {S.}~\bibnamefont
  {Liu}}, \bibinfo {author} {\bibfnamefont {Y.}~\bibnamefont {Song}}, \bibinfo
  {author} {\bibfnamefont {Y.}~\bibnamefont {Feng}}, \bibinfo {author}
  {\bibfnamefont {M.}~\bibnamefont {Huang}}, \bibinfo {author} {\bibfnamefont
  {Z.}~\bibnamefont {Wang}}, \emph {et~al.},\ }\bibfield  {title} {\bibinfo
  {title} {Three-dimensional hidden phase probed by in-plane magnetotransport
  in kagome metal csv3sb5 thin flakes},\ }\href@noop {} {\bibfield  {journal}
  {\bibinfo  {journal} {arXiv preprint arXiv:2405.04059}\ } (\bibinfo {year}
  {2024})}\BibitemShut {NoStop}%
\bibitem [{\citenamefont {Rojo}\ and\ \citenamefont
  {Leggett}(1991)}]{Leggett-prl91}%
  \BibitemOpen
  \bibfield  {author} {\bibinfo {author} {\bibfnamefont {A.~G.}\ \bibnamefont
  {Rojo}}\ and\ \bibinfo {author} {\bibfnamefont {A.~J.}\ \bibnamefont
  {Leggett}},\ }\bibfield  {title} {\bibinfo {title} {Sign of the coupling
  between t-violating ground states in second-order perturbation theory},\
  }\href {https://doi.org/10.1103/PhysRevLett.67.3614} {\bibfield  {journal}
  {\bibinfo  {journal} {Phys. Rev. Lett.}\ }\textbf {\bibinfo {volume} {67}},\
  \bibinfo {pages} {3614} (\bibinfo {year} {1991})}\BibitemShut {NoStop}%
\bibitem [{\citenamefont {Zhu}\ \emph {et~al.}(2013)\citenamefont {Zhu},
  \citenamefont {Aji},\ and\ \citenamefont {Varma}}]{Varma-prb13}%
  \BibitemOpen
  \bibfield  {author} {\bibinfo {author} {\bibfnamefont {L.}~\bibnamefont
  {Zhu}}, \bibinfo {author} {\bibfnamefont {V.}~\bibnamefont {Aji}},\ and\
  \bibinfo {author} {\bibfnamefont {C.~M.}\ \bibnamefont {Varma}},\ }\bibfield
  {title} {\bibinfo {title} {Ordered loop current states in bilayer graphene},\
  }\href {https://doi.org/10.1103/PhysRevB.87.035427} {\bibfield  {journal}
  {\bibinfo  {journal} {Phys. Rev. B}\ }\textbf {\bibinfo {volume} {87}},\
  \bibinfo {pages} {035427} (\bibinfo {year} {2013})}\BibitemShut {NoStop}%
\end{thebibliography}%
\end{document}